\documentclass{aa}  

\usepackage{graphicx}
\usepackage{txfonts}
\usepackage[normalem]{ulem}
\usepackage{xcolor}
\begin{document}

  \title{What can be learnt from UHECR anisotropies observations}

\subtitle{Paper III: Update with new data and Galactic magnetic fields models}

   \author{D. Allard
          \inst{1},
          J. Aublin, 
          B. Baret
         \and E. Parizot
          }

   \institute{Université Paris Cité, CNRS, Laboratoire Astroparticule et Cosmologie, F-75013 Paris, France\\
             \email{julien.aublin@apc.in2p3.fr}
             }

   \date{Received ... ; accepted ... }


 \abstract
{
Recent measurements have revealed several anisotropy signals in the arrival directions of ultra-high-energy cosmic rays (UHECRs). At large angular scales, the Pierre Auger Observatory has reported a significant dipole modulation in right ascension, while at intermediate angular scales, localized flux excesses have been identified by both the Auger and Telescope Array collaborations. These observations were investigated in the first two papers of this series.
}
{
We examine the implications of these anisotropy measurements and assess to what extent they can be used to constrain the origin of UHECRs and the astrophysical or physical parameters of viable source scenarios.
}
{
As in the first two papers of this series, we generate realistic UHECR sky maps for a wide range of astrophysical models consistent with current spectral and composition constraints, assuming that UHECR sources trace the distribution of galaxies in the Universe. We update our previous studies by incorporating the most recent models of the Galactic magnetic field and apply the same large- and intermediate-scale anisotropy analyses as those used by the Auger Collaboration to simulated datasets with current experimental exposure.
}
{
The main novelty of this third paper is the improved compatibility between simulations and data, in particular regarding the reconstructed dipole direction, when using several of the recently proposed Galactic magnetic field models. Despite this progress, our main conclusions remain unchanged: although the observed anisotropies are compatible with an extragalactic origin of UHECRs, present data and magnetic-field uncertainties do not allow strong constraints to be placed on the nature, spatial distribution, or density of UHECR sources.
}
{
Further progress in the interpretation of UHECR anisotropies will require improved constraints on cosmic magnetic fields, advances in source modeling, and significantly larger experimental exposures.
}


   \maketitle
%

\section{Introduction}
After almost 20 years of activity, the Pierre Auger observatory (\citet{AugerObs}, hereafter Auger)  has shown that the ultra-high-energy cosmic-ray (UHECR) arrival direction distribution is genuinely anisotropic above 8~EeV. This observation relies on various analyses released in the past few years. First, a Rayleigh analysis \citep{Linsley1975} performed on the UHECR data set above 4~EeV revealed a dipole modulation of the arrival directions above 8~EeV with a confidence level larger than the 5$\sigma$ discovery threshold (\citet{AugerDip2017} and \citet{AugerDip2024} for the lastest update). Furthermore, a power spectrum decomposition of the same dataset did not show so far any significant anisotropy signal beyond the dipole mode  $\ell=1$ (\citet{AugerMulti2017} and \citet{AugerDip2024} for the lastest update). Second, at higher energies, namely above 32~EeV, searches for flux excesses, either blind or targeted \citep{AugerAni2015}, as well as a likelihood-based study of the correlation between the UHECR skymap and some astrophysical catalogs \citep{AugerSFG2018} have yielded some strong hints of an isotropic signal at intermediate angular scales with a post trial significance of the order of $\sim4\sigma$. Importantly, the targeted search which results in the most significant signal is in the direction of the nearby radiogalaxy CenA and the maximum significance of the blind search is also located very close ($\sim 2^\circ$) to this direction (see \citep{Golup2023} and \citep{AugerAniso2022} for the latest updates).

In the first two papers of this series (\citet{Paper I,Paper II}, hereafter Paper I and Paper II), we modelled UHECR anisotropies assuming that the sources trace the cosmic distribution of galaxies. The source spectrum and mass composition were chosen to reproduce, after propagation, the energy spectrum and composition measured by Auger. Paper I addressed large-scale anisotropies, reproducing the Rayleigh analyses and the angular-power spectrum on simulated datasets. We found that both the amplitude of the dipole and its evolution with energy are easily reproduced within extragalactic source models across a broad and degenerate parameter space—including source properties, extragalactic and Galactic magnetic-field models, and magnetic-field coherence lengths. These observables therefore provide only limited constraints on the nature, distribution, or density of the true UHECR sources. By contrast, the direction of the dipole is much harder to reproduce with current Galactic magnetic-field (GMF) models, especially with the Jansson–Farrar model (\citet{JF2012a,JF2012b}, hereafter JF12). For high source densities (e.g. $\gtrsim 3\times 10^{-4},\mathrm{Mpc^{-3}}$), the predicted dipole directions are incompatible with the Auger observation. Agreement is achieved only in rare cases at low source densities ($\lesssim 10^{-4},\mathrm{Mpc^{-3}}$), which implies substantial cosmic variance and strongly reduces the discriminating power between a galaxy-tracing source population and a randomly distributed one.

When comparing the predictions made for various GMF models, we also made an important observation regarding the importance of magnification and demagnification effects. One of the models we tested, proposed by \citep{Sun2008, Sun2010} and revised following the analysis of the Planck satellite data \citep{GMFPlanck2016}, hereafter referred to as the Sun+Planck GMF model, predicts a strong demagnification of vast zones of the sky over a wide range of UHECR rigidities, in particular in the region of the Virgo cluster. This behaviour contrasts with the predictions with the JF12 model\footnote{In paper I and paper II, we also used a revised version of the JF12 model based on \citep{GMFPlanck2016}, which we referred to as the JF12+Planck model. The quantitative differences between the original and the revised versions are small, and none of the conclusions of these two studies depend on which one is used.}. Given the importance and proximity of the Virgo cluster, whether this specific region is or is not demagnified obviously has an impact on the resulting orientation of the main dipolar modulation of the UHECR sky. However, by comparing the anisotropy predictions for the two GMF models,
we found that the demagnification of the Virgo cluster expected in the Sun+Planck model can be almost perfectly mimicked within the JF12 model by introducing a bias in the UHECR source distribution relative to that of galaxies—specifically, by assuming that UHECR production is disfavoured in rich galaxy clusters. This highlights the importance of magnification and demagnification effects caused by the structure of the GMF (which also underlies some of our findings on intermediate-scale anisotropies discussed below) and shows that interpreting UHECR anisotropies in terms of the nature and distribution of the sources requires a very accurate knowledge of the strength and spatial structure of the GMF.

The intermediate-scale anisotropies above 32 EeV were discussed in Paper II. We found that the amplitudes reported by Auger can again be reproduced relatively easily in our simulations, even for high source densities of order $10^{-3},\mathrm{Mpc^{-3}}$. Both the excess in the direction of Cen A and the location of the blind-search maximum predicted by the simulations are compatible with the observations once the expected positional scatter (of order $\sim 20$–$25^\circ$ at current Auger statistics and anisotropy levels) is taken into account.

The origin of the UHECR events responsible for the flux excess around Cen A, for a given source-distribution hypothesis, depends sensitively on the assumed GMF model. With the JF12 model, we found that the Virgo cluster provides a strong and dominant contribution to the blind-search maximum. This is not the case for the Sun+Planck model, in which the Virgo region is demagnified. Under that GMF model, the excess typically arises from the combined contribution of many sources, each accounting individually for at most $\sim$20\% of the total excess. In addition, the Auger observation is generally best reproduced when Cen A and/or NGC 4945 and/or M83 are included among the assumed sources.

Regarding the Auger maximum-likelihood analysis, the $p$-value obtained for our simulated datasets, as well as the preference for the starburst/star-forming galaxy catalog (hereafter SBG), can be reproduced for various assumptions on the UHECR source distribution, again depending on the GMF model. For the JF12 model, the observed preference for the SBG catalog is more likely to appear in simulations where UHECR sources are absent (or disfavoured) in rich galaxy clusters. In contrast, for the Sun+Planck model, the preference is reproduced without invoking any such bias relative to the galaxy distribution. This behaviour again results from the different magnification/demagnification patterns predicted by the two GMF models.

These findings reinforce the conclusion that uncertainties in the structure and strength of the GMF severely limit our ability to interpret the UHECR sky at the highest energies. Paper II also examined the energy evolution of the anisotropy signal. We noted that applying blind and targeted searches at energies below 32 EeV is expected to reveal more significant signals for models that reproduce the Auger composition trend above the ankle. The measurement of the energy evolution of an anisotropy signal (say, from 10~EeV to the highest energies) was shown to provide important constraints on its origin. However, detecting anisotropies above 80 EeV at meaningful significance will likely require exposures roughly an order of magnitude larger than those currently available.


Interestingly, several new models for the coherent component of the GMF have been proposed over the past year by three different groups: \citet{Xu2024} (hereafter the XH24 model), \citet{Unger2024} (hereafter the UF23 models), and \citet{KTS2024} (hereafter the KST24 model). We note that UF23 does not correspond to a single coherent GMF configuration, but to a suite of models based on different, physically motivated functional forms. In Paper II, we briefly discussed the UF23 family and pointed out that, in stark contrast with the predictions of the JF12 model, most UF23 configurations imply a strong demagnification of the Virgo cluster region for rigidities below $\sim 10,\mathrm{EV}$—a key behaviour more reminiscent of the Sun+Planck model.


In this paper, we revisit the interpretation of the large-scale and intermediate-scale anisotropies reported by Auger, updating our previous studies by producing realistic simulated sky maps using the newly proposed GMF models mentioned above, and accounting for the increased Auger statistics accumulated since Papers I and II. In the next section, we briefly review the models and the procedure used to generate the simulated datasets, and we summarize the anisotropy analyses applied to the resulting UHECR sky maps. Sections 3 and 4 present our predictions for various combinations of astrophysical model parameters and compare them with Auger observations. We then discuss our results and conclude.

\section{UHECR skymap simulations and analyses}


In Paper I, the ingredients of our astrophysical model were presented in detail in Sects.~2–5, including the physical parameters and the numerical tools used to propagate cosmic rays through photon backgrounds and cosmic magnetic fields. We refer the reader to that paper for full details (see also \citealt{BRDO2014}) and briefly review here the main astrophysical inputs of our simulations.

\subsection{Model parameters}



Constructing realistic UHECR datasets at Earth requires several assumptions regarding: (i) the cosmic-ray output at each source, including composition, spectral shape, and possible time evolution; (ii) the spatial distribution of sources; and (iii) the cosmic magnetic fields through which the particles propagate.
To limit the number of free parameters, we assume a single, universal source composition for all sources. The relative abundances of nuclear species, the spectral shape, and the maximum rigidity correspond to those of Model~A (Table~1 of Paper I), which was shown to reproduce the UHECR spectrum and mass composition measured by Auger.

\subsection{Source distribution}


For the UHECR source distribution, we assume that sources follow the distribution of galaxies. Individual realizations are drawn from the 2MRS catalog \citep{Huchra2012} using the ``mother catalog'' approach described in detail in Paper I and summarized here. The mother catalog corresponds to the highest-density, volume-limited catalog we could construct from 2MRS (namely $7.6\times 10^{-3},\mathrm{Mpc^{-3}}$). It is obtained by applying a luminosity cut in the $K_\mathrm{s}$ band at $10^{10},L_\odot$ (with $L_\odot$ the solar luminosity in this band), significantly below both the luminosity of the Milky Way and the characteristic galaxy luminosity parameter $L^\star$ of the Schechter luminosity function \citep{Schechter1976}. With this cut, the 2MRS catalog is complete out to $D_{\rm max}\simeq 40$,Mpc.
Beyond this distance, we complete the mother catalog using the 3D matter distribution of the large-scale structure simulations (LSSS) of \citet{LSSS2018}, which were constrained by the Cosmicflows-2 peculiar-velocity catalog \citep{CF22014}. Once the mother catalog is assembled, we generate many realizations (typically 300 in the examples below) for a given source density by randomly sub-sampling the catalog, selecting each galaxy with a probability equal to the ratio of the desired source density to the mother-catalog density.


We note in passing that a model based solely on the constrained LSSS (as adopted, e.g., in \citealt{Ding2021,Bister2025}) may seem simpler to implement than our hybrid approach. However, a pure-LSSS strategy is not suited to our purposes for several reasons. First, the LSSS describe the full 3D matter distribution, dominated by dark matter, which is known to be less clustered than the galaxy distribution (see \citealt{No2019}). Second, although the LSSS reproduce the statistical properties of the large-scale matter distribution in the local Universe, they are not designed to reproduce the immediate environment of the Milky Way in detail—an essential ingredient for UHECR anisotropies. For this reason, nearby groups containing prominent galaxies such as Cen~A, NGC~253, M82, Circinus, or structures such as the Fornax cluster are not represented with sufficient fidelity. Even the Virgo cluster, while present in the LSSS, is offset by several Mpc from its actual position \citep{LSSS2018}.


Our hybrid approach avoids these caveats by constructing a dense mother catalog based entirely on real galaxies (outside the zone of avoidance close to the Galactic plane) out to distances encompassing the local supercluster. As a consequence, anisotropy predictions based purely on LSSS differ numerically from those obtained here. For a fixed source density and identical physical parameters, anisotropies are generally smaller when using only the LSSS, particularly for multipoles with $\ell>1$. At high source densities (e.g. $\sim 10^{-3},\mathrm{Mpc^{-3}}$), the cosmic variance is overestimated in the pure-LSSS approach because local matter overdensities near the Galaxy are missing, and the predicted dipole direction also shifts. Finally, our approach, being based on real nearby galaxies, naturally allows us to analyze the respective roles of individual sources and of large-scale structures in shaping the overall UHECR anisotropy.

\begin{figure*}
   \centering
   \includegraphics[width=8.5cm]{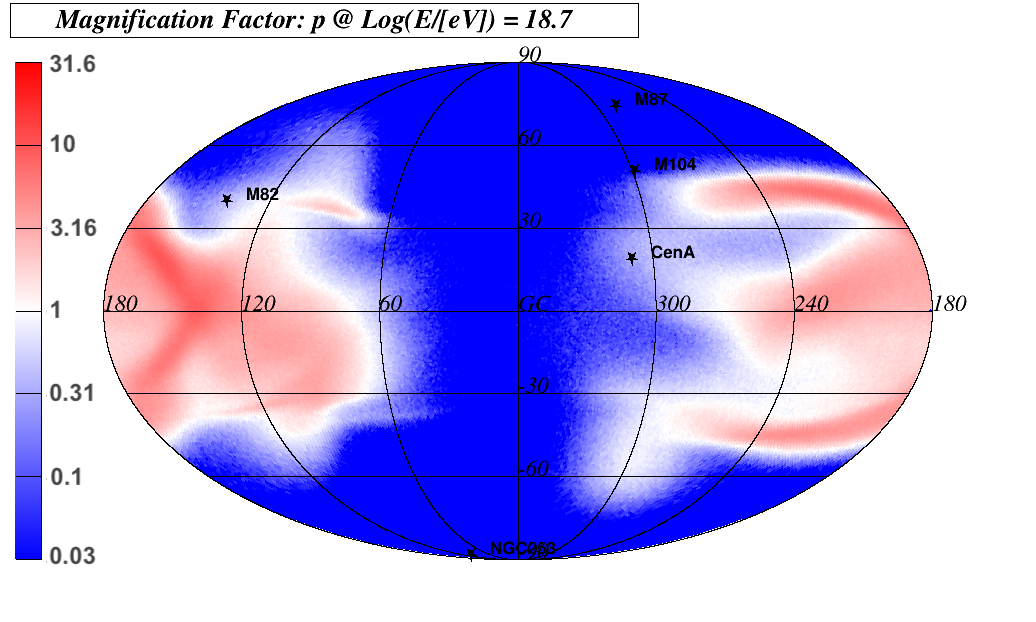}
   \includegraphics[width=8.5cm]{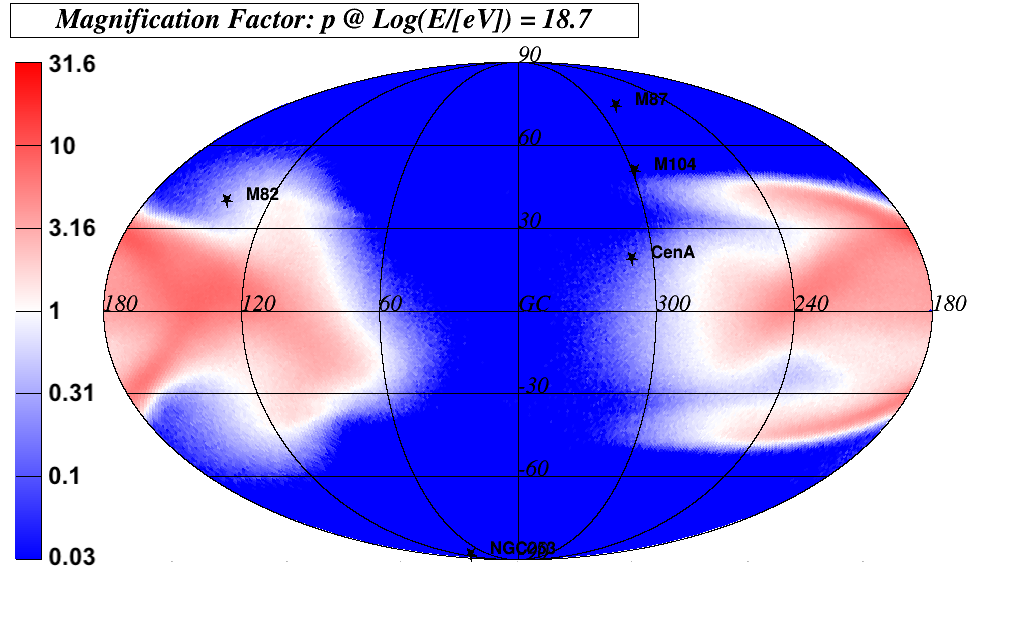}
   \includegraphics[width=8.5cm]{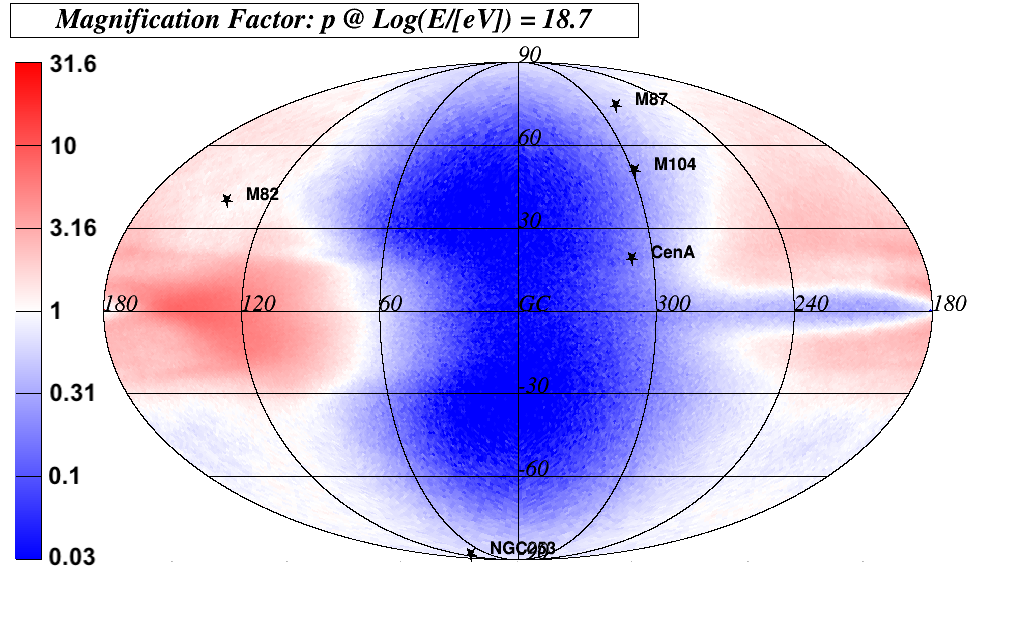}
   \includegraphics[width=8.5cm]{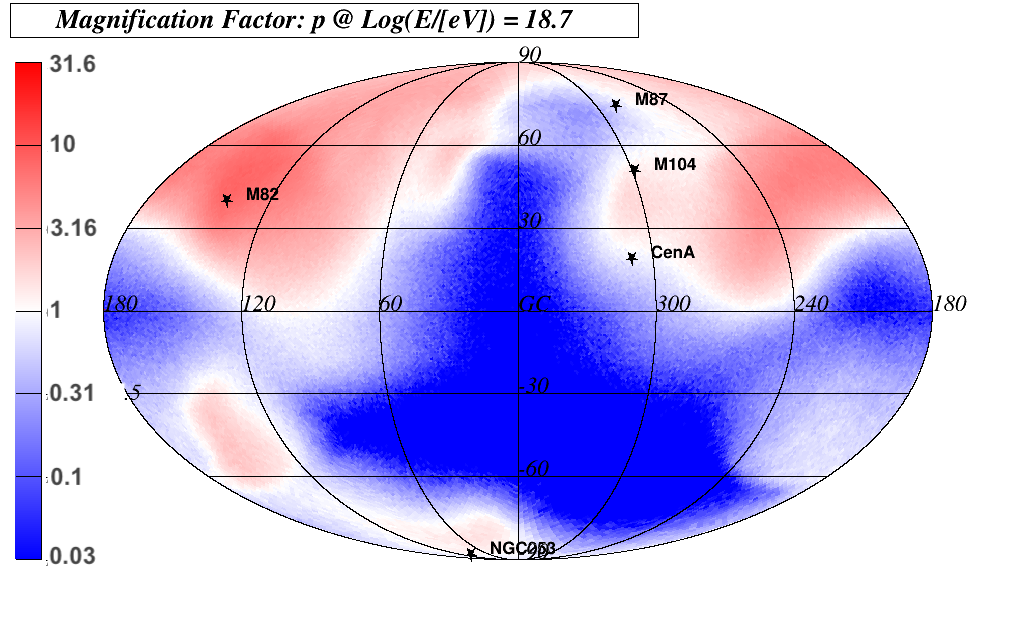}
    \includegraphics[width=8.5cm]{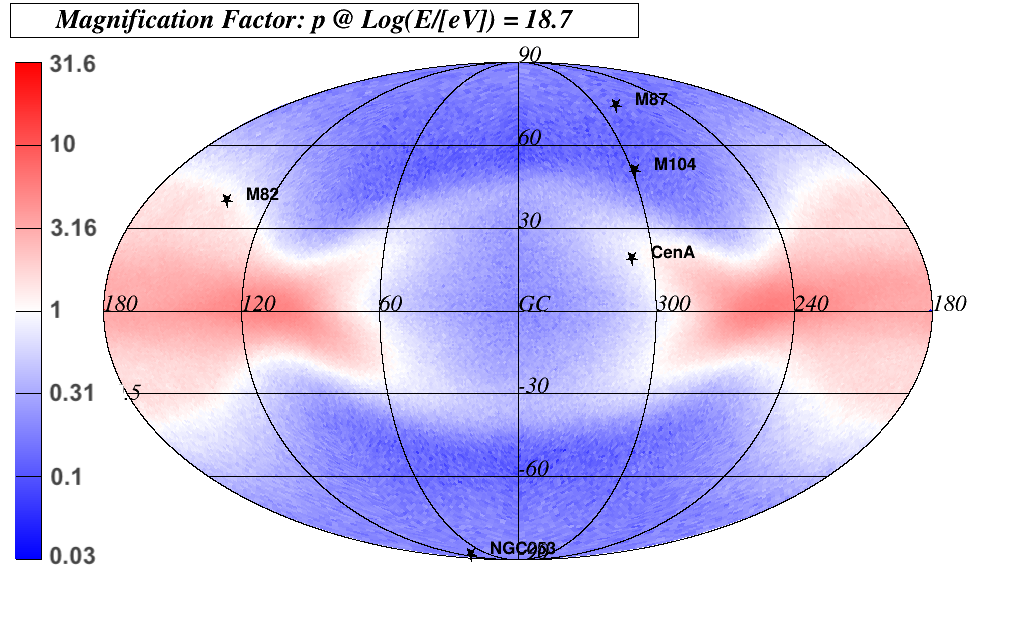}
     \includegraphics[width=8.cm]{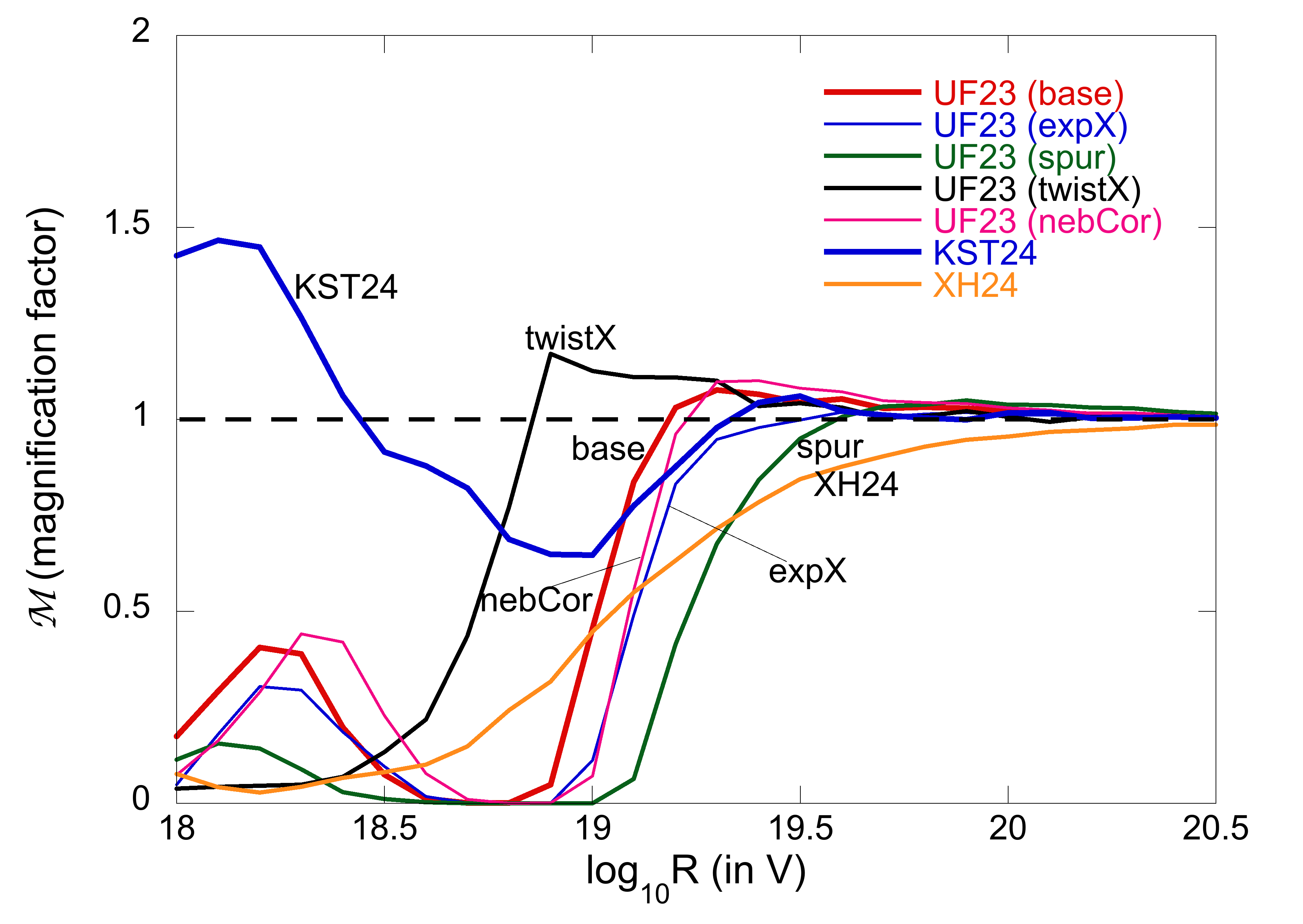}
      \caption{Top-left to bottom-left : Magnification maps in Galactic coordinates for a 5 EV rigidity assuming various GMF models for the regular component coupled to the JF12+Planck turbulent component. The GMF models shown are (from left to right and top to bottom) : UF23 base, UF23 spur, UF23 (twistX), KST24 and XH24. The turbulent GMF coherence lengths assumed are $\lambda_{\rm c}$=50~pc for UF23 base, UF23 spur and XH24, $\lambda_{\rm c}$=100~pc for KST24 and $\lambda_{\rm c}$=200~pc for UF23 (twistX). Bottom-right : rigidity evolution of the magnification factor in the region of the Virgo cluster for various models of the GMF regular component (see legend) coupled to JF12+Planck model for the turbulent component.  
              }
         \label{Magnif}
   \end{figure*}

\subsection{Magnetic fields}


Concerning extragalactic magnetic fields, and in the absence of strong constraints on their intensity, structure, and coherence length outside the cores of rich galaxy clusters, we retain the same simple hypothesis as in Paper I (see discussion in Sect.~3) and Paper II. Namely, we assume a homogeneous, purely turbulent EGMF with various intensities. Most of the cases considered in this paper use an r.m.s. field strength of 1~nG and a coherence length of 200~kpc.

For the GMF, we implemented the recent magnetic-field models mentioned above (UF23, XH24, and KST24) for the regular component of the field, and added a turbulent component similar to the one used in Papers I and II for the JF12+Planck model, as estimated in \citet{GMFPlanck2016}. For this turbulent component, we tested coherence lengths between 20 and 500~pc. In the case of the UF23 suite of GMF models, we implemented all eight models, but only the base, expX, Spur, twistX, and nebCor configurations were combined with a turbulent component and will be discussed in detail in the following.

As in Paper I (see Sect.~3.3 for details), we computed magnification maps for these various GMF models. The magnification maps obtained for a rigidity of 5~EV are shown in Fig.~\ref{Magnif}. The two models displayed in the top panel (UF23 base and Spur) exhibit many similarities (the same is true for the expX and nebCor models, not shown in the figure) and are expected to yield comparable magnification patterns for prominent nearby structures and potential sources such as Cen~A, M82, or the Virgo cluster for cosmic rays with rigidities close to 5~EV. By contrast, the three other GMF models displayed (UF23 twistX, KST24, and XH24) show markedly different behaviours.

As we strongly emphasized in Papers I and II, the predicted magnification or demagnification of specific regions of the sky (and its evolution with rigidity)—in particular that of the Virgo cluster—is a key ingredient for understanding and interpreting the observed anisotropy patterns. To better quantify the differences between the models, we computed the rigidity evolution of the magnification factor in the region of the Virgo cluster.\footnote{More precisely, the magnification factors were calculated for the HEALPix pixel containing the center of the Virgo cluster, using a grid with NSide=4, whose angular size is comparable to that of Virgo as seen from Earth.} The result is shown in the bottom-right panel of Fig.~\ref{Magnif} for the models mentioned above (see legend). This figure is very similar to Fig.~20 of Paper II, except that the Sun+Planck and JF12+Planck models have been replaced by the KST24 and XH24 models, and the UF23 models are now combined with a turbulent component (which has only a small quantitative impact).

The UF23 models predict a strong demagnification of the Virgo region below $\sim 10^{19}$~V, except for the twistX model, which shows an earlier recovery above $\sim 4$~EV. The XH24 model also predicts a strong demagnification of this part of the sky, whereas the KST24 model yields only a mild demagnification between $\sim 3$ and $10$~EV, similar to what was found for the JF12+Planck model in Papers I and II. We note, however, that although the rigidity dependence of the magnification factor in the Virgo region is qualitatively similar for JF12+Planck and KST24, the corresponding angular deflections differ significantly, as we will discuss below.

\begin{figure*}
   \centering
   \includegraphics[width=8.2cm]{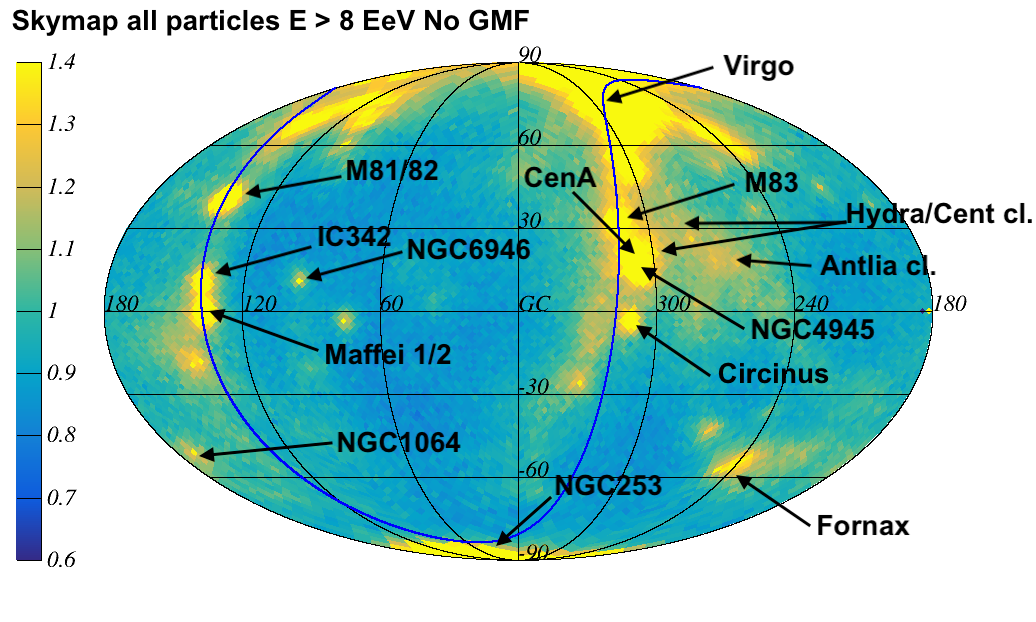}
   \includegraphics[width=8.5cm]{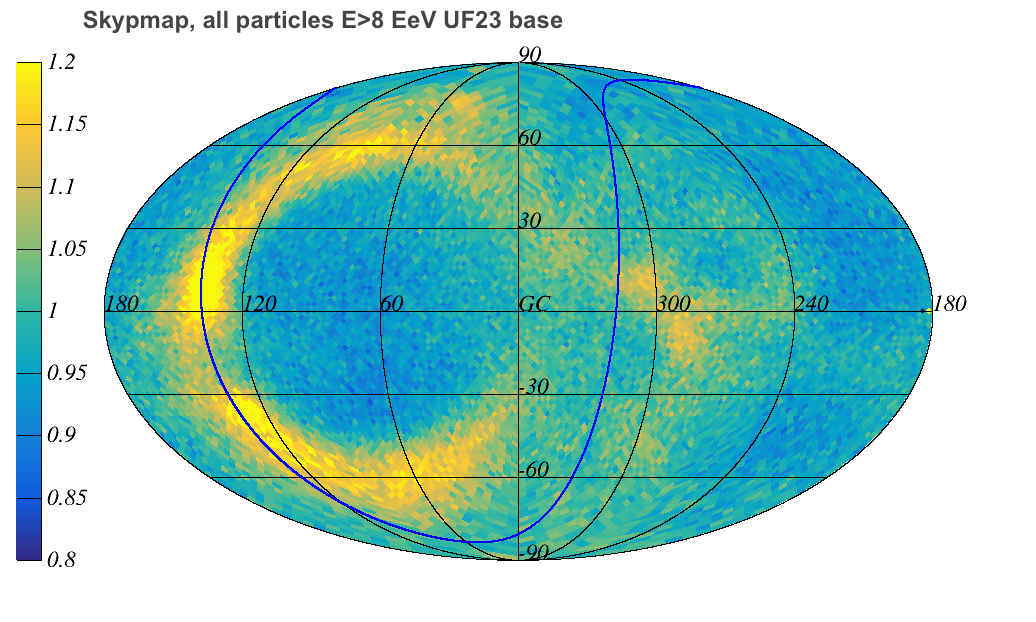}
   \includegraphics[width=8.5cm]{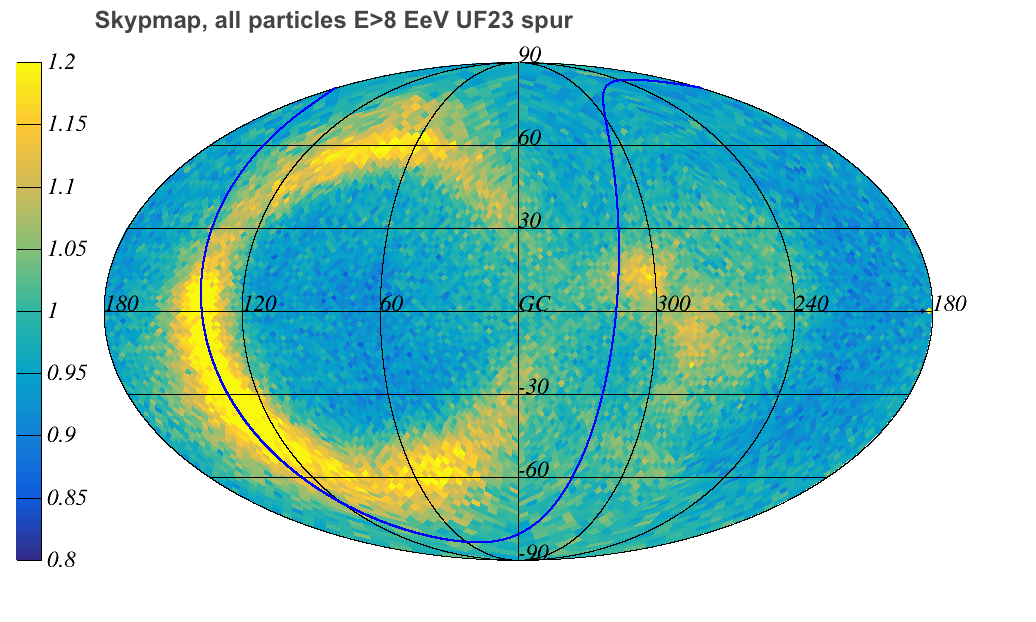}
   \includegraphics[width=8.5cm]{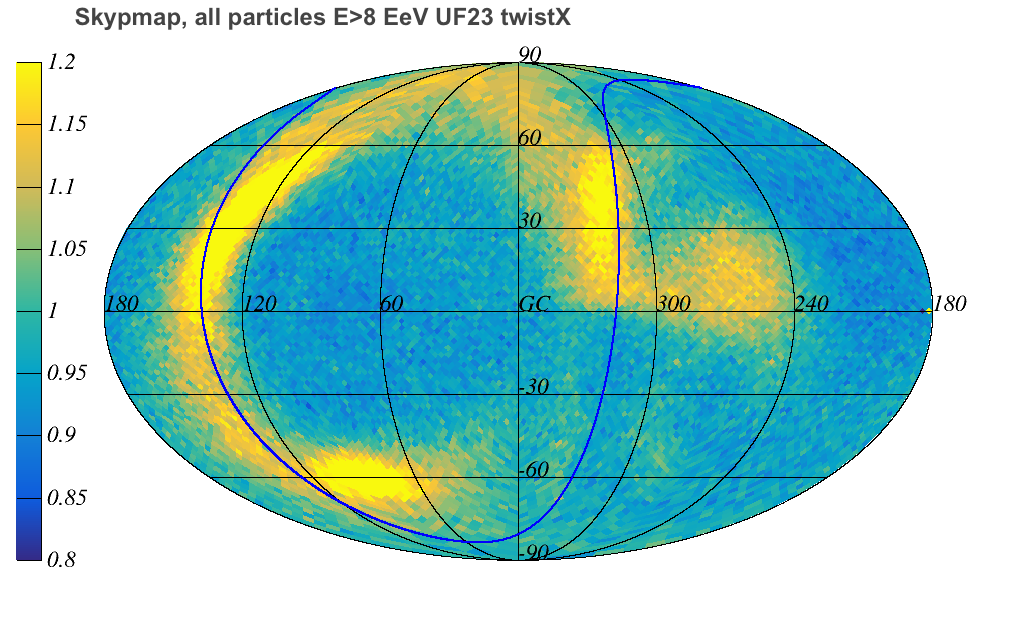}
    \includegraphics[width=8.5cm]{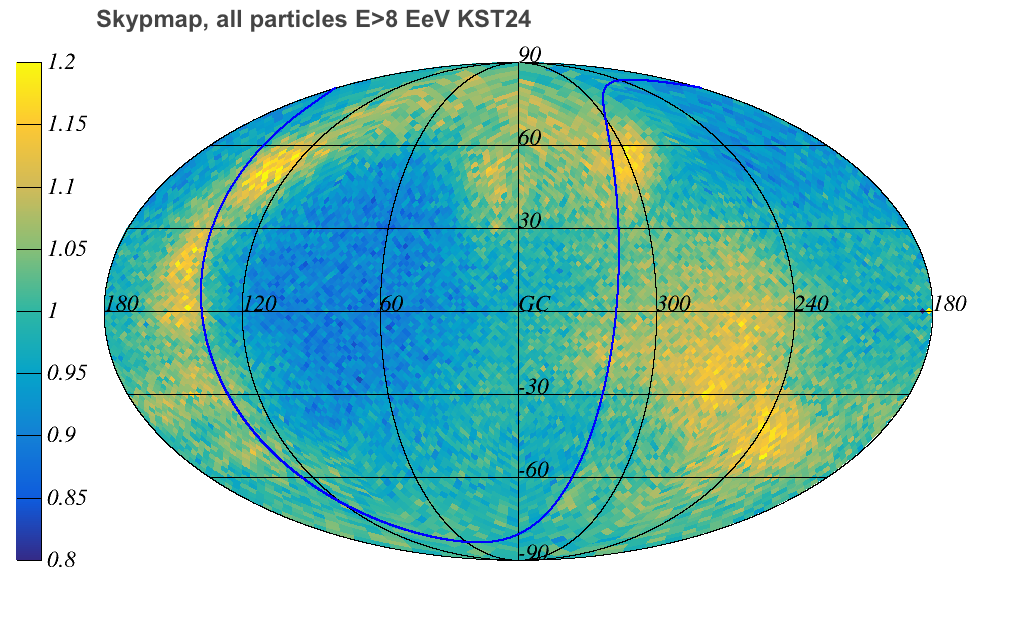}
     \includegraphics[width=8.5cm]{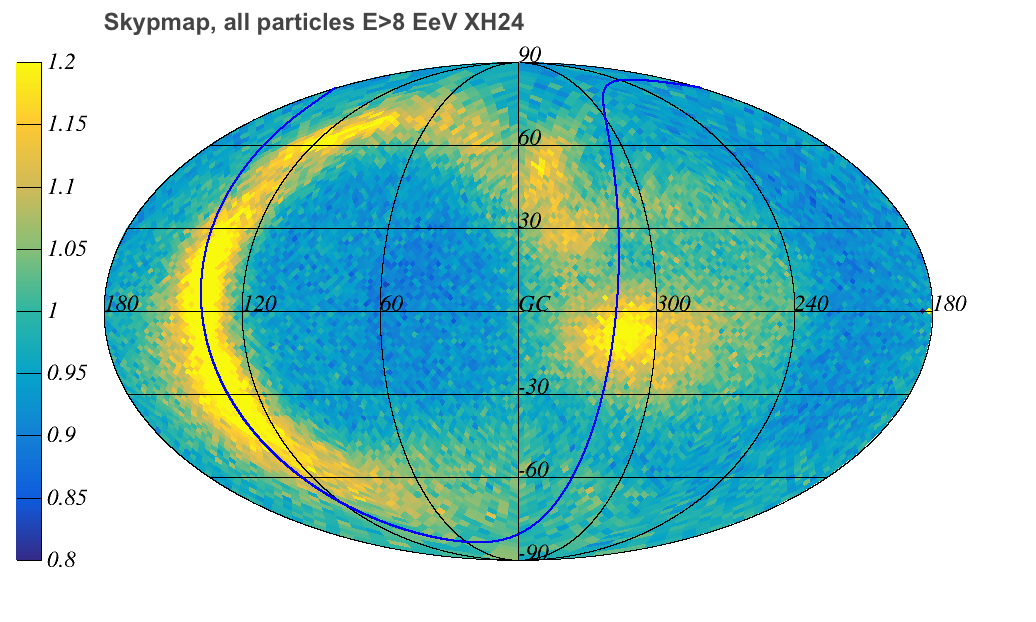}
      \caption{Full sky density maps of UHECR arrival directions in Galactic coordinates, the assumed source distribution is that of the mother catalog ($\rho\simeq8\times 10^{-3}\,\rm Mpc^{-3}$). The color scale is linear and the value 1 represent the average density calculated over the whole skymap. For all the maps a 1~nG EGMF with $\lambda_{\rm c}$=200~kpc is assumed, the various panels show the UHECR skymaps obtained for various assumptions on the GMF model, from left to right and from top to bottom : No GMF (the labels show various prominent galaxies and structures of the nearby Universe), UF23 base, UF23 spur, UF23 twistX, KST24 and XH24. The coherence lengths of the turbulent component of the GMF are the same as those mentioned in Fig.~\ref{Magnif}. 
              }
         \label{SkyMaps}
   \end{figure*}

\subsection{UHECR skymaps}

Using the physical assumptions reviewed above and the numerical tools presented in Paper~I (Sects.~2 and 3, and references therein), we constructed full-sky UHECR maps above 8~EeV with a very large number of simulated trajectories, ensuring that statistical variance is negligible. The source distribution is drawn from the mother catalog, and we generate skymaps for various hypotheses regarding the GMF.

The top-left panel of Fig.~\ref{SkyMaps} shows the UHECR skymap expected in the absence of a GMF, assuming a 1~nG EGMF with coherence length $\lambda_{\rm c}=200$~kpc. The most prominent nearby galaxies and galaxy clusters are indicated by arrows and labels. The following panels display the skymaps obtained for the different GMF models. Even in this idealized configuration, with uniform full-sky exposure and effectively infinite statistics, the maps illustrate the difficulty of interpreting UHECR anisotropies. While all maps are clearly anisotropic, the locations of the regions with the highest concentrations of arrival directions depend sensitively on the GMF model and show no simple correspondence with the underlying UHECR source distribution.

As already noted when discussing Fig.~\ref{Magnif}, the skymaps produced with the UF23 base and Spur models (and similarly with the expX and nebCor models, not shown) are very similar, although differences appear upon closer inspection. Only two of the GMF models—UF23 twistX and KST24—predict a significant contribution from sources within the Virgo cluster to the overall anisotropy. These contributions, however, differ markedly from one another. In the KST24 case, UHECRs originating from the Virgo cluster account for a substantial fraction of the concentration of arrival directions in the southern Galactic hemisphere.

\subsection{Anisotropy analyses}


To evaluate the compatibility between model predictions and observations, one must take into account the partial sky coverage and limited statistics of current UHECR experiments. In the following, we consider only the Auger anisotropy analyses, since these are the only ones that have reached, or are close to reaching, the $5\sigma$ discovery threshold. For our simulated skymaps we assume the latest Auger exposures, namely $123{,}000~\mathrm{km^2\,sr\,yr}$ for the large-scale anisotropy analyses above 4~EeV and $135{,}000~\mathrm{km^2\,sr\,yr}$ for the highest-energy analyses above 32~EeV. These exposures are respectively $\sim$1.6 and $\sim$1.3 times larger than those assumed in Papers I and II. As discussed in those papers, such moderate increases in statistics are not expected to significantly modify the conclusions; the main new elements in the present study arise instead from the recently released GMF models.


For the study of large-scale anisotropies, we reproduce on our simulated datasets (above 4~EeV) the same analyses applied to Auger data: the Rayleigh dipole analysis \citep{AugerDip2017, AugerDip2024}; the combined dipole+quadrupole Rayleigh analysis (formalism in \citealt{AugerDipQuad2015}, latest update in \citealt{AugerDip2024}); and the power-spectrum decomposition \citep{AugerMulti2017}. At higher energies, to investigate small- and intermediate-scale anisotropies, we reproduce the blind and targeted (Cen~A) searches \citep{AugerAni2015}, the likelihood analysis \citep{AugerSFG2018}, and the recent search for excesses near the supergalactic plane (SGP) \citep{AugerSGP2025}. Our implementation of these analyses was described in Papers I and II. We note that the SGP analysis is essentially a straightforward extension of the blind-search (BS) method, which we briefly summarise below.

\begin{figure*}
   \centering
   \includegraphics[width=8.5cm]{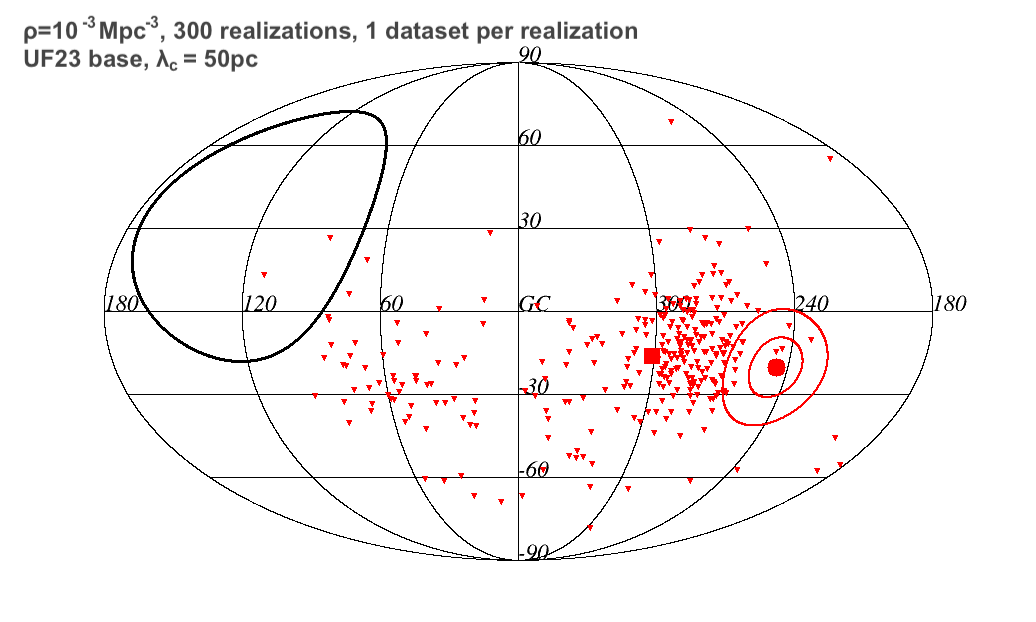}
   \includegraphics[width=8.5cm]{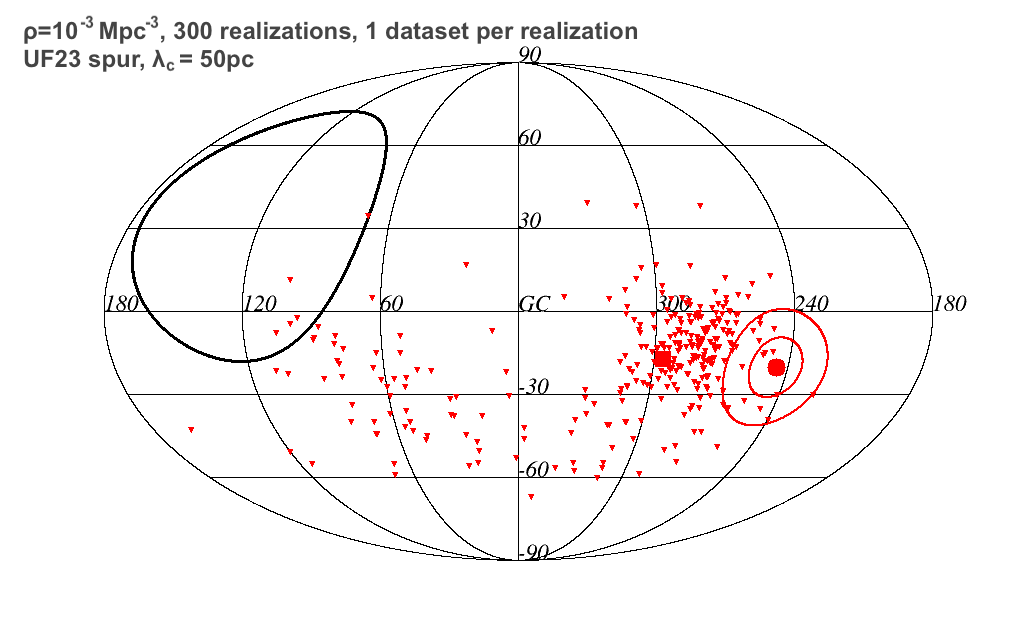}
   \includegraphics[width=8.5cm]{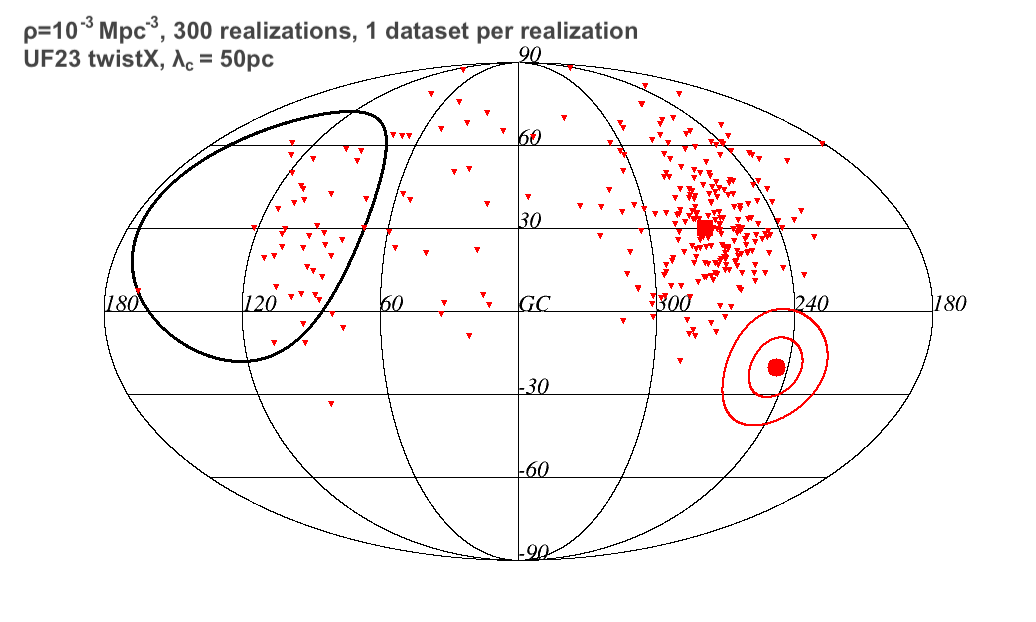}
   \includegraphics[width=8.5cm]{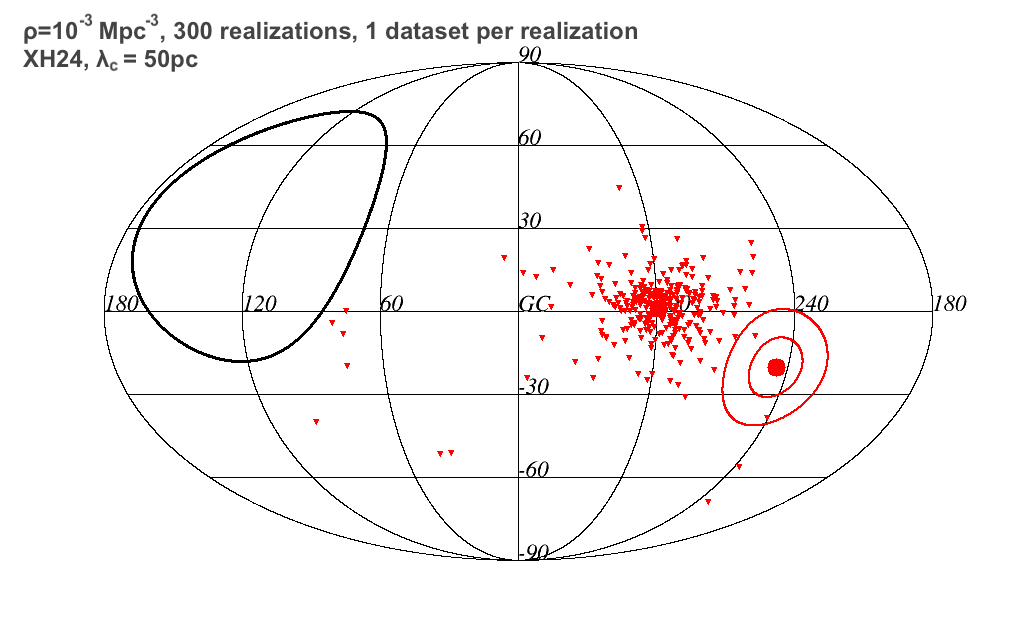}
    \includegraphics[width=8.5cm]{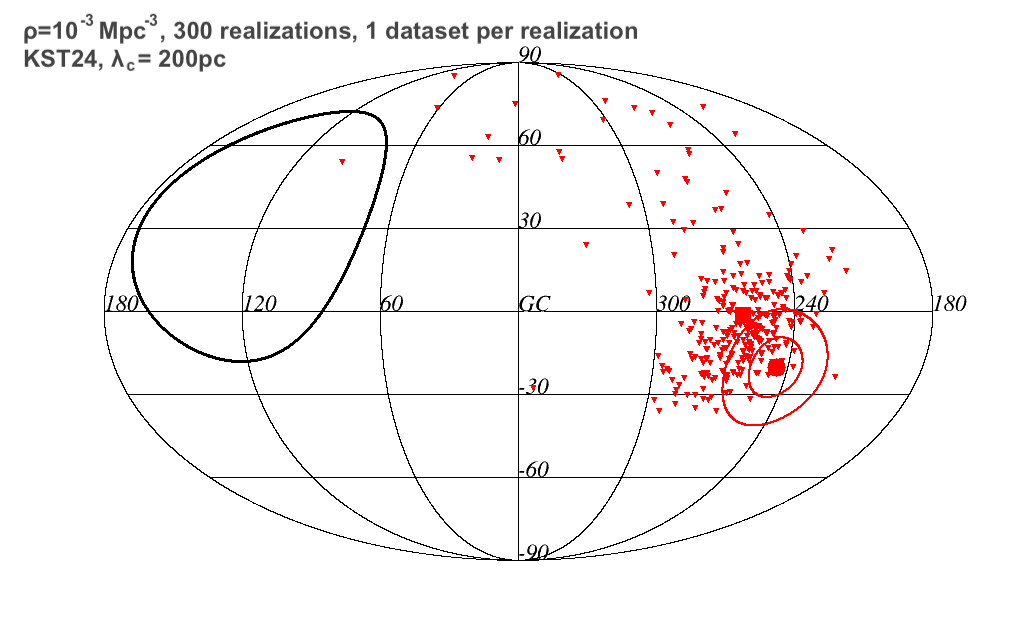}
     \includegraphics[width=8.5cm]{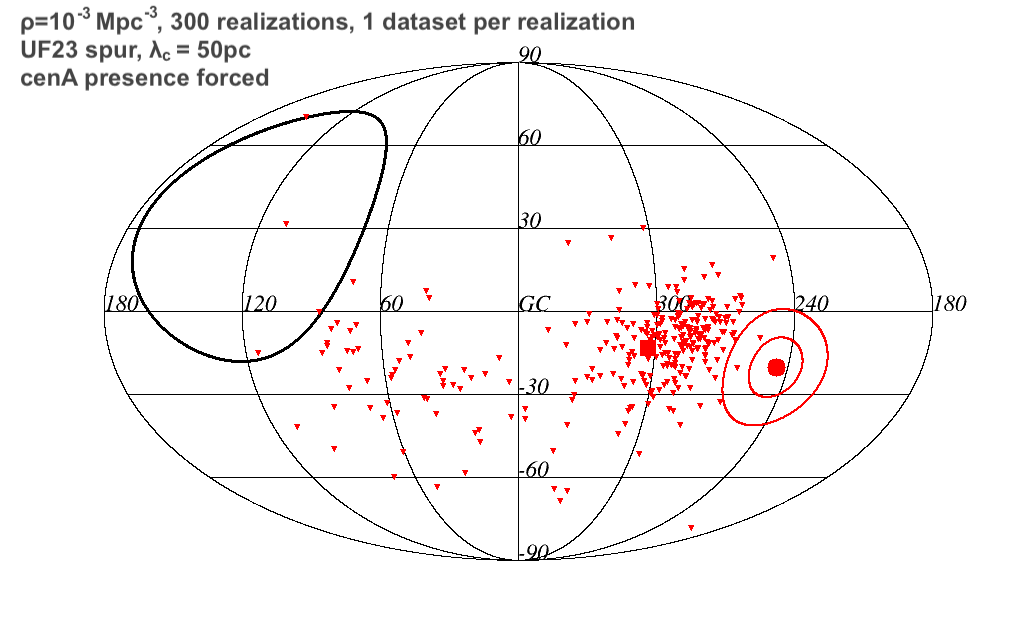}
     
      \caption{Location of the dipole reconstructed for each of the 300 different source distributions drawn from the mother catalog. The barycenter of the predictions is indicated with a large square marker. Auger data are shown with a large red circle marker surrounded by its 1 and 2$\sigma$ error ellipses. Each panel represents a different combination of assumed source distribution and GMF models (see legend). The map shown are in galactic coordinates. 
              }
         \label{dipoleLoc}
   \end{figure*}

\begin{figure*}
   \centering
     \includegraphics[width=8.5cm]{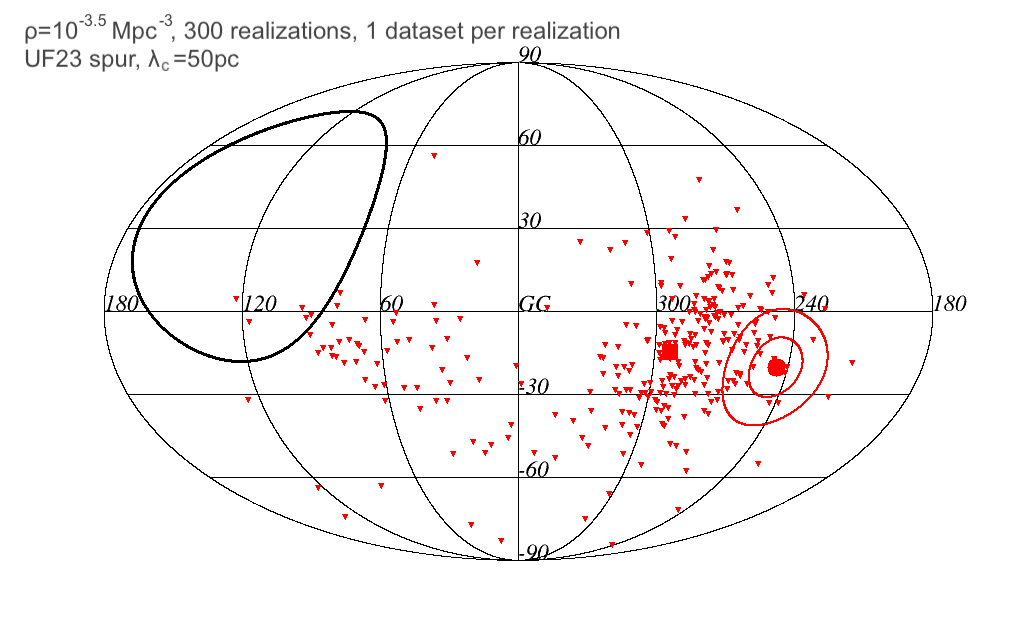}
     \includegraphics[width=8.5cm]{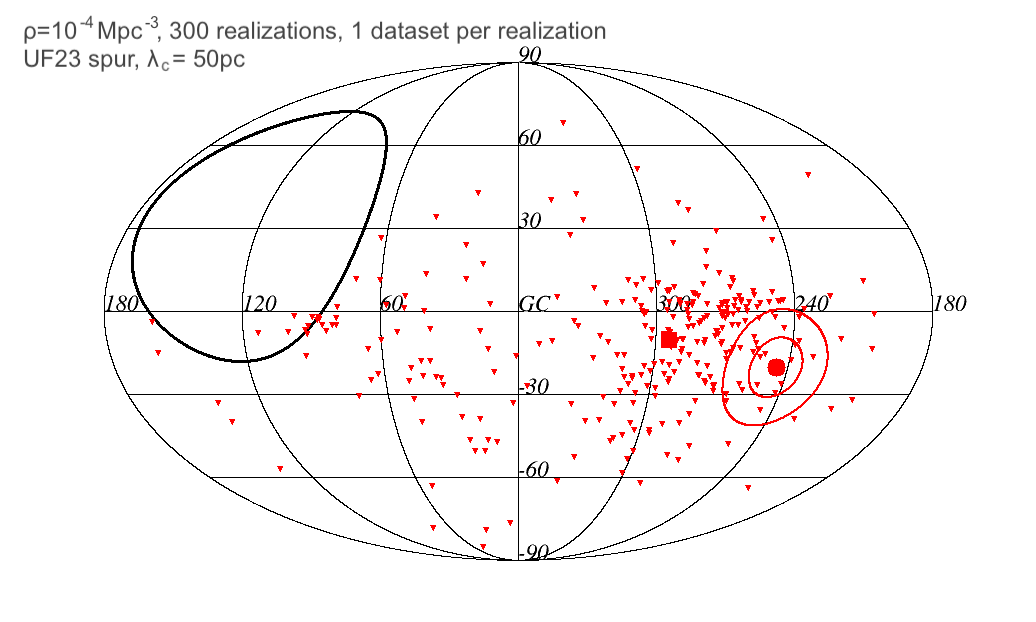}
      \caption{Same as Fig.~\ref{dipoleLoc}, the two displayed cases correspond to lower source density hypotheses (see legend).  
              }
         \label{dipoleLoc2}
   \end{figure*}


Regarding TA anisotropy analyses, the TA hotspot initially reported in \citet{TASpot2014} has not increased in significance in recent years \citep{TAICRC2021}. One proposed interpretation of this potential signal was an association with the starburst galaxy M82 (e.g. \citealt{He16, Pfeffer16}). We simply note here that none of the scenarios we investigated support such an association: for none of the recent GMF models tested do we predict a maximum excess in the region of the sky where TA reports the hotspot when M82 is assumed to be among the nearby UHECR sources.


\begin{figure*}
   \centering
   \includegraphics[width=8.5cm]{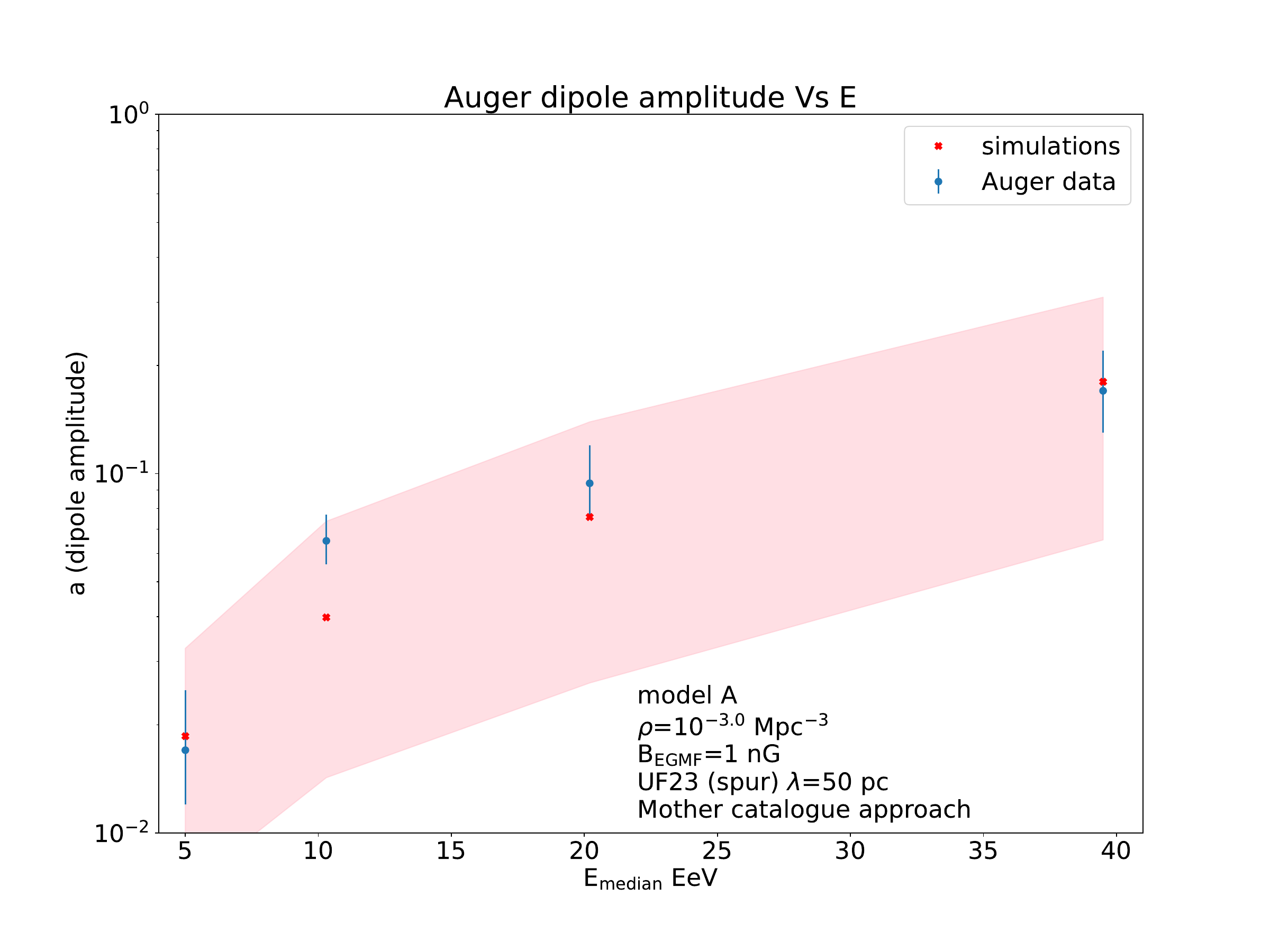}
   \includegraphics[width=8.5cm]{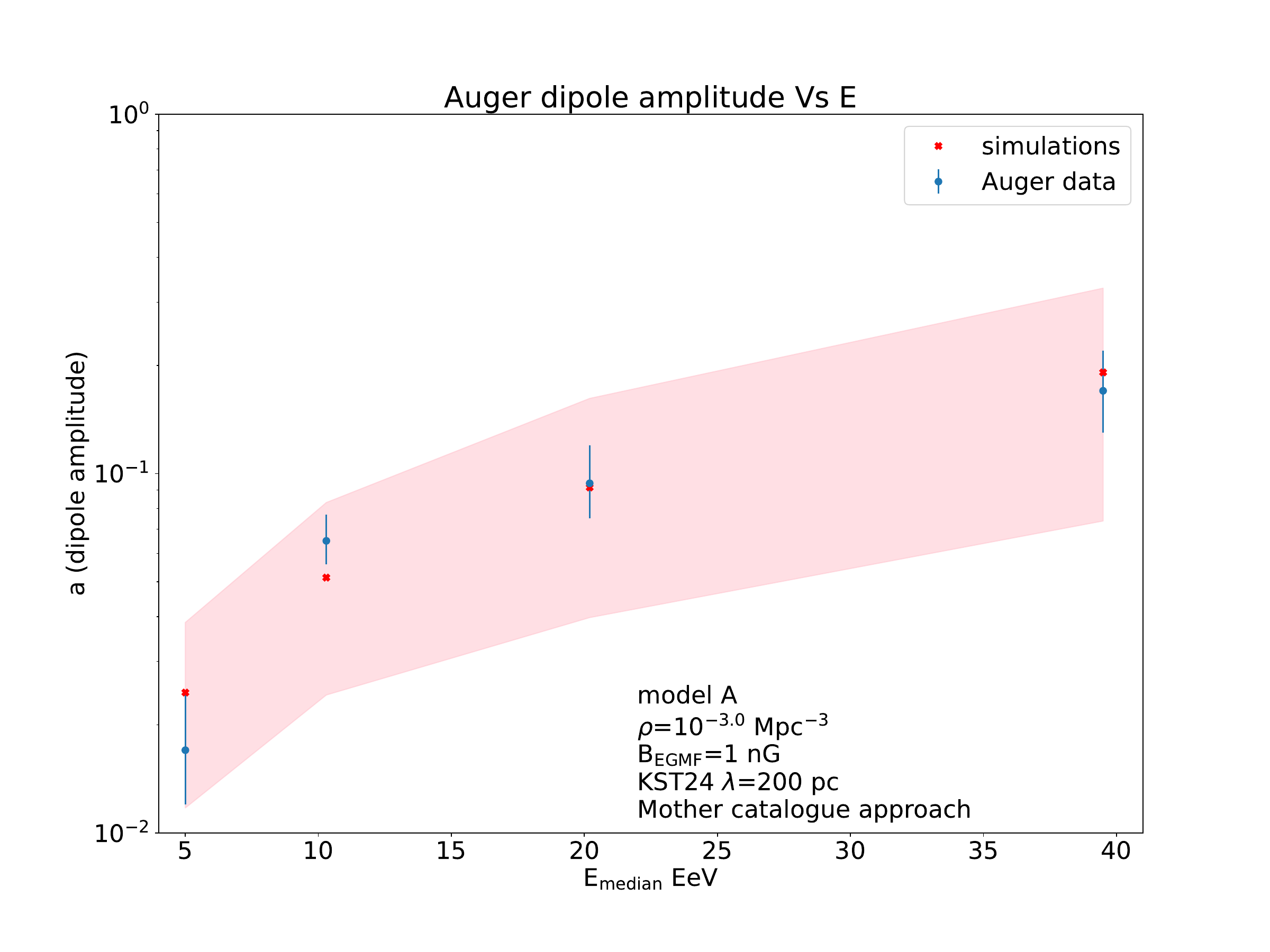}
   \includegraphics[width=8.5cm]{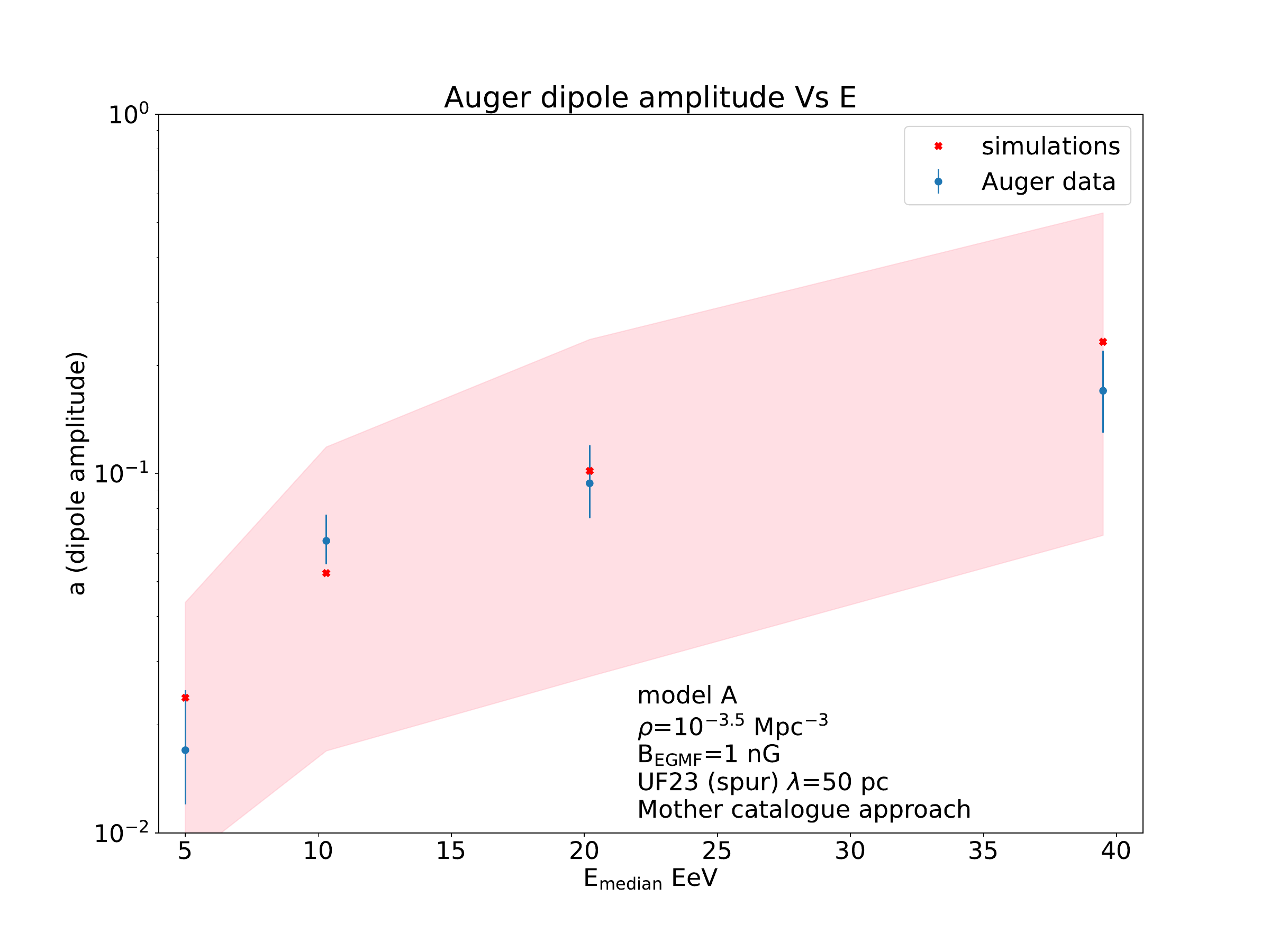}
   \includegraphics[width=8.5cm]{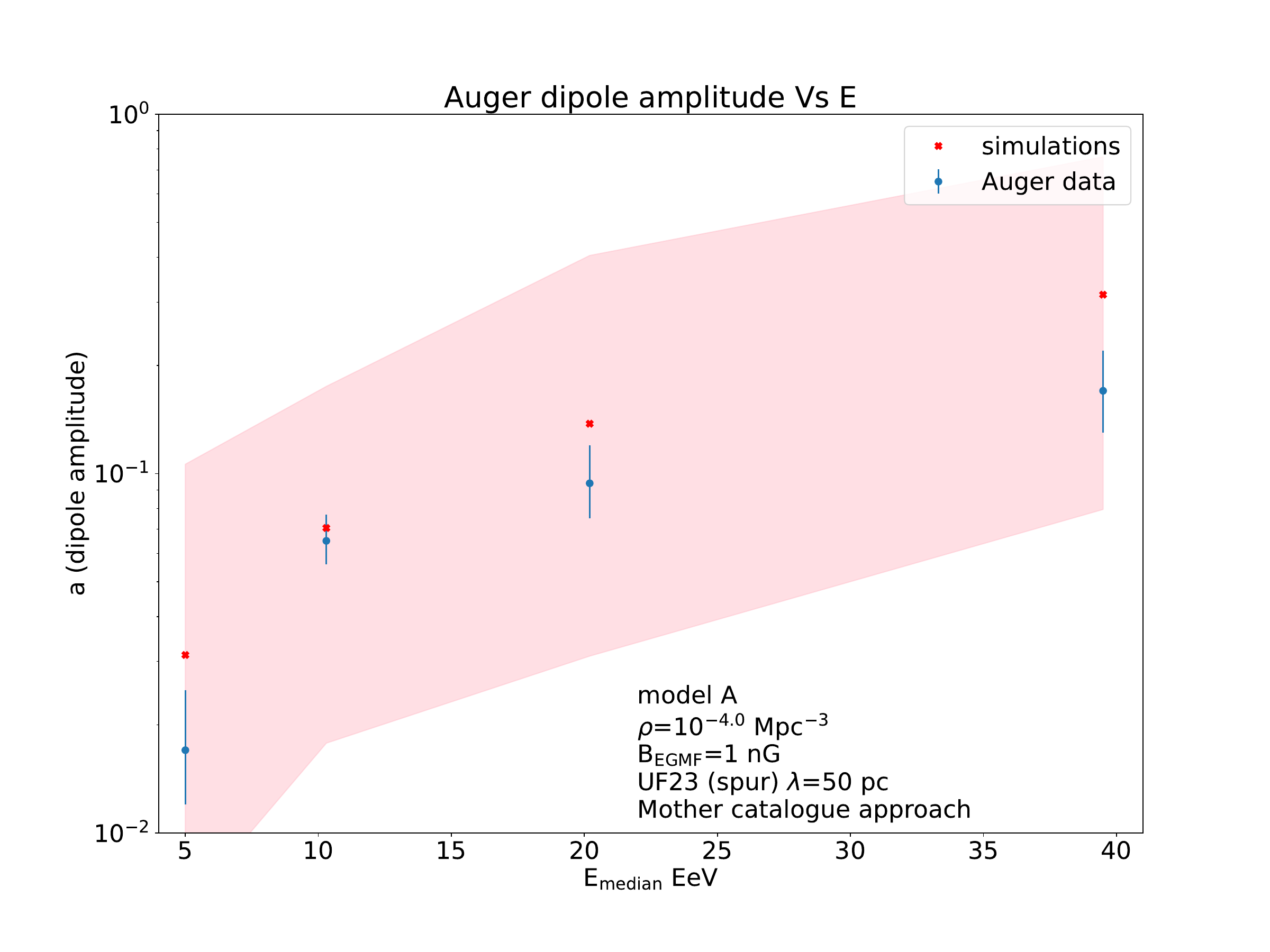}

      \caption{Energy evolution of the dipole amplitude predicted for our simulation and compared with Auger data (shown in blue with their error bars). The red marker shows the mean  value (calculated over the 300 different realizations) obtained for the simulations and the shaded area shows the range in which 90\% of the simulations are found. The four panels correspond to some of the models displayed in Figs.~\ref{dipoleLoc} and \ref{dipoleLoc2} (see legend and text).
              }
         \label{DipAmp}
   \end{figure*}

\begin{figure*}
   \centering
   \includegraphics[width=7.cm]{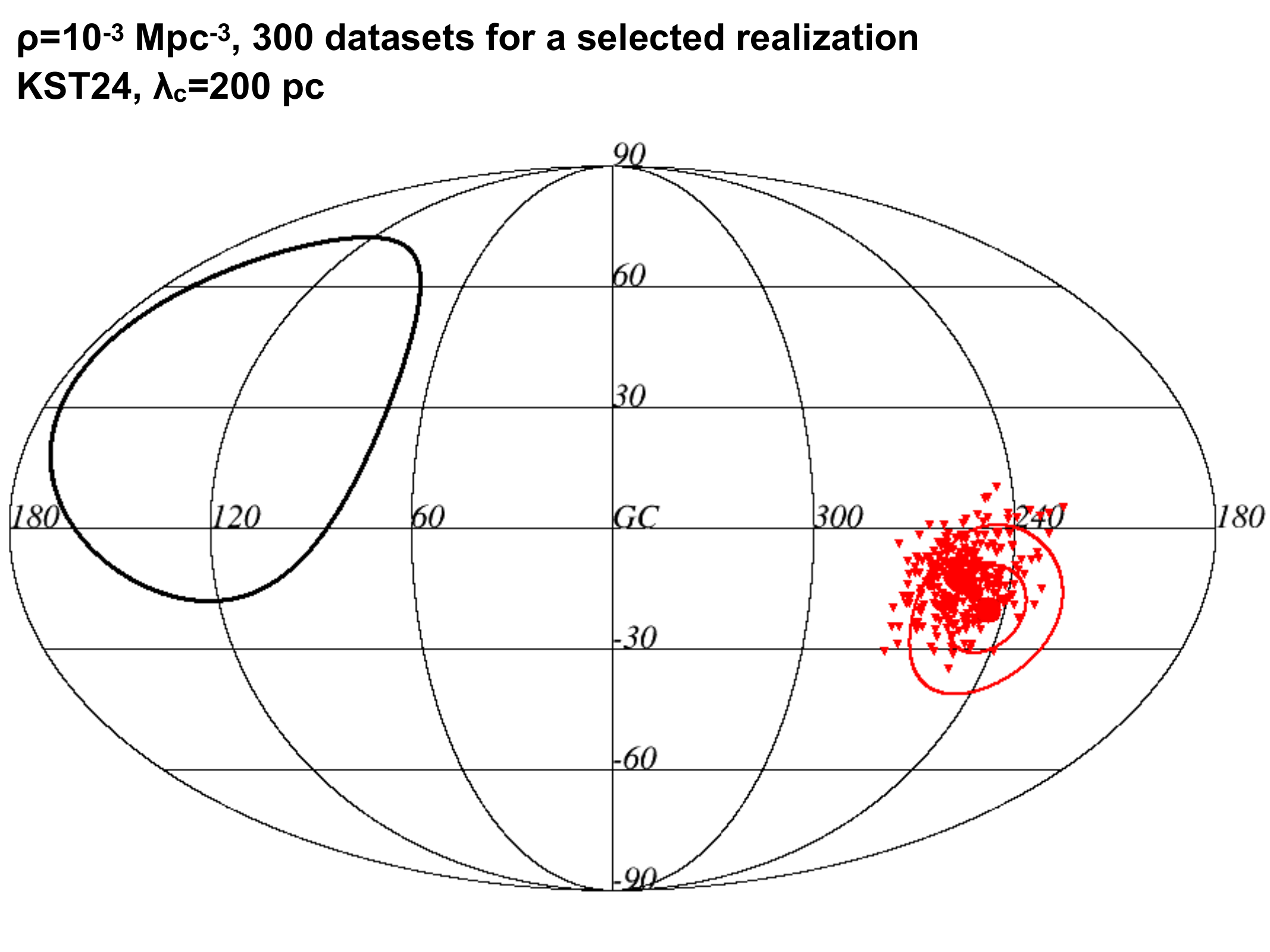}
   \includegraphics[width=7.5cm]{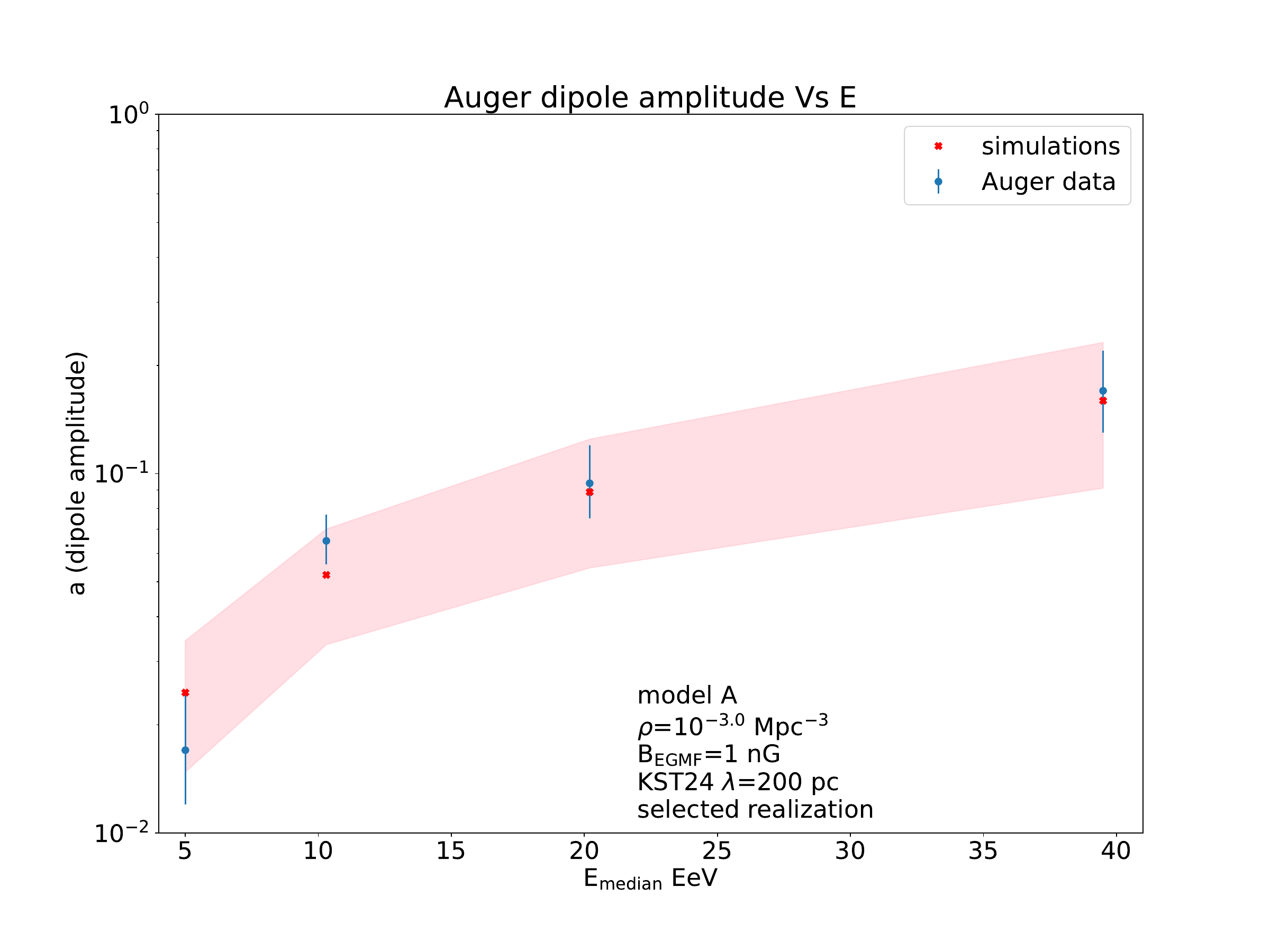}
   \includegraphics[width=7.cm]{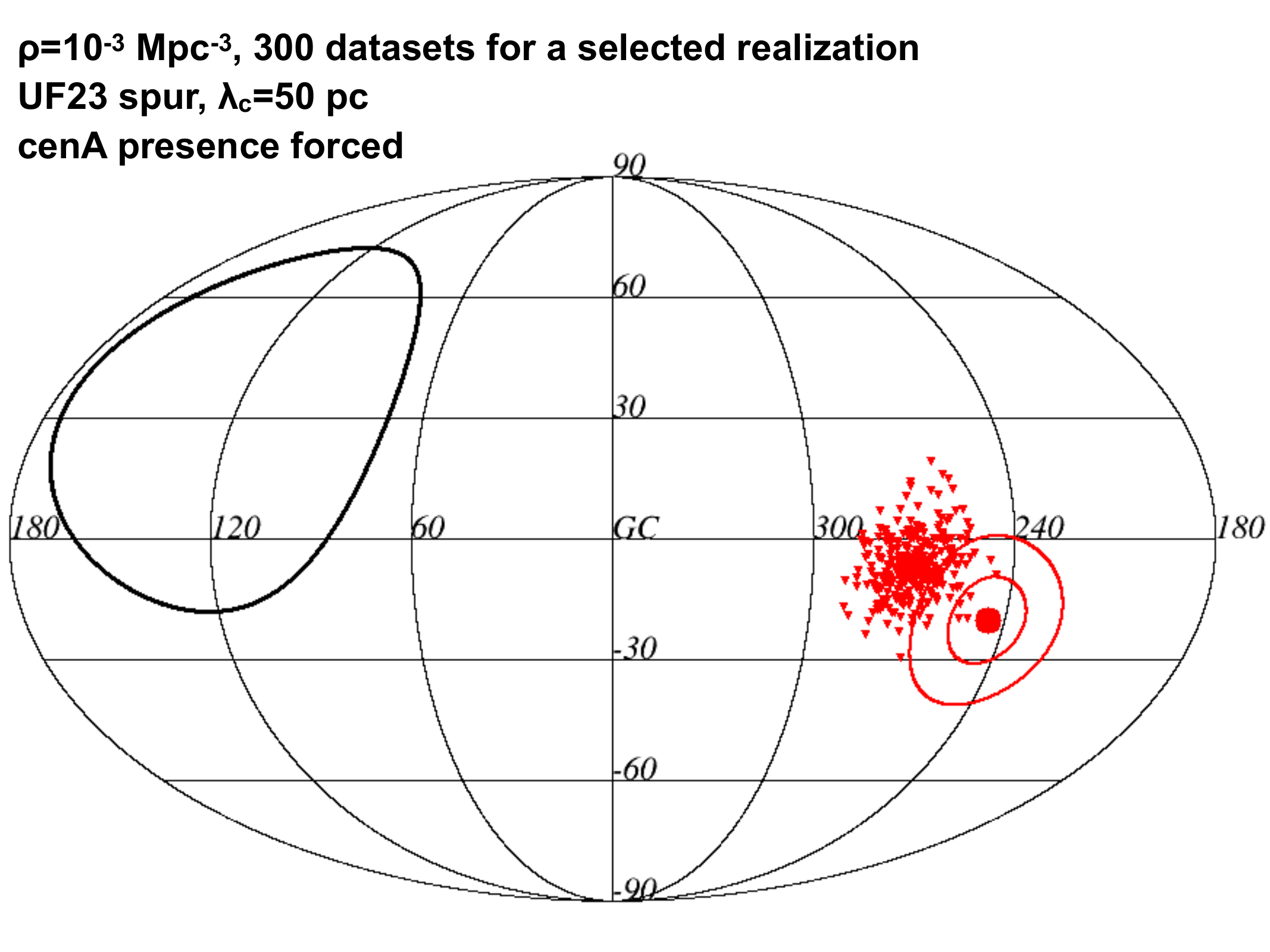}
   \includegraphics[width=7.5cm]{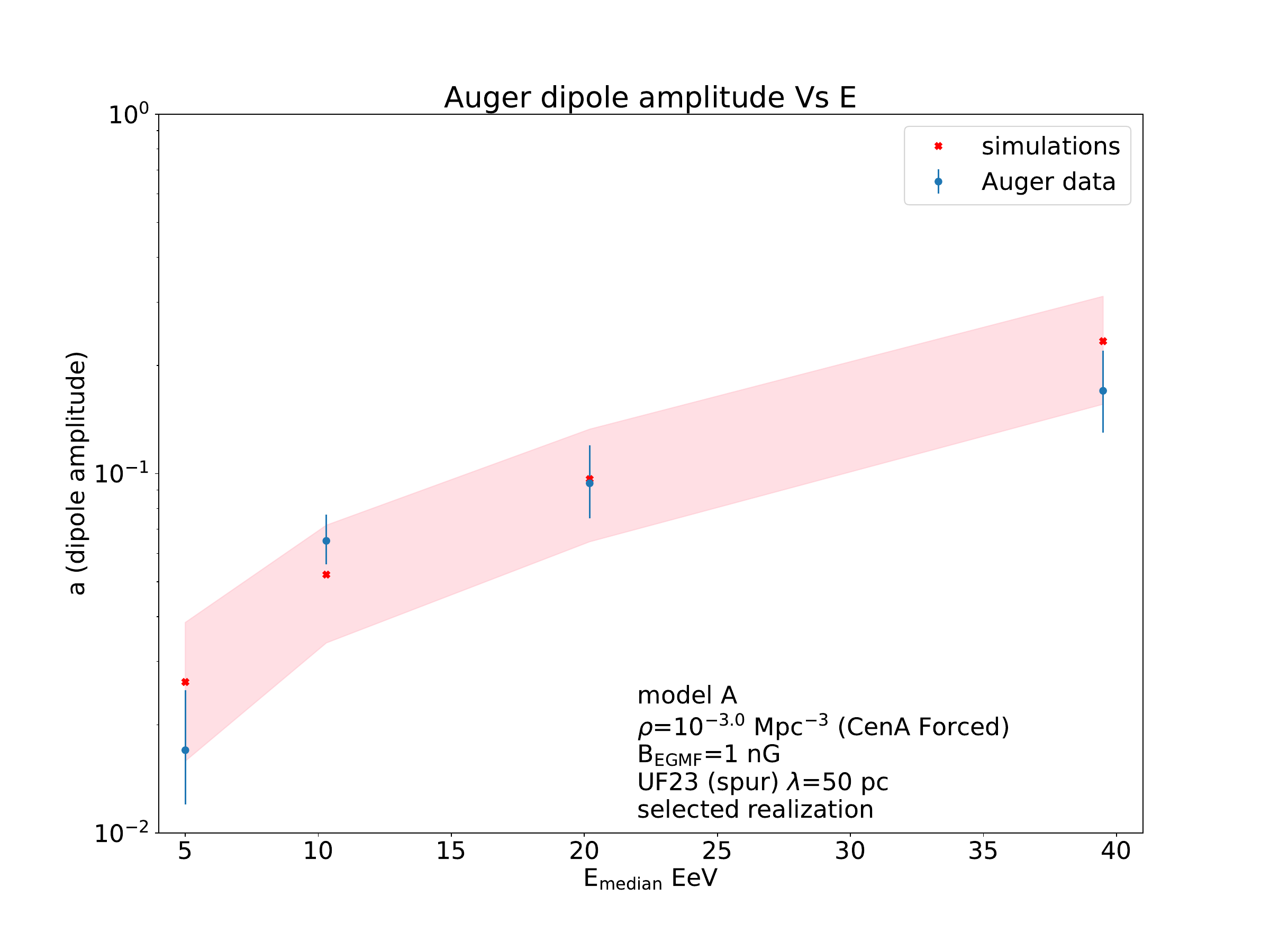}
   
   \includegraphics[width=7.cm]
   {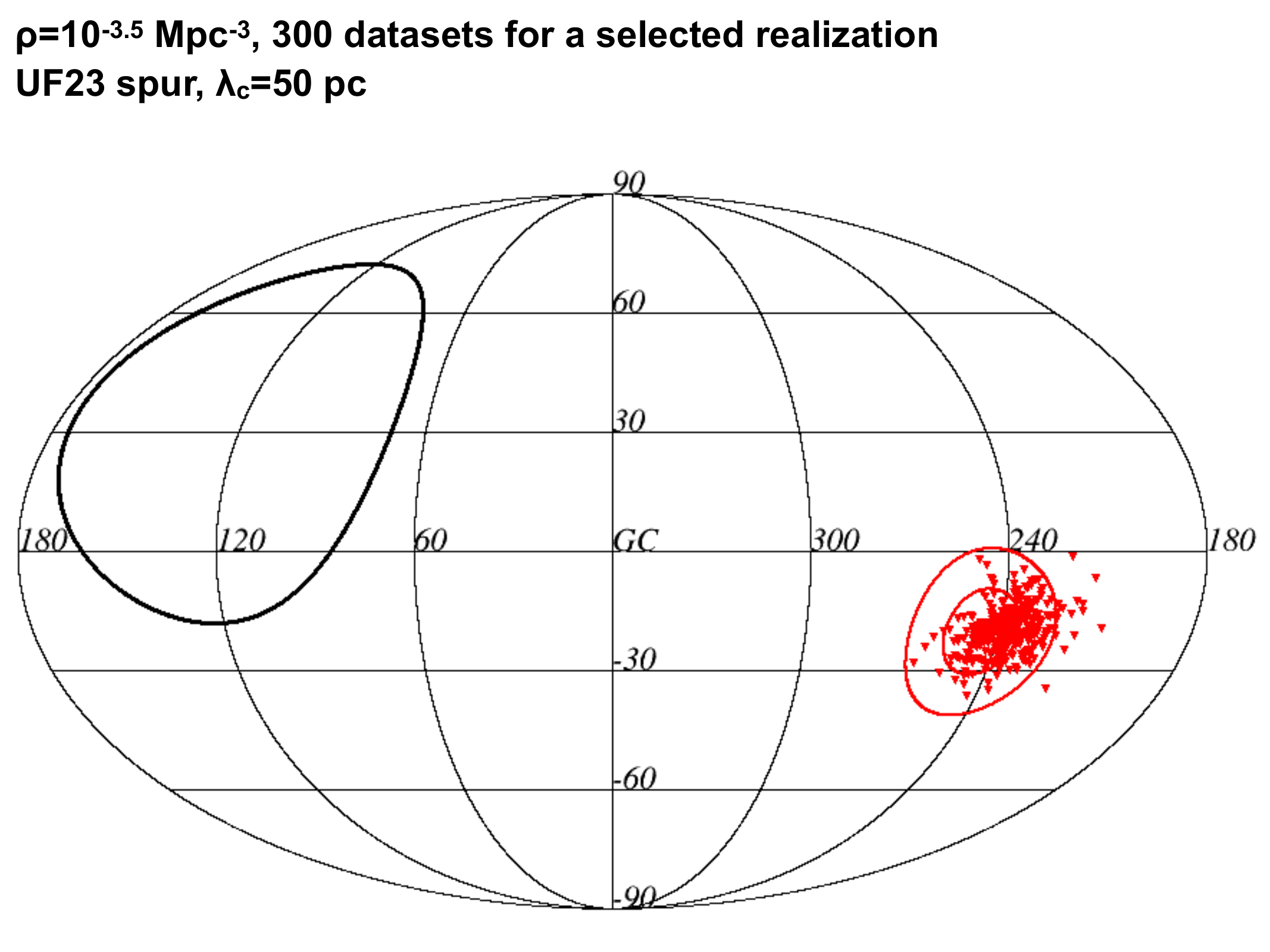}
   \includegraphics[width=7.5cm]{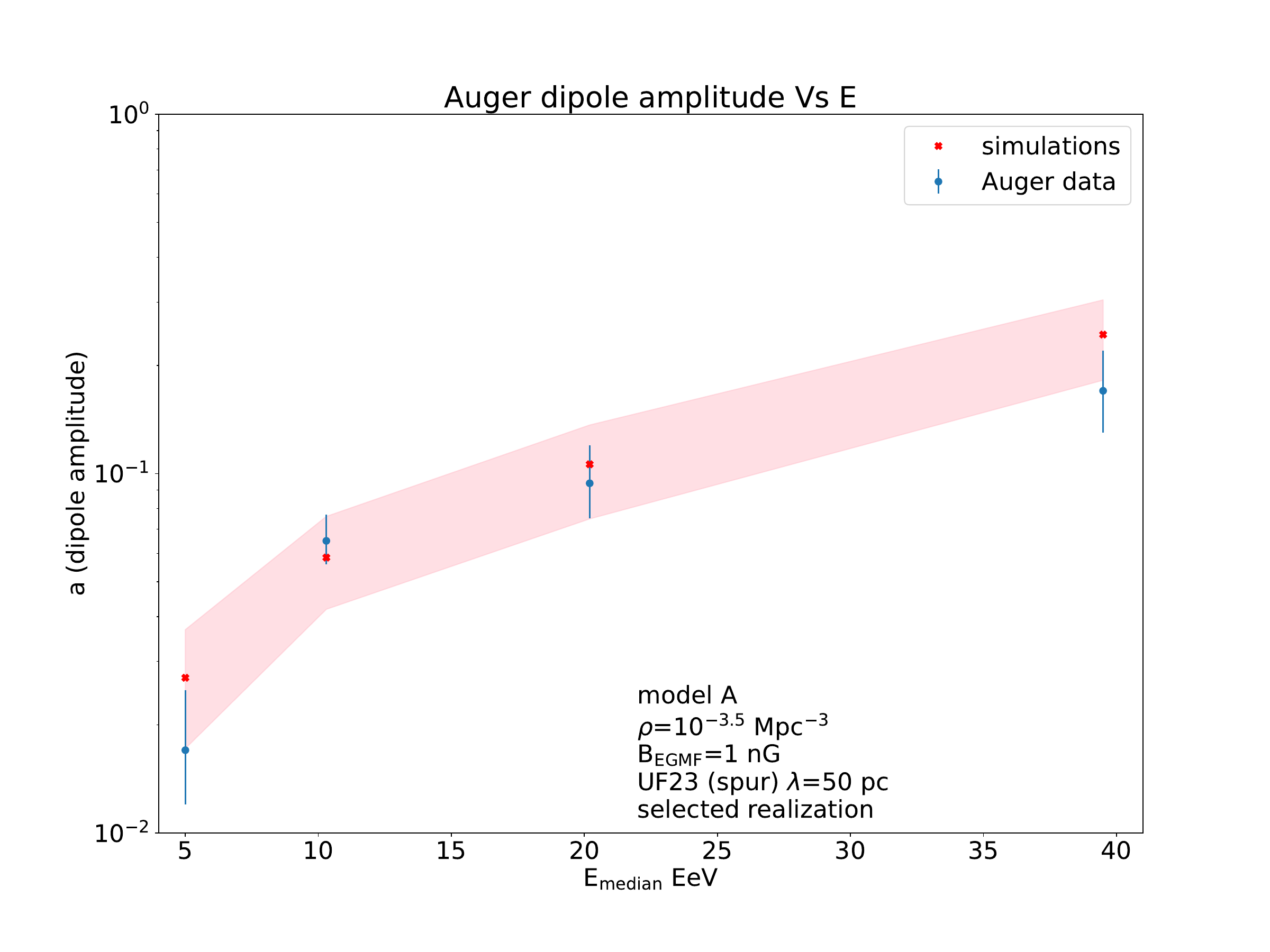}
   \includegraphics[width=7.cm]{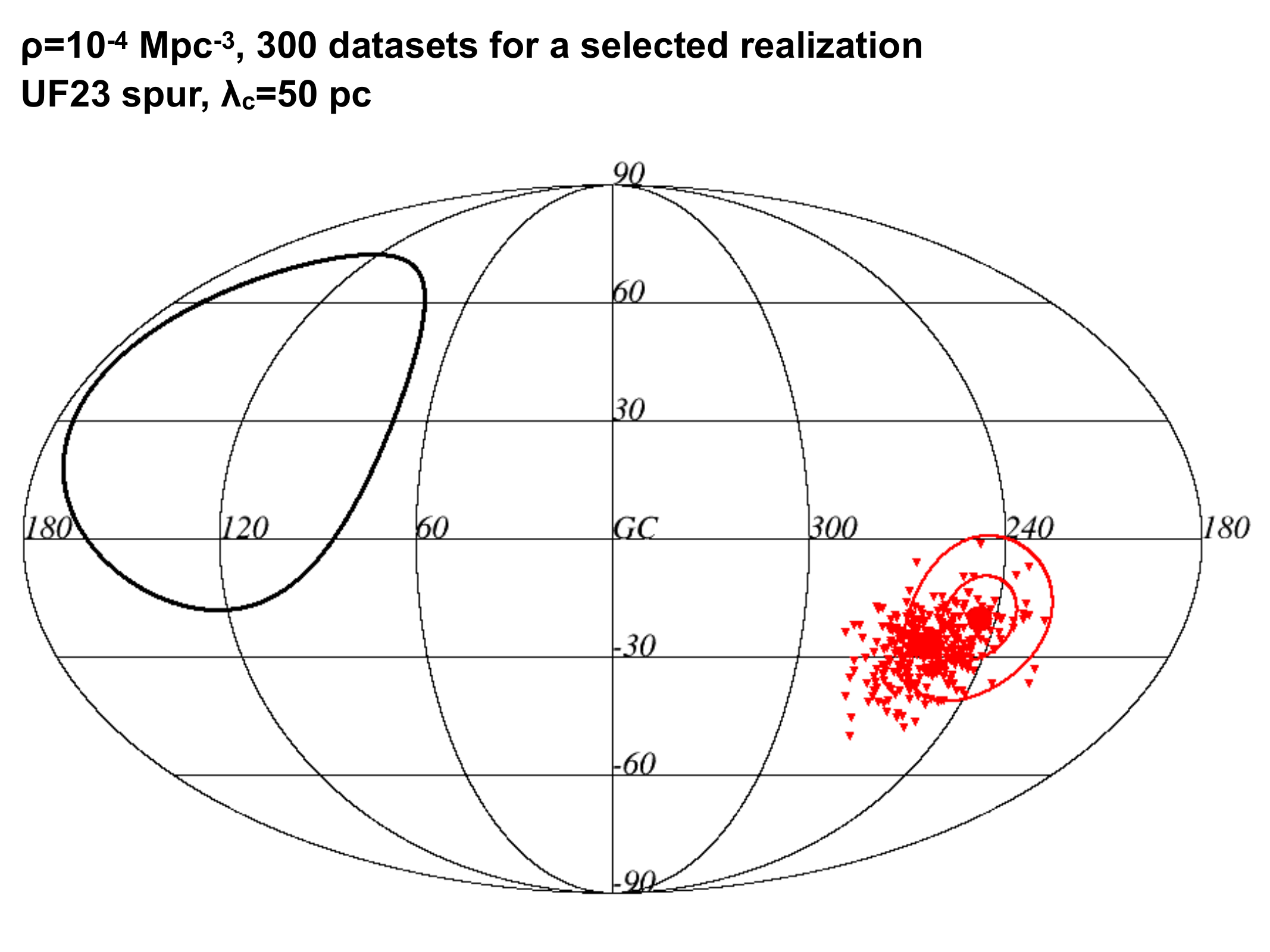}
  \includegraphics[width=7.5cm]{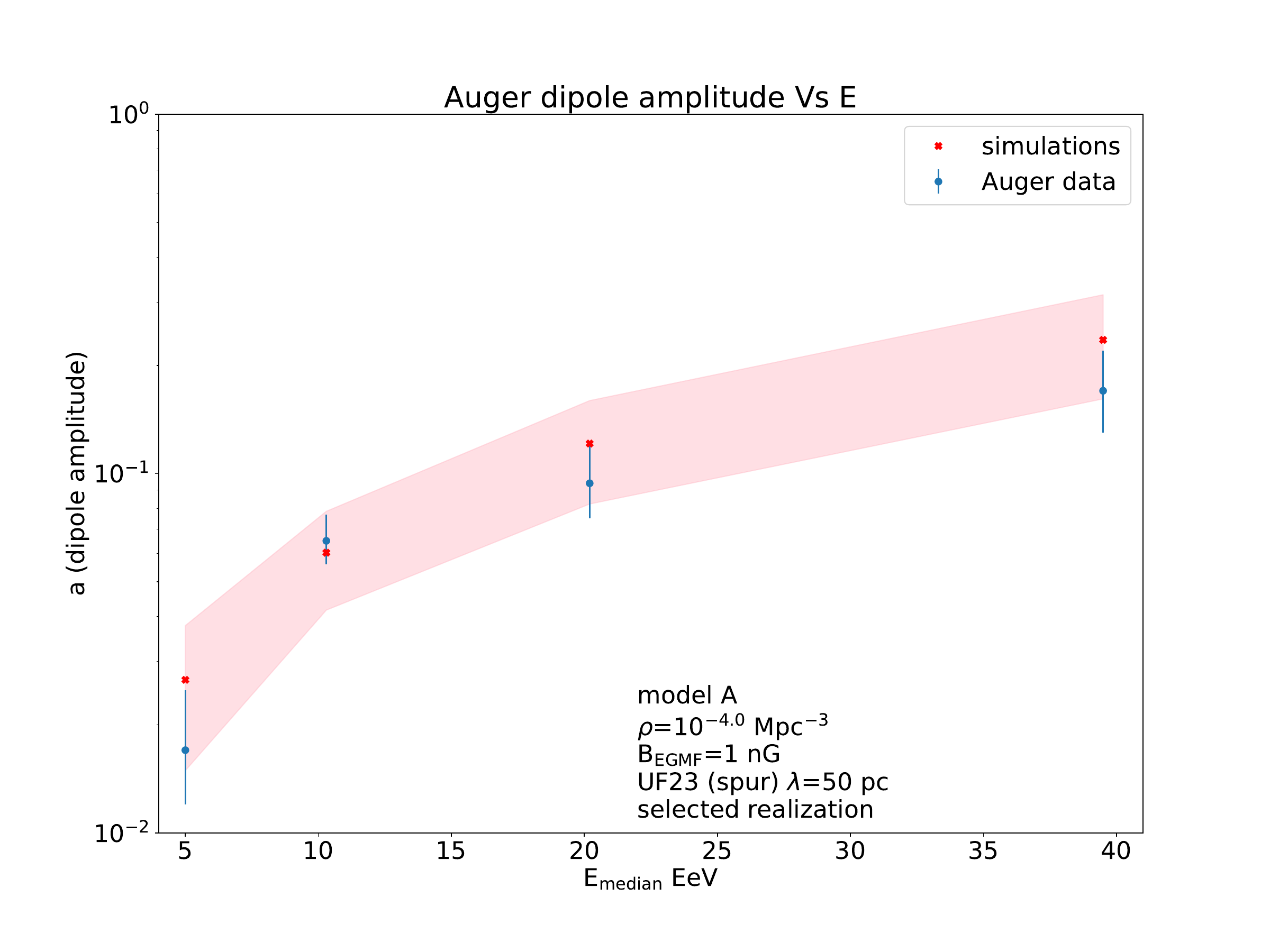}
  
      \caption{Skymaps of the dipole location (in galactic coordinates) and energy evolution of the dipole amplitudes reconstructed for 300 independent datasets generated for a selected realization of models I to IV (from top to bottom, see text and legend). On the skymaps, the triangular markers show the reconstructed dipole location for each individual Auger size dataset generated, the barycenter of the distribution is indicated with a large squared marker while the Auger data are shown with a large circle. For the energy evolution of the dipole amplitude, the mean value and the dispersion ($90\%$) of the simulations (calculated over the 300 datasets) are shown in red and the Auger data in blue.  
              }
         \label{Selected_location}
   \end{figure*}


To assess the compatibility of our predictions with Auger observations, we used our mother-catalog approach. For a given assumed source density, we generated 300 realizations of the source distribution by randomly sampling the mother catalog; for each realization, one dataset with Auger statistics and exposure was produced. Throughout the next two sections, Model~A for the source spectrum and composition, as well as a 1~nG EGMF, are assumed in combination with the GMF models introduced above.


\section{Large scale anisotropy analyses}

\subsection{Dipole analysis}

\subsubsection{Dipole direction: cosmic variance and GMF dependence}

We begin with the Auger dipole analysis, which remains the only UHECR anisotropy measurement above 8~EeV with a post-trial significance exceeding $5\sigma$. In Paper~I we showed that, under the assumption that UHECR sources follow the galaxy distribution, the most difficult aspect to reproduce was the direction of the observed dipole (in contrast with the energy evolution of its amplitude, which is easier to match). The introduction of more recent GMF models may, however, modify this conclusion, particularly since most UF23 configurations predict magnification/demagnification patterns quite different from those of JF12, including in the Virgo region.

Figure~\ref{dipoleLoc} shows the dipole directions reconstructed for 300 simulated datasets above 8~EeV for each astrophysical model, assuming a source density of $10^{-3}\,\mathrm{Mpc^{-3}}$. The panels differ only by the GMF model adopted (except for the bottom-right panel, where the presence of Cen~A is forced in every realization instead of appearing randomly in only one realization out of seven; the UF23 Spur model is used in both cases). In all models the predictions are widely dispersed across the sky. This dispersion is dominated by the cosmic variance, with statistical fluctuations contributing only secondarily (see below). Some realizations yield dipole directions far from the barycenter of the ensemble (indicated by the large square markers). These correspond to specific configurations of nearby sources located in highly magnified regions of the sky, which can strongly attract the reconstructed dipole direction. This occurs for instance in the UF23 base and Spur models when galaxies such as Maffei~1/2 or IC~342 are included in a realization (see Figs.~\ref{Magnif} and \ref{SkyMaps}).

The most striking result from Fig.~\ref{dipoleLoc} is the very good agreement obtained with the KST24 model (bottom-left panel): a significant fraction of realizations lie within the $2\sigma$ contour of the Auger dipole. In contrast, for the UF23 base and Spur models (top-left and top-right panels; similar results are found for expX and nebCor), the Auger dipole lies near the edge of the predicted distributions. Nevertheless, these predictions remain noticeably closer to the Auger direction than those obtained in Paper~I with the JF12 model.

Forcing Cen~A to be included in every realization (bottom-right panel) produces results very similar to the UF23 Spur case without this constraint. This shows that, for such a high source density—and because the UF23 Spur model does not strongly magnify Cen~A in the relevant rigidity range—the presence or absence of Cen~A does not significantly affect the dipole direction.

The UF23 twistX model behaves differently: owing to its distinct rigidity-dependent magnification near the Virgo region, it yields predictions incompatible with the Auger dipole for the corresponding astrophysical model. Its behaviour is qualitatively similar to that of JF12, with the dipole drawn toward the Galactic north due to the strong contribution of Virgo sources. Interestingly, in the KST24 model, Virgo also contributes substantially to the expected anisotropy, but the associated deflections shift the arrival directions into the southern Galactic hemisphere, explaining the very different level of agreement with Auger.

The XH24 model yields predictions similar to those of the Sun+Planck model and provides a less satisfactory match to the Auger dipole direction than KST24 or the UF23 suite (except twistX). However, we note that the current XH24 implementation lacks a vertical component of the regular GMF. Adding, for example, the $z$ component of the UF23 base model leads to strong changes in the predicted dipole direction, suggesting that the XH24 model may require completion before being fully assessed for UHECR anisotropy studies.

Lower source densities are shown in Fig.~\ref{dipoleLoc2}, where the cases $10^{-3.5}\,\mathrm{Mpc^{-3}}$ (left panel) and $10^{-4}\,\mathrm{Mpc^{-3}}$ (right panel) are displayed for the UF23 Spur GMF model. As expected, the spread of the predicted dipole directions increases as the source density decreases, while the barycenter of the distribution remains essentially unchanged relative to the $10^{-3}\,\mathrm{Mpc^{-3}}$ case. The larger cosmic variance makes it easier for some realizations to fall within the $2\sigma$ region of the Auger dipole: for source densities of $10^{-4}\,\mathrm{Mpc^{-3}}$, about $10\%$ of the realizations lie within $2\sigma$ for both the UF23 Spur and base models.

For most versions of the UF23 suite, the spread of dipole locations appears significantly more compatible with the Auger data than what we found using the JF12 model in Paper~I, under the hypothesis that UHECR sources trace the galaxy distribution. However, the observed dipole direction is not by itself indicative of such a scenario. Because of the large cosmic variance and the strong dependence on the GMF model, it is not possible to use the dipole direction to constrain potential biases of the UHECR source distribution with respect to that of galaxies.

Even when adopting a radically different source-distribution hypothesis—namely drawing sources from an isotropic and homogeneous distribution—about $5\%$ of the realizations yield a dipole direction within $2\sigma$ of the Auger measurement for the UF23 Spur GMF model (for source densities of $10^{-3}$ or $10^{-4}\,\mathrm{Mpc^{-3}}$). This further illustrates that the dipole direction has limited discriminating power in its current form, owing to the combined effects of cosmic variance and GMF-model uncertainties.

\subsubsection{Dipole amplitude and its energy evolution}

The various GMF models we tested reproduce the dipole amplitude measured by Auger, as well as its evolution with energy, for source densities of $10^{-3}\,\mathrm{Mpc^{-3}}$. This was already found in Paper~I for the previous generation of GMF models, and remains true for the new models considered here. This is illustrated in the top-left and top-right panels of Fig.~\ref{DipAmp}, which show the cases of the UF23 Spur and KST24 models; the other GMF models yield very similar results. The spread of the predictions reflects the combined effect of statistical and cosmic variance, with the latter being dominant.

In all cases, the coherence length $\lambda_{\rm c}$ of the turbulent GMF component was tuned to reproduce the observed dipole amplitude. As already noted in Paper~I, smaller coherence lengths are generally required for GMF models that predict a strong demagnification of the Virgo region (e.g. the UF23 suite and XH24).

The effect of lower source densities is shown in the bottom-left and bottom-right panels of Fig.~\ref{DipAmp}, corresponding to $10^{-3.5}$ and $10^{-4}\,\mathrm{Mpc^{-3}}$, respectively. The impact of the increased cosmic variance is clearly visible, particularly for the $10^{-4}\,\mathrm{Mpc^{-3}}$ case. The mean dipole amplitude (averaged over the 300 realizations) increases as the source density decreases. This behaviour is generic: it affects the dipole amplitude as well as essentially all anisotropy observables discussed in this paper. At the same time, the enlarged cosmic variance makes it possible to obtain not only realizations with relatively high anisotropy, but also some that are nearly isotropic, depending on whether the realization includes unusually nearby sources.

The compatibility between the simulated dipole amplitude (and its energy evolution) and the Auger measurement over a broad range of source densities demonstrates that this observable alone cannot strongly constrain the UHECR source density. Similar agreement can also be obtained when sources are sampled from an isotropic and homogeneous distribution. We therefore emphasise that the main information extracted from the observed dipole amplitude and its energy evolution is their consistency with a wide range of extragalactic-source scenarios, both in terms of source density and spatial distribution.

\subsubsection{Selected realizations}

Having discussed both the dipole direction and amplitude for the updated GMF models, the main new result with respect to Paper~I is the improved compatibility with the Auger dipole direction for some of the GMF models—most notably KST24 and several members of the UF23 family. This raises the question of whether the realizations that best reproduce the Auger dipole could simultaneously reproduce the other large-scale anisotropy observables (power spectrum and Rayleigh dipole+quadrupole), as well as the small- and intermediate-scale observables at higher energies (likelihood, blind, and targeted searches).

To address this, we selected, for each astrophysical model, the realization whose dipole direction lies closest to the Auger value among those falling within the $2\sigma$ contour of the data. For each selected realization, we generated 300 Auger-sized datasets to quantify the statistical variance of the predictions. In the following, we consider the four combinations of source density and GMF model listed below:
\begin{itemize}
    \item Model~I: $10^{-3}\,\mathrm{Mpc^{-3}}$ with KST24,
    \item Model~II: $10^{-3}\,\mathrm{Mpc^{-3}}$ with Cen~A forced and UF23 Spur,
    \item Model~III: $10^{-3.5}\,\mathrm{Mpc^{-3}}$ with UF23 Spur,
    \item Model~IV: $10^{-4}\,\mathrm{Mpc^{-3}}$ with UF23 Spur.
\end{itemize}

Figure~\ref{Selected_location} shows, for each model, the dipole locations and the energy evolution of the dipole amplitude obtained from the 300 datasets generated for the selected realization. In this case, all variance arises solely from statistical fluctuations, since the source distribution is fixed. As a result, the spread is much smaller than in the earlier figures where cosmic variance was included.

From the point of view of the dipole amplitude, all four models are compatible with the Auger data across the full energy range. The same holds for the dipole direction in all but one case: Model~II, for which the agreement is marginal. Only about $2\%$ of the realizations lie farther from the prediction barycenter than the Auger value. With a larger pool of source-distribution realizations (more than the 300 used here), it is likely that a realization more closely matching Auger could be found. However, such refinement would not materially affect the conclusions that follow. Model~II remains only marginally compatible with the dipole direction, which is also the case for other UF23 models. It is nevertheless valuable for the discussion of small- and intermediate-scale anisotropies at higher energies (see below).

\begin{figure*}
    \centering
   \includegraphics[width=8.5cm]{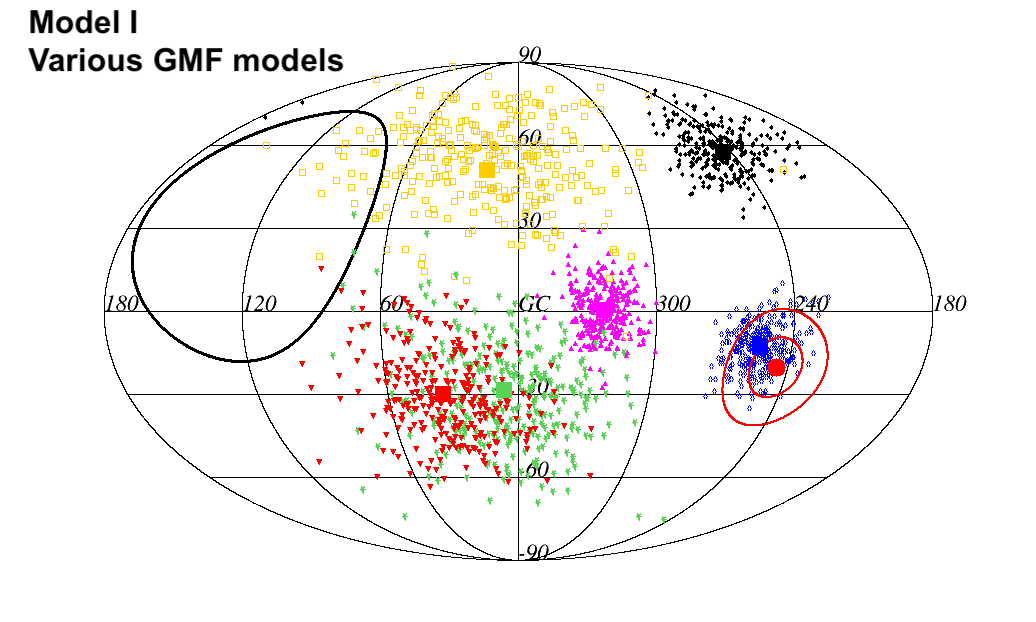}
   \includegraphics[width=8.5cm]{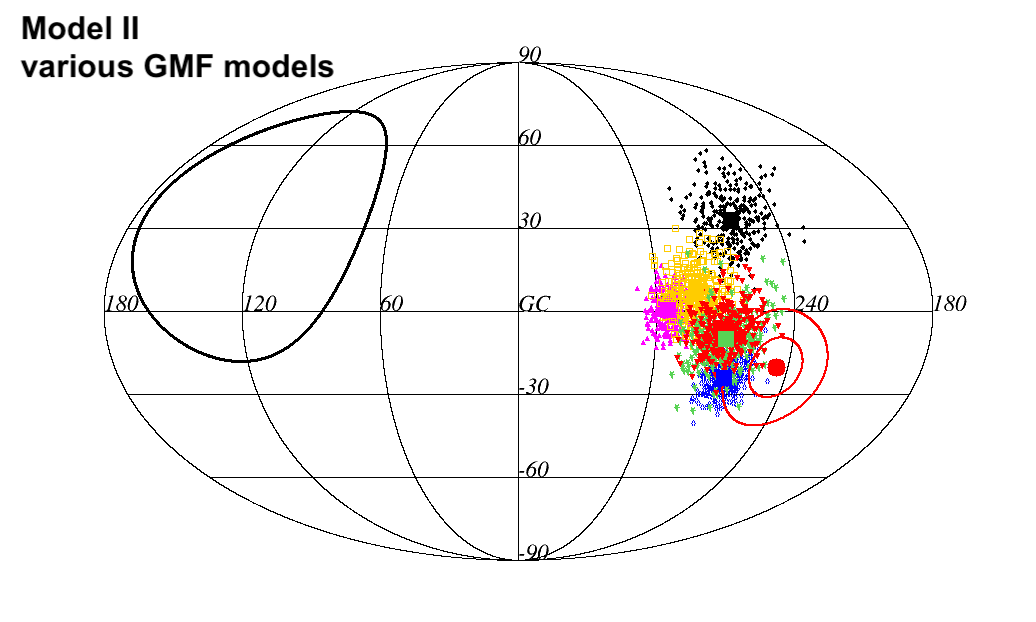}
   \includegraphics[width=8.5cm]{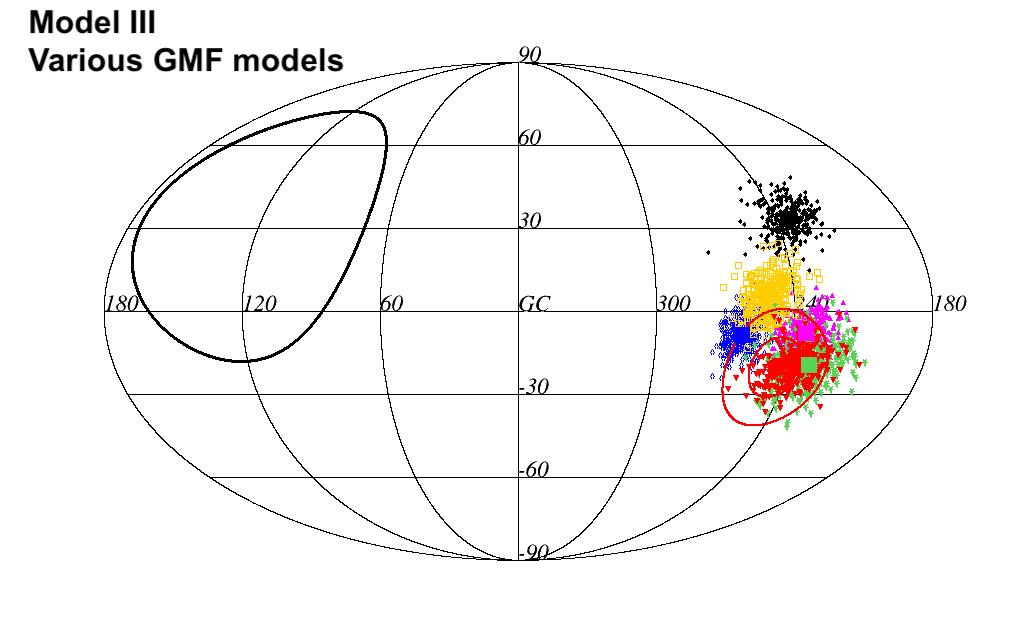}
   \includegraphics[width=8.5cm]{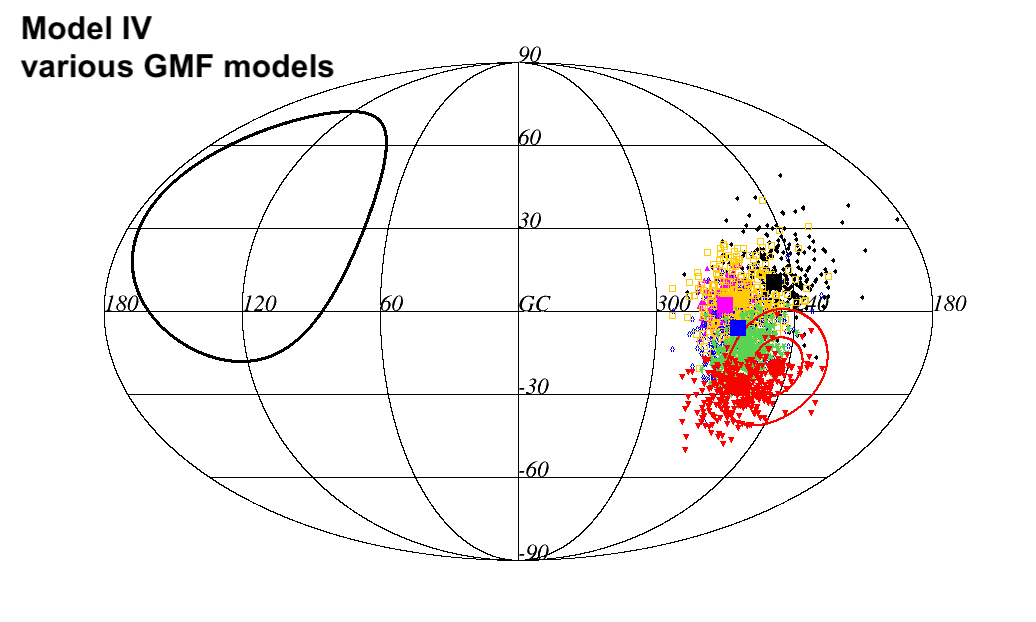}
  
      \caption{Skymaps of the dipole locations reconstructed for 300 independent datasets generated for a selected realization of models I to IV (from top to bottom and left to right), the various colors show the locations obtained for various GMF models (see text). Small markers show the location reconstructed for each individual dataset, the large squared markers show the barycenters of the distributions. The maps shown are in galactic coordinates.
              }
         \label{Selected_real}
   \end{figure*}

Since the selected realizations were chosen to optimise the dipole direction for a given GMF model, it is instructive to examine how the predicted dipole direction changes when only the GMF model is varied while keeping the same source distribution. This comparison is shown in Fig.~\ref{Selected_real} for models~I–IV. The GMF models tested include UF23 Spur (red), UF23 base (green), UF23 twistX (yellow), KST24 (blue), XH24 (magenta), and JF12+Planck (black). The UF23 expX and nebCor models were also evaluated but are omitted from the figure because they largely overlap with the Spur and base predictions.

For Models~II, III and IV, the differences in dipole direction introduced by switching between GMF models exceed significantly the dispersion associated with statistical noise only, for a fixed GMF model. However, the UF23 models (except twistX) tend to cluster closely: the angular separations between their prediction barycenters are typically comparable to, or smaller than, the width of the individual distributions (about $25$–$30^\circ$ in most cases).

Model~I shows more pronounced differences. The source distribution selected to match the KST24 dipole direction produces predictions that lie far from those of the other GMF models. This arises because, for the UF23 models, the dipole direction in this realization is unusually influenced by a specific configuration of nearby sources. We note that several other realizations exist (see Fig.~\ref{dipoleLoc}, bottom-left panel) that match the Auger dipole direction under KST24 and would not exhibit such extreme GMF-to-GMF variations; any of these could have been selected instead.

Overall, this exercise demonstrates that the fine tuning required to identify a particular source distribution that reproduces the Auger dipole direction—made possible by the large cosmic variance of the predictions—remains strongly dependent on the assumed GMF model.

\subsection{Other large-scale anisotropy analyses}

Besides the Rayleigh dipole analysis discussed above, the Auger Collaboration has performed additional large-scale anisotropy studies, namely the Rayleigh method extended to the quadrupole term and the power-spectrum decomposition. We apply these analyses to the selected realizations introduced in the previous subsection. (Including full cosmic variance would naturally broaden the distributions, but the qualitative conclusions would remain unchanged.)

\subsubsection{Rayleigh quadrupole}

\begin{figure*}
   \centering
   \includegraphics[width=8.5cm]{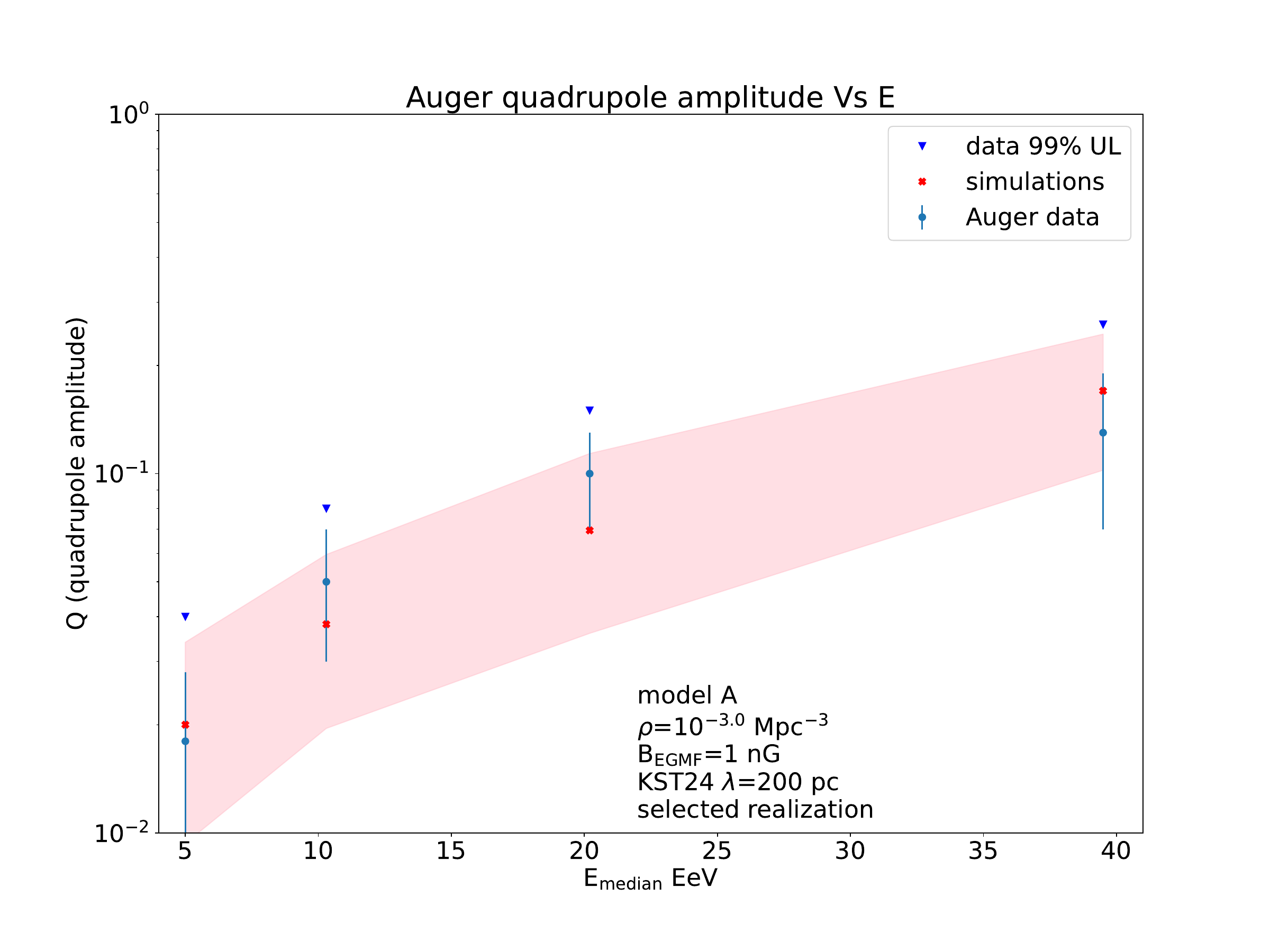}
   \includegraphics[width=8.5cm]{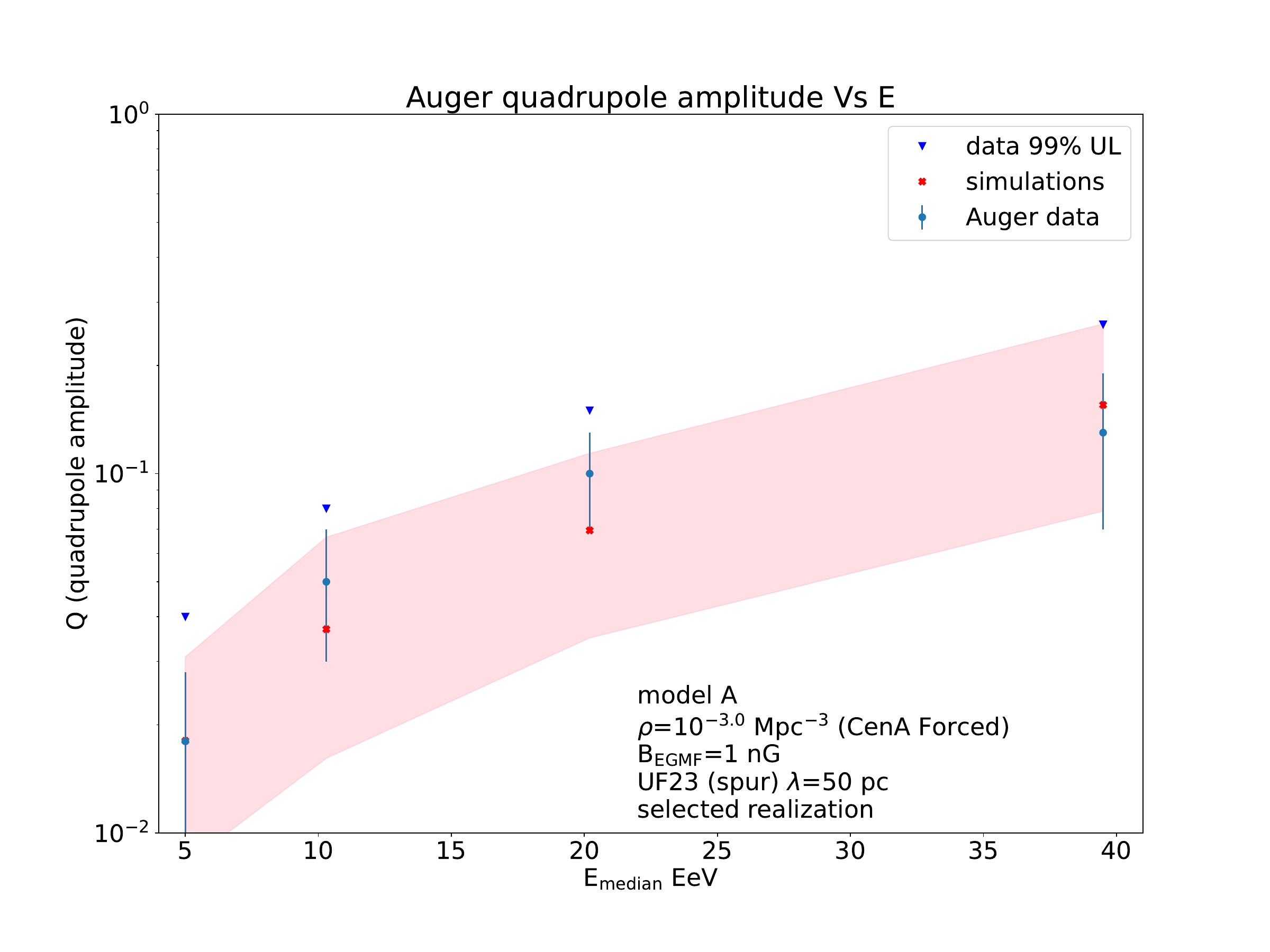}
   \includegraphics[width=8.5cm]{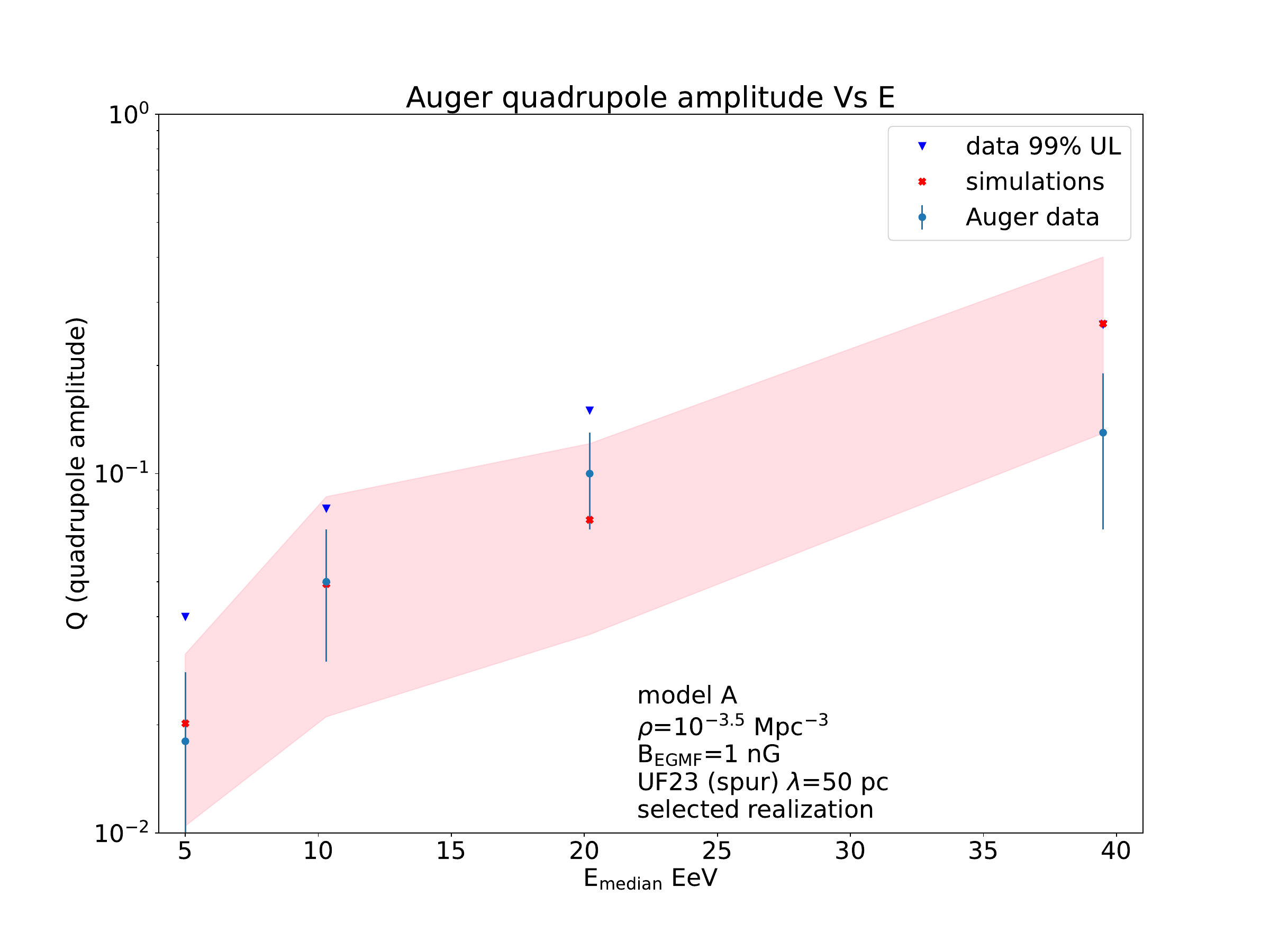}
   \includegraphics[width=8.5cm]{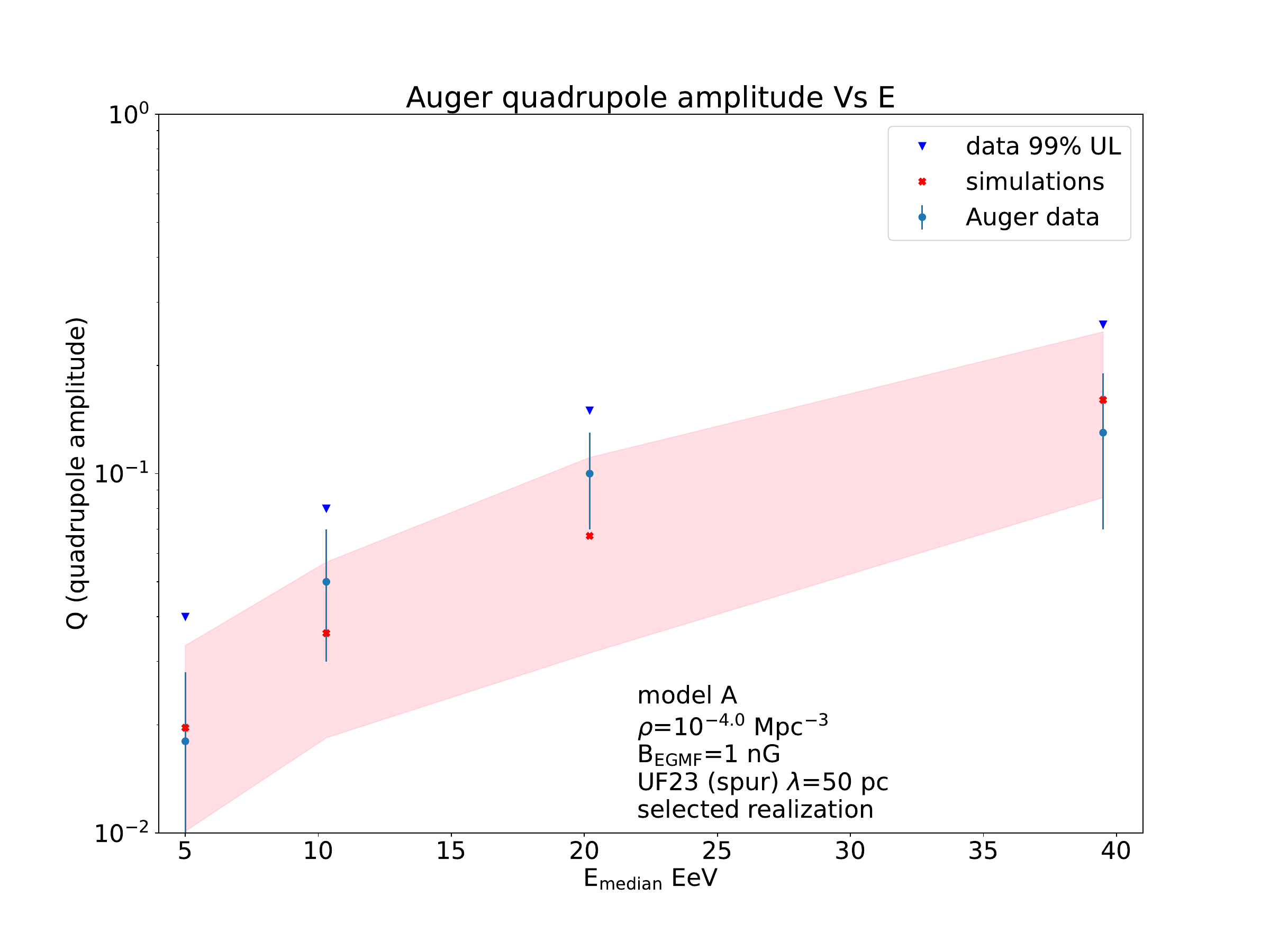}

      \caption{Energy evolution of the quadrupole amplitudes reconstructed (applying the Rayleigh analysis extended to the quadrupole term) for 300 independent datasets generated for a selected realization of models I to IV. The mean value and the dispersion ($90\%$) of the simulations (calculated over the 300 datasets) are shown in red and the Auger data in blue (mean value and $99\%$ upper limit).
              }
         \label{QuadAmp}
   \end{figure*}

The quadrupole amplitudes reconstructed with the Rayleigh dipole+quadrupole method (last updated in \citealt{AugerDip2024}) are shown in Fig.~\ref{QuadAmp} for Models~I–IV. In all four cases, the predictions are compatible with the Auger data across the full energy range. This is true despite the fact that the dipole directions of some models show only marginal agreement with the data: the quadrupole amplitude appears to be a much less discriminating observable.

\subsubsection{Power-spectrum multipoles}

The Auger Collaboration has also reported power-spectrum coefficients $C_\ell$ for the first few multipoles \citep{AugerDip2024}. The dipole and quadrupole amplitudes $a$ and $Q$ can be related to the coefficients $C_1$ and $C_2$ through:
\begin{equation}
\label{eu_eqn}
\begin{aligned}
a &= \frac{3}{2\sqrt{\pi}}\,C_1, \\
Q &= \sqrt{\frac{25\,C_2}{6\pi}}.
\end{aligned}
\end{equation}

These relations provide additional, independent estimates of $a$ and $Q$, which we compare with the Rayleigh-based estimates and with Auger data. As discussed in Paper~I, higher-order multipoles that might in principle be expected when sources trace the galaxy distribution are efficiently suppressed by GMF deflections, largely independently of the specific GMF model used.

\begin{figure*}
   \centering
   \includegraphics[width=8.5cm]{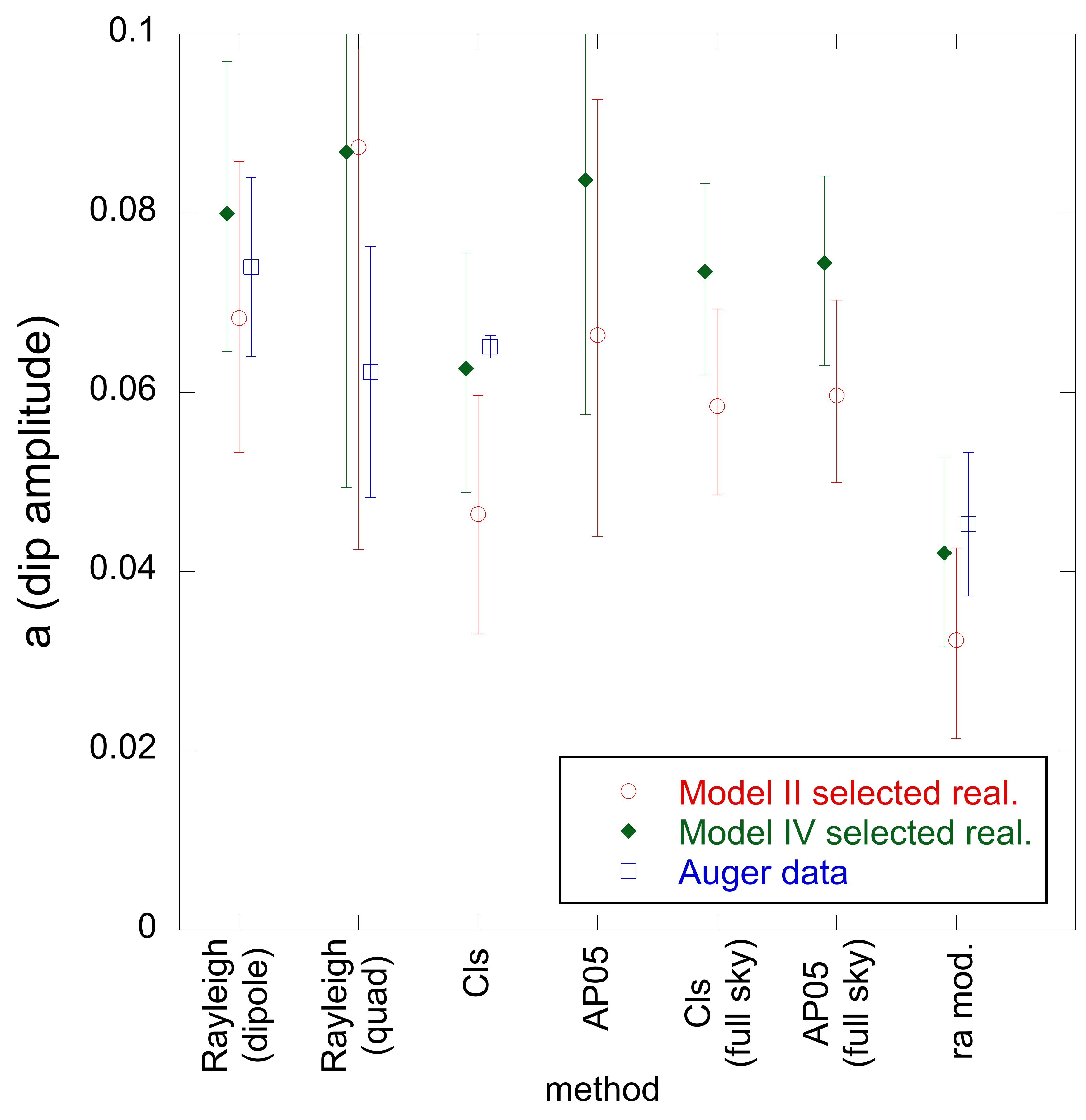}
   \includegraphics[width=8.5cm]{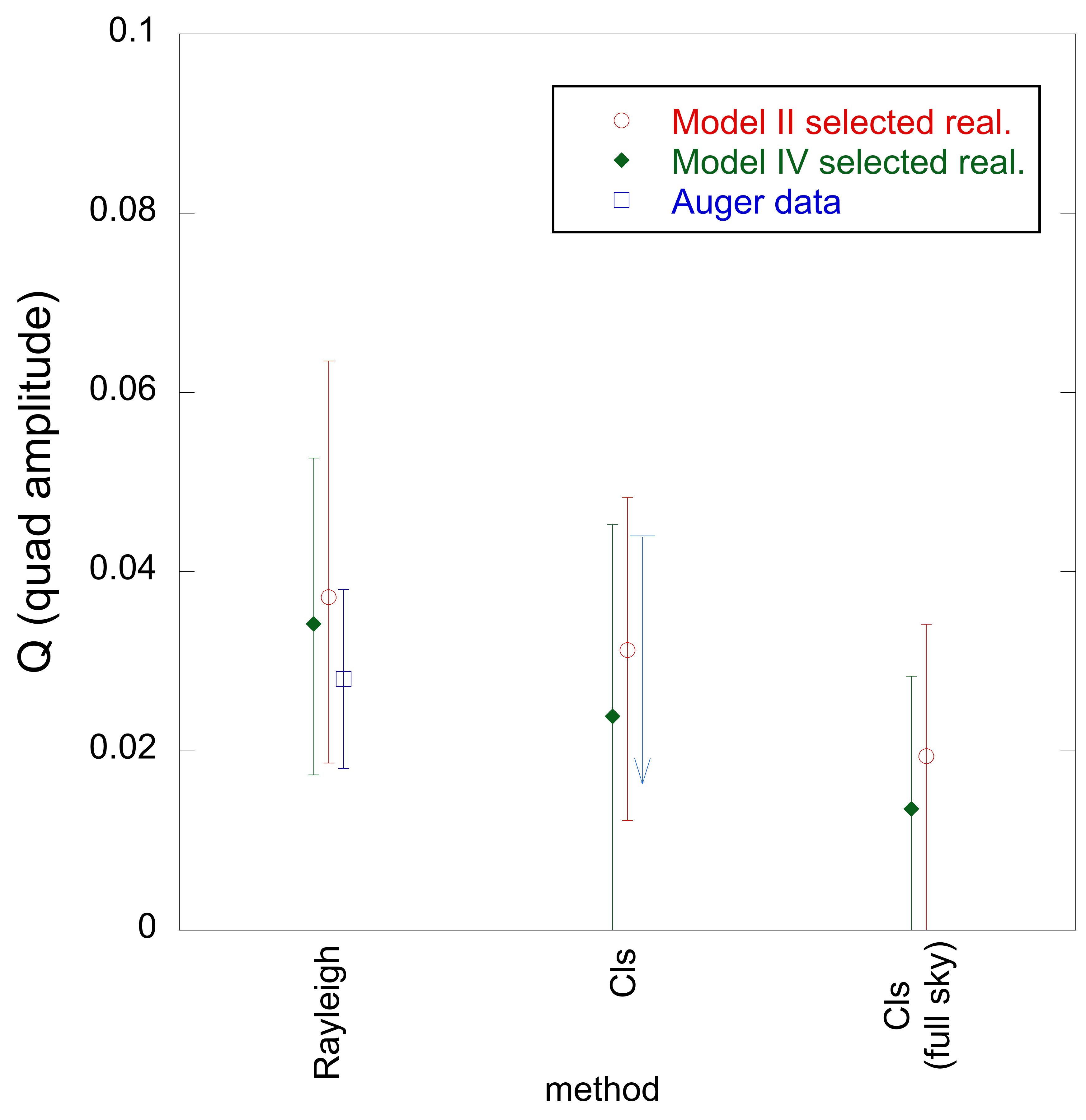}

      \caption{Summary of the estimates of the dipole (left) and quadrupole (right) amplitudes with various methods (see text) considering UHECRs with $E>8$~EeV. The simulations are shown in red and green for of model II and IV respectively the error bars show the range in which $90\%$ of the simulations are found, the Auger data are shown in blue whenever available the associated error bars refer to the errors on the various measurements.  
              }
         \label{DipQuadCompare}
   \end{figure*}

A summary of the reconstructed dipole and quadrupole amplitudes for events with $E > 8$~EeV is shown in Fig.~\ref{DipQuadCompare}. For the dipole, we show: (i) the Rayleigh dipole estimate, (ii) the Rayleigh dipole+quadrupole estimate, (iii) estimates derived from $C_1$ for partial-sky (Auger-like) and full-sky exposures, and  
(iv) the 3D dipole reconstruction of \citet{Aublin2005} (AP05), again for both partial and full-sky coverage. The amplitude of the right-ascension dipole modulation, which dominates the observed anisotropy in the Auger data, is also shown.  

For the quadrupole, the Rayleigh estimate and the estimates derived from $C_2$ (for both partial- and full-sky exposure) are compared with the corresponding Auger measurements. The error bars for the simulations indicate the interval containing $90\%$ of the 300 datasets, whereas the error bars on the Auger data represent measurement uncertainties.

Only the cases of Models~II and IV are displayed in Fig.~\ref{DipQuadCompare}, but Models~I and III behave very similarly. For Models~I and II ($10^{-3}\,\mathrm{Mpc^{-3}}$), only the upper $\sim 2\%$ of the simulated datasets reproduce the right-ascension modulation amplitude or the $C_1$ value reconstructed from Auger data, indicating a slight tension with the observations. All other dipole and quadrupole estimates show good compatibility.

For the lower source-density cases (Model~IV is shown, Model~III yields nearly identical results), all estimates—Rayleigh, power spectrum, AP05, partial and full sky—are mutually consistent and lie within the range allowed by the Auger data.

\subsubsection{Summary of large-scale constraints}

Overall, the large-scale anisotropy observations can be reasonably well reproduced by simulations in which UHECR sources follow the galaxy distribution, provided that some degree of fine tuning is allowed and cosmic variance is taken into account. This agreement is good for the lower source-density cases considered. For source densities around $10^{-3}\,\mathrm{Mpc^{-3}}$, the agreement is more marginal but still significantly better than what we obtained in Paper~I using the earlier JF12 GMF model, especially regarding the dipole direction. 

However, the strong dependence of the predictions on the GMF model and the large intrinsic cosmic variance prevent robust conclusions from being drawn about the true UHECR source distribution based solely on the Auger large-scale anisotropy results. In particular, these results do not allow us to constrain the source density, to identify possible biases relative to the galaxy distribution, or to distinguish between galaxy-tracing and alternative source scenarios.

 \begin{figure*}
    \centering
   \includegraphics[width=7.5cm]{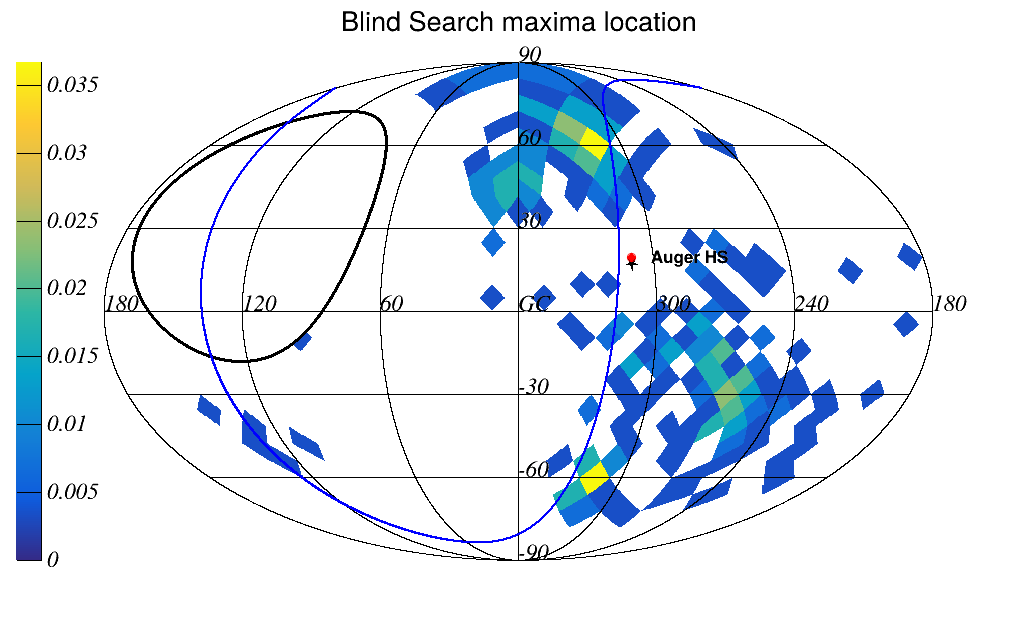}
   \includegraphics[width=7.5cm]{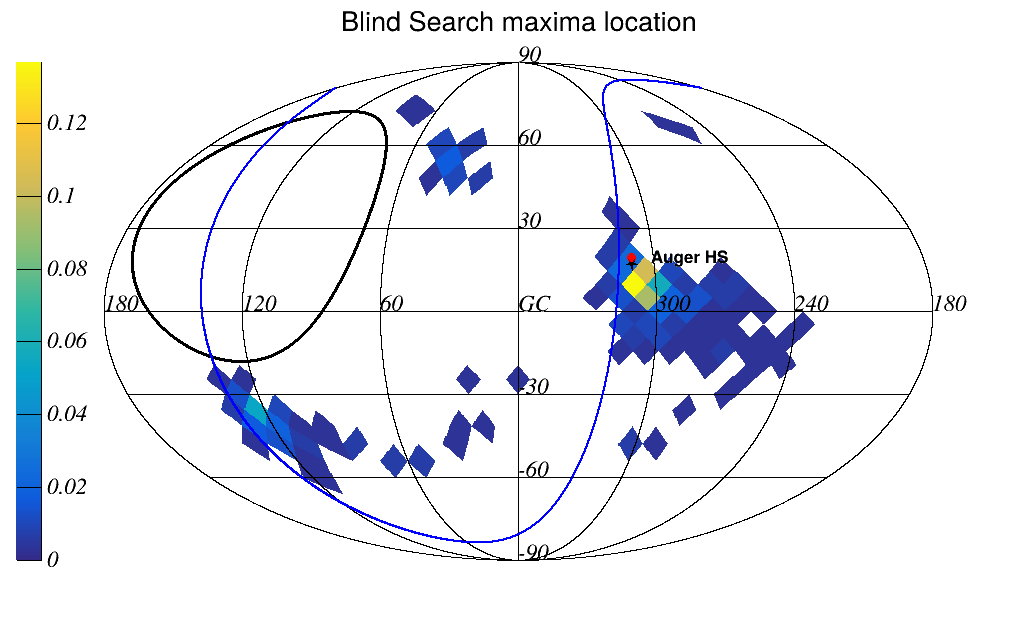}
   \includegraphics[width=7.5cm]{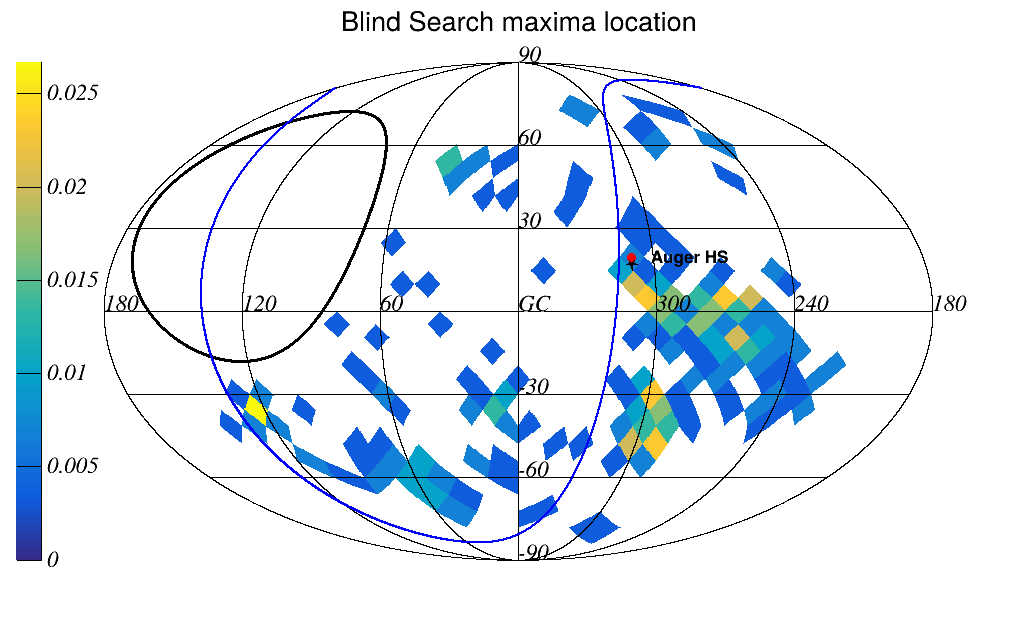}
   \includegraphics[width=7.5cm]{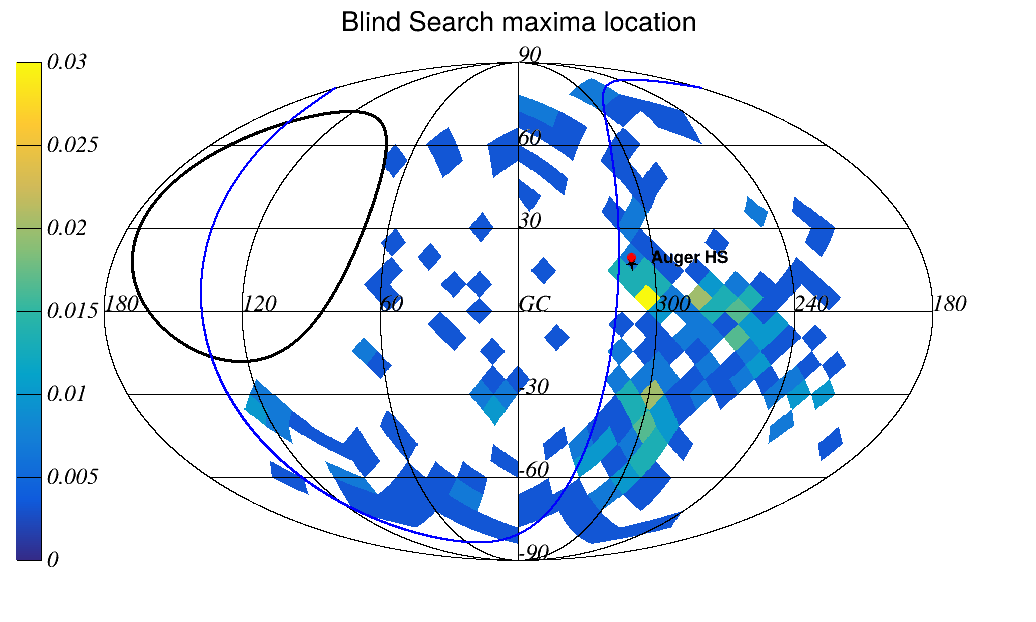}
  
      \caption{Distribution of the locations (in galactic coordinates) of the blind search maximum obtained from the generation 300 realizations of the assumed source distribution. The models I to IV are shown on the different panels from left to right and from top to bottom. 
              }
         \label{BS_loc_mc}
   \end{figure*}

\begin{figure}
    \centering
   \includegraphics[width=7.5cm]{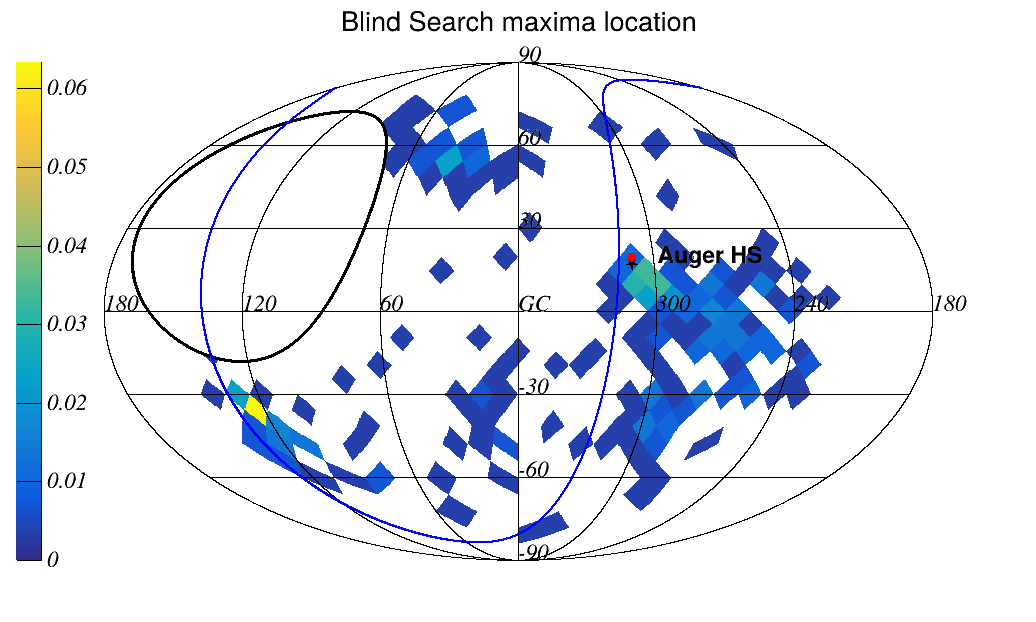}
   \includegraphics[width=6.5cm]{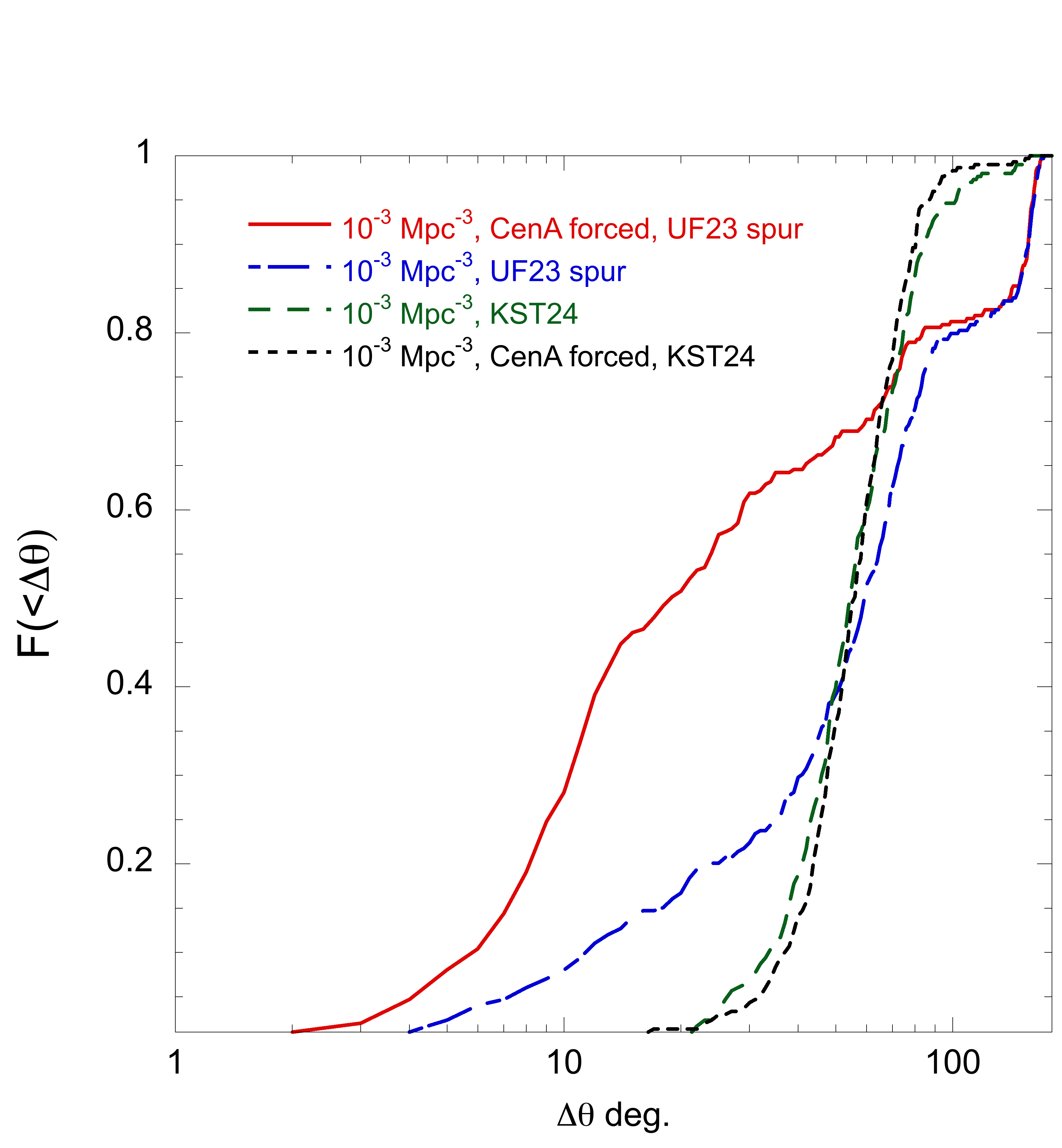}
  
      \caption{Top : distribution of the locations of the blind search maximum obtained from 300 realizations assuming a source density of $10^{-3}\rm\, Mpc^{-3}$ and the UF23 spur GMF model. Bottom : Cumulative distribution build over 300 realizations of the angular distance between the predicted location of the BS maximum and the one observed by Auger for different models (see legend and text).
              }
         \label{BS_loc_mc2}
   \end{figure}

\begin{figure*}
    \centering
   \includegraphics[width=8.5cm]{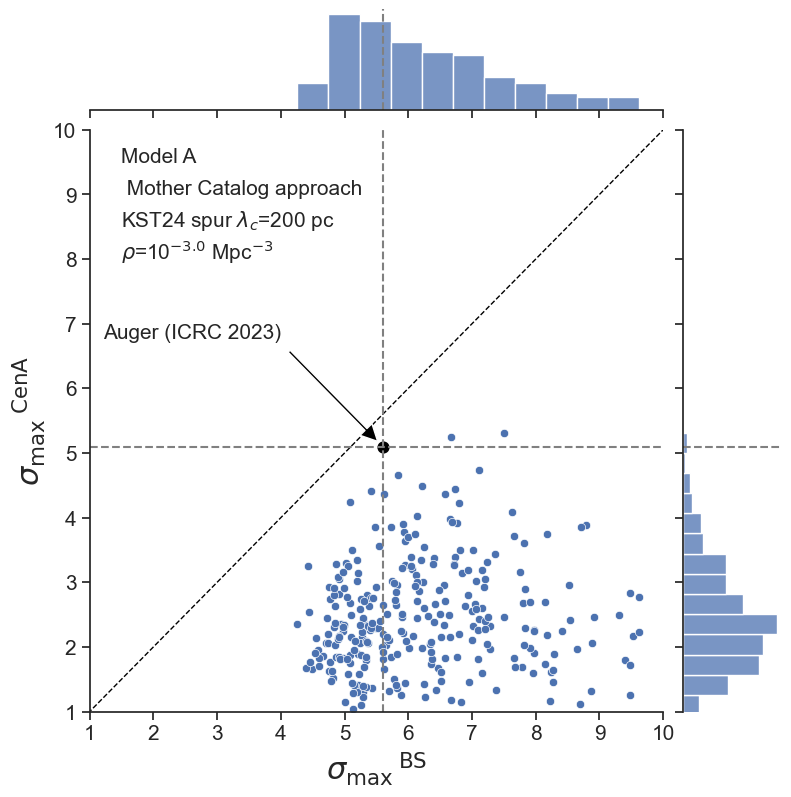}
   \includegraphics[width=8.5cm]{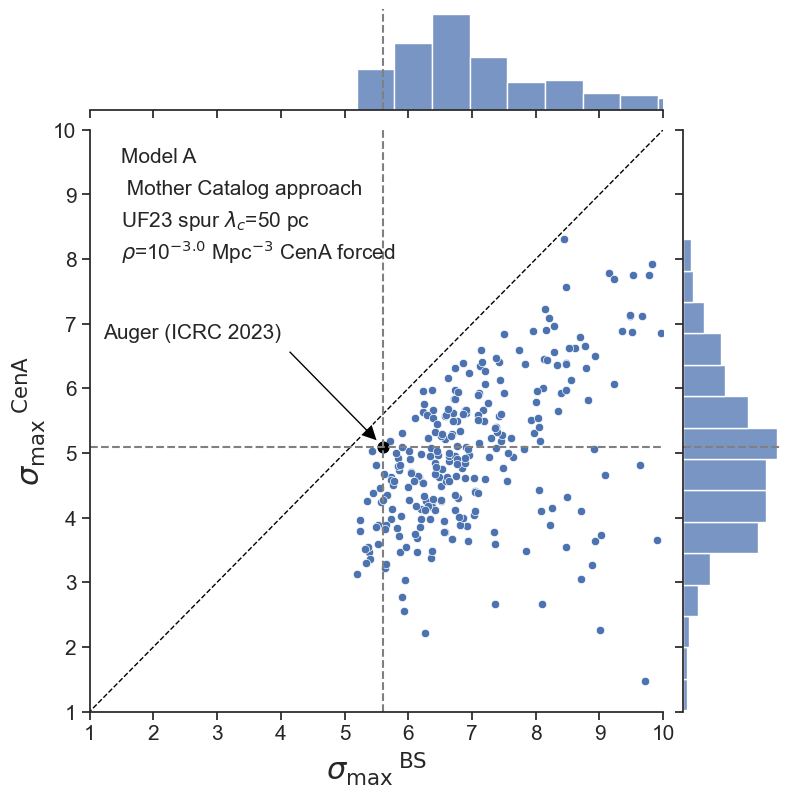}
   \includegraphics[width=8.5cm]{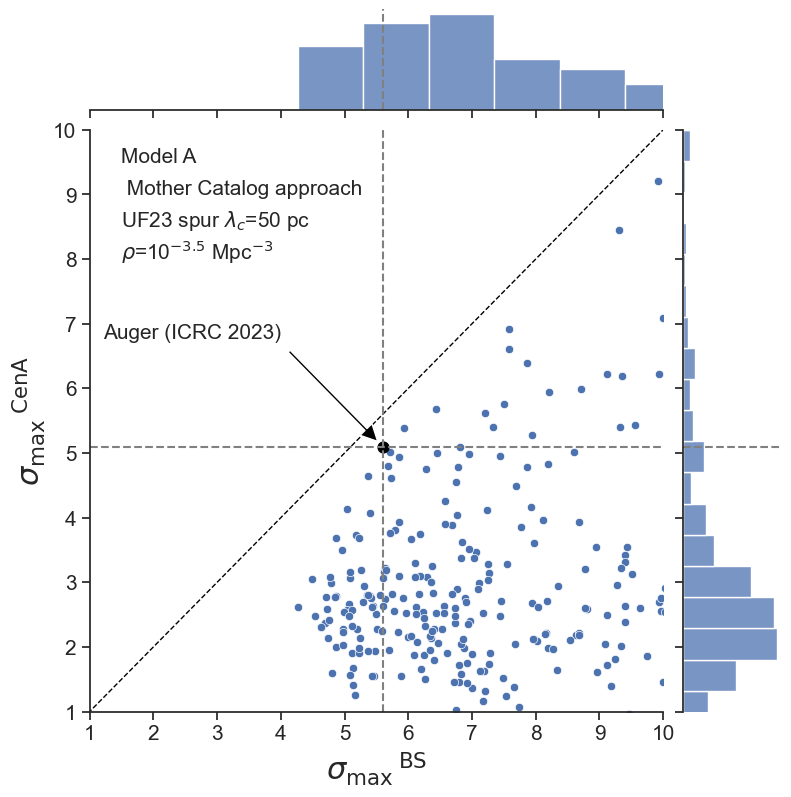}
   \includegraphics[width=8.5cm]{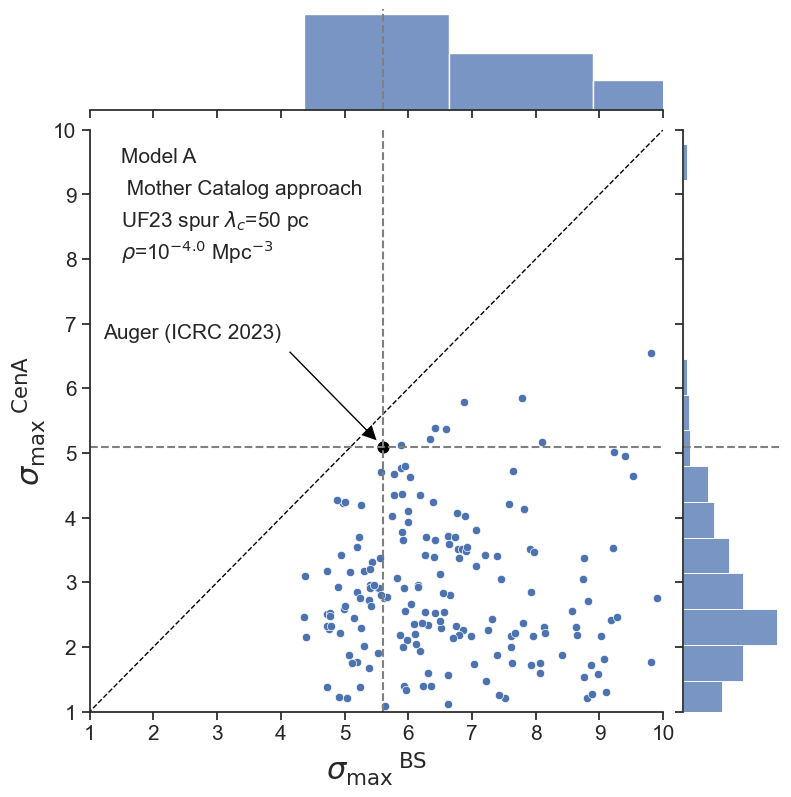}
  
      \caption{Scatter plot of $\sigma_{\rm max}^{\rm BS}$ versus $\sigma_{\rm max}^{\rm CenA}$ obtained from the generation of 300 realizations of the assumed source distribution. The models I to IV are shown on the different panels from left to right and from top to bottom. The individual distributions of the different quantities plotted are shown on top of the coordinate axis.
              }
         \label{BS_MC}
   \end{figure*}

\begin{figure*}
    \centering
   \includegraphics[width=8.5cm]{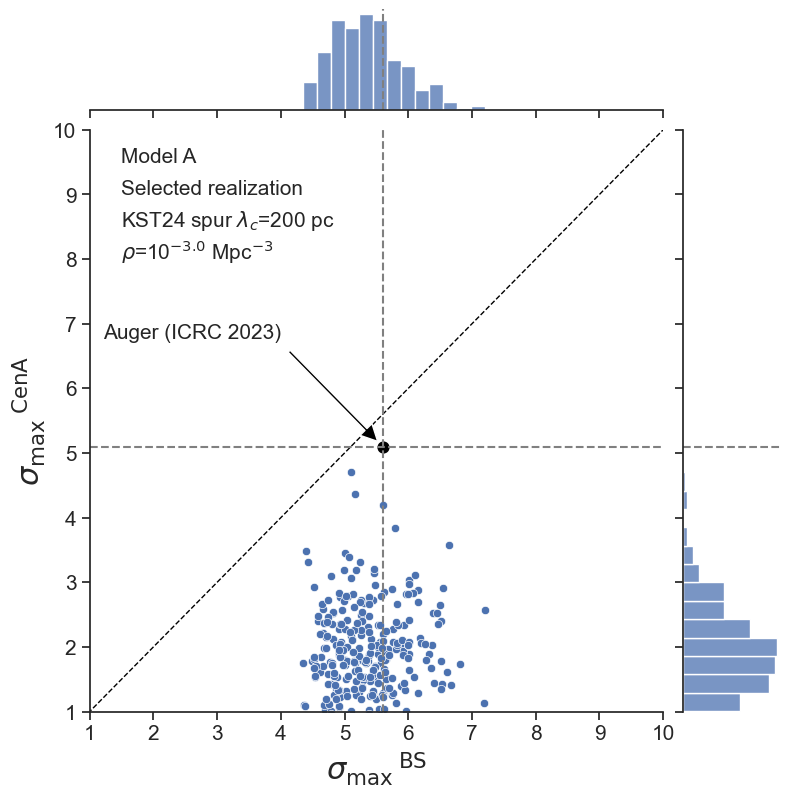}
   \includegraphics[width=8.5cm]{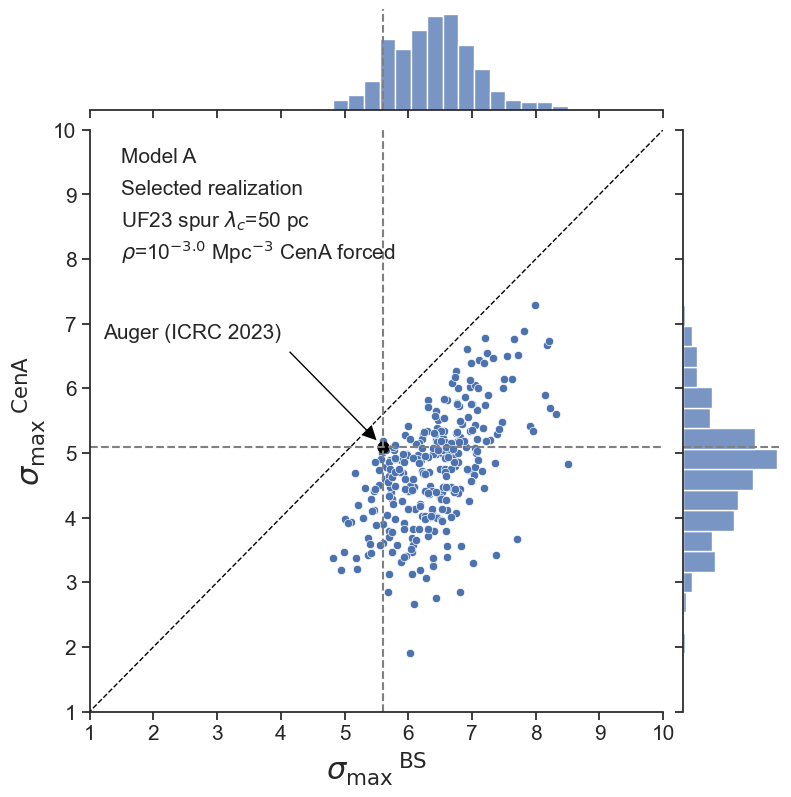}
   \includegraphics[width=8.5cm]{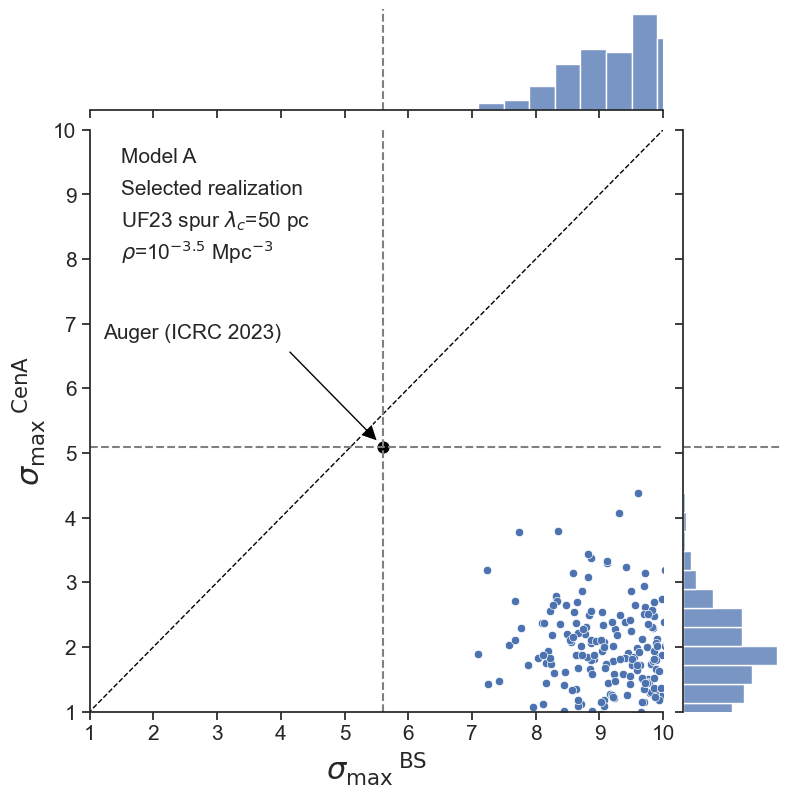}
   \includegraphics[width=8.5cm]{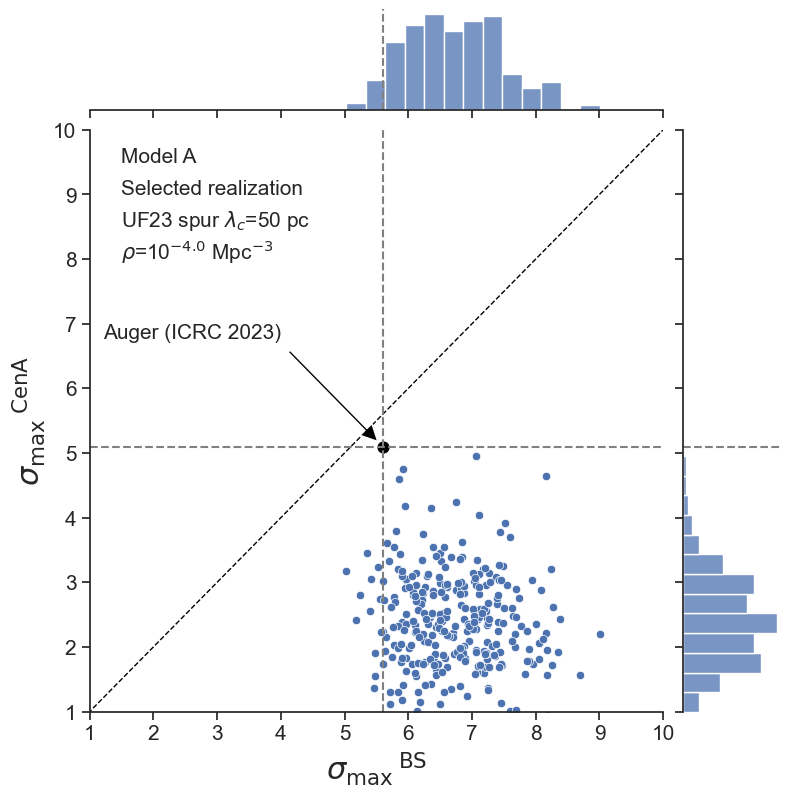}
  
      \caption{Scatter plot of $\sigma_{\rm max}^{\rm BS}$ versus $\sigma_{\rm max}^{\rm CenA}$ obtained from the generation of 300 datasets for the selected realization (on the basis of the location of the reconstructed dipole) of the assumed source distribution. Models I to IV are shown on the different panels from left to right and from top to bottom. The individual distributions of the different quantities plotted are shown on top of the coordinate axis.
              }
         \label{BS_Select}
   \end{figure*}


\section{Small and intermediate scale anisotropies at the highest energies}

\subsection{Blind and targeted searches}

\subsubsection{Cosmic variance}

We begin our discussion of small- and intermediate-scale anisotropies with the blind search (BS) and the targeted search toward Cen~A. The analysis follows exactly the procedure used in Paper~II, reproducing the scan over angular windows and dataset energy thresholds described in \citet{AugerAni2015}. In this section we adopt an exposure of $135\,000~\mathrm{km^2\,sr\,yr}$, identical to that of the most recent Auger update \citep{Golup2023}.

We restrict our attention to those of the models discussed above for which the dipole direction was found to be in marginal or satisfactory agreement with the Auger data. For each of these models we consider again the 300 independent realizations of the source distribution (and one Auger-like dataset for each realization) in order to quantify cosmic variance.

Figure~\ref{BS_loc_mc} shows the celestial distribution (in Galactic coordinates) of the BS maxima extracted from the 300 realizations for each model. The BS maximum reported by Auger is indicated by a black star (“Auger HS”), while the location of Cen~A is shown by a red circle. The four combinations of GMF models and source densities considered (from left to right, top to bottom) correspond to Models~I–IV, we remind here their characteristics:
\begin{itemize}
    \item Model~I: $10^{-3}\,\mathrm{Mpc^{-3}}$ with KST24,
    \item Model~II: $10^{-3}\,\mathrm{Mpc^{-3}}$ with UF23 Spur and Cen~A forced,
    \item Model~III: $10^{-3.5}\,\mathrm{Mpc^{-3}}$ with UF23 Spur,
    \item Model~IV: $10^{-4}\,\mathrm{Mpc^{-3}}$ with UF23 Spur.
\end{itemize}


For the BS maximum location, Model~II shows a much better compatibility with the Auger data than the other models: a significant fraction ($\sim 50\%$) of the realizations predict a BS maximum within $20^\circ$ of the Auger value.\footnote{$20^\circ$ corresponds to the estimate we made in Paper~II for the uncertainty on the reconstructed BS-maximum position for Auger-size datasets, given the measured anisotropy strength (see Sect.~4.2 of Paper~II).} This better compatibility is clearly linked to forcing the presence of Cen~A in every realization. To quantify this effect, we compare Model~II with a variant in which Cen~A is not forced. This variant, shown in the top panel of Fig.~\ref{BS_loc_mc2}, uses the same GMF model (UF23 Spur) but allows Cen~A to be included or not in each realization according to the usual random sampling; we refer to this case as Model~V.\footnote{Model~V differs from Model~I only by the GMF model (UF23 Spur instead of KST24); the realizations of the source distribution are otherwise identical in both models.} In Model~V, BS maxima close to the Auger direction can still be obtained, but with a lower probability, as illustrated by the cumulative distributions of angular distance shown in the lower panel of Fig.~\ref{BS_loc_mc2}. When Cen~A is not forced, the realizations yielding a BS maximum near the Auger value are predominantly those for which Cen~A and/or NGC~4945 happen to be selected. Forcing Cen~A reduces the cosmic variance and concentrates the BS maxima in a region relatively close to Cen~A.

For the other models displayed in Fig.~\ref{BS_loc_mc}, the agreement with the Auger BS maximum is less satisfactory. Models~III and IV, which assume lower source densities, do produce some realizations with a BS maximum close to the Auger position due to the larger cosmic variance in these cases. It can be noted that the resulting distributions of BS maxima resemble those obtained for the dipole directions in Fig.~\ref{dipoleLoc2}. 
In the case of model I none of the BS maxima found appear to lie close to the Auger maximum. The use of the KST24 GMF model associated with the assumed source distribution of model I thus appears incompatible with the data. With this GMF model, forcing the presence of Cen~A in the source distribution does not significantly alter this conclusion: as shown by the almost identical cumulative distributions in the lower panel of Fig.~\ref{BS_loc_mc2}, no realization in either case yields a BS maximum within $20^\circ$ of the Auger value, and both are clearly less compatible with the data than Models~II and~V.

In addition, Fig.~\ref{BS_MC} displays the scatter plots of the BS maximum significance, $\sigma_{\rm max}^{\rm BS}$, versus the maximum significance obtained in the targeted search toward Cen~A, $\sigma_{\rm max}^{\rm CenA}$, for the same models as in Fig.~\ref{BS_loc_mc}. The interpretation of these plots follows directly from the discussion of the BS-maximum locations. Concerning $\sigma_{\rm max}^{\rm BS}$, all models predict values compatible with the Auger measurement ($\sigma_{\rm max}^{\rm BS}\simeq 5.6$), although the Auger value lies toward the lower edge of their respective distributions. As expected, the spread of $\sigma_{\rm max}^{\rm BS}$ increases as the source density decreases, reflecting the larger cosmic variance.

Reproducing the pair $(\sigma_{\rm max}^{\rm BS},\,\sigma_{\rm max}^{\rm CenA})$ measured by Auger is significantly more challenging, as already noted in Paper~II. The fact that Auger finds $\sigma_{\rm max}^{\rm CenA}\simeq 5.1 \lesssim \sigma_{\rm max}^{\rm BS}$ implies that a flux excess nearly as strong as the BS maximum occurs specifically in the direction of Cen~A, which is difficult to reproduce in many astrophysical realizations. 
Model~II provides by far the best agreement: the observed value of $\sigma_{\rm max}^{\rm CenA}$ lies near the centre of the corresponding distribution for this model. The remaining models show a much poorer agreement, particularly Model~I, for which the predicted values of $\sigma_{\rm max}^{\rm CenA}$ are typically much lower than those measured by Auger. This reflects the fact that, under the KST24 GMF model, strong flux excesses are not expected in the direction of Cen~A, regardless of whether Cen~A (or NGC\,4945) is present in the source list.

For Models~III and~IV, the Auger value of $\sigma_{\rm max}^{\rm CenA}$ lies in the tail of the predicted distribution. Compatible pairs $(\sigma_{\rm max}^{\rm BS},\,\sigma_{\rm max}^{\rm CenA})$ can still be obtained, thanks to the very large cosmic variance associated with the lower source densities.

\begin{figure}
    \centering
   \includegraphics[width=8.5cm]{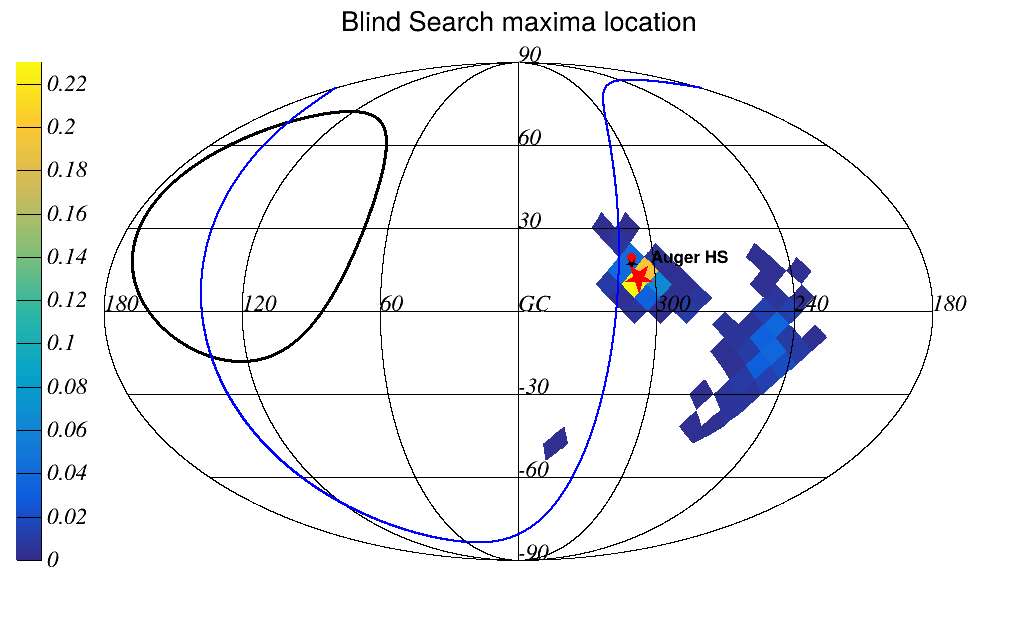}
   \includegraphics[width=7.5cm]{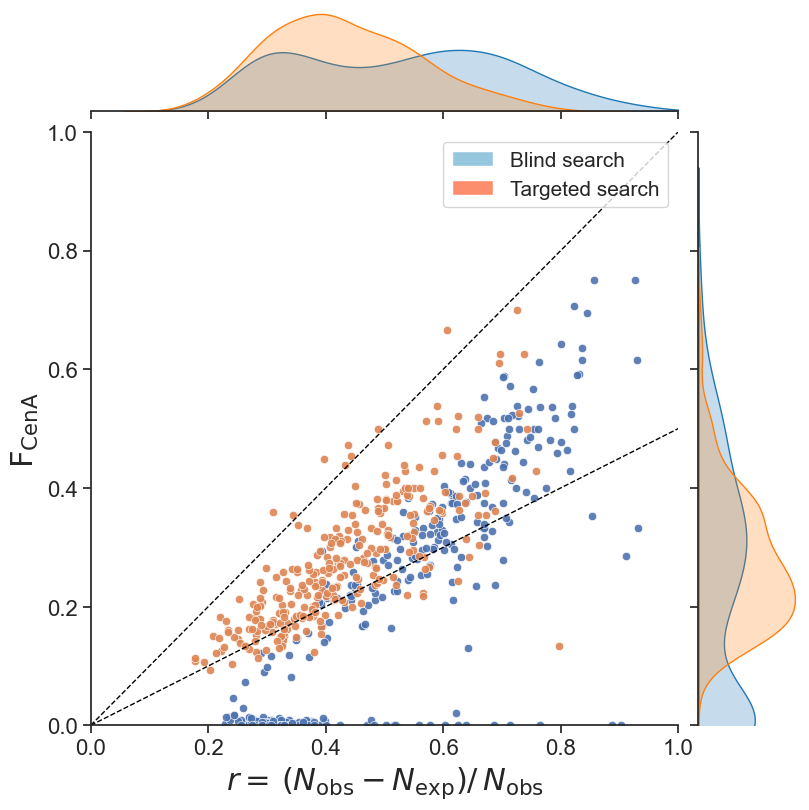}
  
      \caption{Top : distribution of the location of the blind search maximum obtained from 300 datasets generated for the selected realization of model II. Bottom : Scatter plot for the fractional flux excess $r$ in the angular window of the maximum excess versus the fractional contribution of CenA,  $F_{\rm CenA}$ in that window for the blind (red dots) and targeted (blue squares) searches. Dashed lines showing $F_{\rm CenA}=r$ and  $F_{\rm CenA}=1/2\times r$ are displayed to guide the eye. The individual distributions of the different quantities plotted are shown on top of the coordinate axis.
              }
         \label{BS_loc_selected}
   \end{figure}

\begin{figure}
    \centering
   \includegraphics[width=8.5cm]{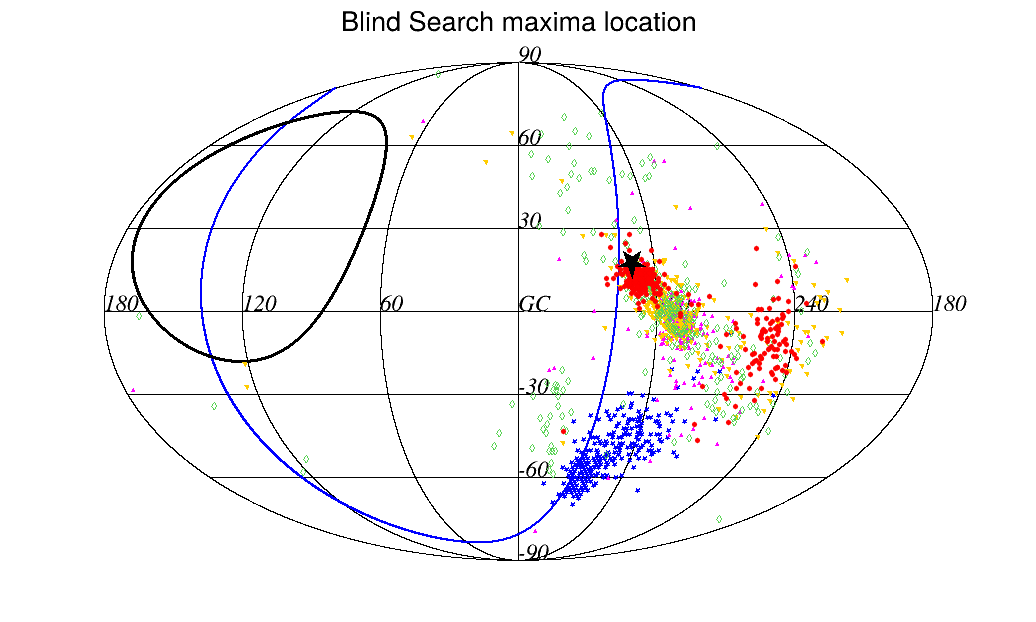}
   \includegraphics[width=7.cm]{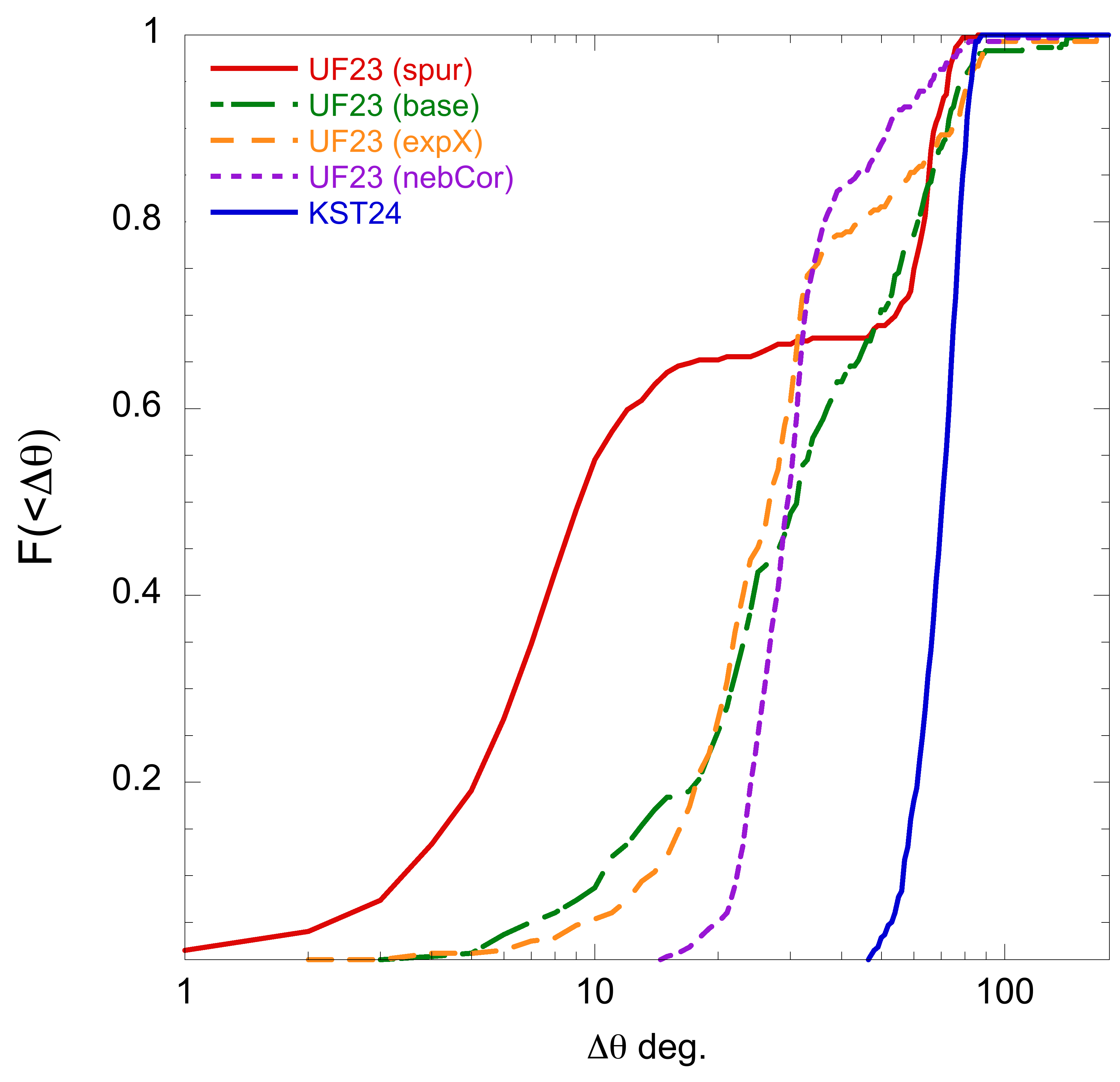}
  
      \caption{Top : locations of the BS maximum found for each of the 300 datasets generated for the selected realization of the source distribution chosen for model II, the Auger BS maximum is shown with a large black star. Various GMF models are represented with different colors and markers : UF23 spur (red dots), UF23 base (green open diamonds), UF23 expX (orange down triangles), UF23 nebCor (magenta up triangles), KST24 (blue stars). Bottom : corresponding cumulative fonctions of the angular distance between the prediction BS maximum and the one observed by Auger. The different GMF models are shown with same colors as in the top panel.  
              }
         \label{BS_loc_selected2}
   \end{figure}

\subsubsection{Selected realizations}

While the large cosmic variance of the predictions allows one to reproduce the observed anisotropy in some realizations of a given model, doing so simultaneously for the dipole and for the BS/targeted searches is considerably more constraining. We therefore tested whether the realizations selected in Sect.~4.1.4—i.e. those optimising agreement with the Auger dipole direction—also reproduce the small- and intermediate-scale anisotropies.

Figure~\ref{BS_Select} shows the scatter plots of $\sigma_{\rm max}^{\rm BS}$ versus $\sigma_{\rm max}^{\rm CenA}$ for the 300 datasets generated for the selected realizations of Models~I–IV. As expected from the discussion above, Model~I fails to reproduce the data: the flux excesses in the direction of Cen~A are far too small. The same holds for the selected realizations of Models~III and IV: although these realizations reproduce the large-scale anisotropies (dipole direction and amplitude), they do not produce BS or targeted-search signals consistent with Auger. In these cases the BS maximum generally lies near the dipole direction, which is far from Cen~A.

Only Model~II achieves a consistent match with the Auger values of both $\sigma_{\rm max}^{\rm BS}$ and $\sigma_{\rm max}^{\rm CenA}$. This makes Model~II the only one among our selected realizations that simultaneously reproduces the large-scale anisotropies and the BS/targeted-search results.

One might consider constructing additional realizations for lower source densities ($10^{-3.5}$ or $10^{-4}\,\mathrm{Mpc^{-3}}$) with Cen~A forced, in the hope that the enhanced cosmic variance could yield better agreement. However, this is not viable. At such low source densities, Cen~A—being very nearby—would overwhelmingly dominate the UHECR flux for the standard-candle sources assumed here, producing anisotropies far larger than those observed, even with an EGMF as large as $10$~nG. Moreover, a dominant source would attract the dipole toward Cen~A,contradicting Auger observations, and reduce the cosmic variance. Avoiding this would require additional hypotheses, such as a strong and structured EGMF in the local universe or significant GMF demagnification of the Cen~A region; none of the GMF models explored here exhibit the latter property.

The distribution of BS maxima for the selected realization of Model~II is shown in the upper panel of Fig.~\ref{BS_loc_selected}. About $60\%$ of the datasets have BS maxima within $15^\circ$ of the Auger value, and the "infinite-statistics" maximum (red star) lies within $10^\circ$. Approximately one third of the datasets exhibit BS maxima near the dipole direction at lower energies. This behaviour reflects a trend already emphasized in Paper~II (Sects.~4 and~8): in the type of models considered here, regions producing flux excesses around 8–10~EeV tend to remain relatively prominent at higher energies.

Regarding the parameters of the BS maximum, the preferred angular window $\psi$ and threshold energy $E_{\rm th}$ follow the patterns described in Paper~II: maxima tend to occur at the largest angular scales ($\psi \simeq 30^\circ$) and at threshold energies close to the lower edge of the scan. This trend is somewhat weaker in the selected realizations of Model II than in the cases studied in Paper~II: roughly one third of the datasets yield $\psi \leq 15^\circ$.

Concerning the composition of the flux excess in terms of source contributions, the bottom panel of Fig.~\ref{BS_loc_selected} displays the scatter plot of the fractional flux excess
\begin{equation}
    r=\frac{N_{\rm obs}-N_{\rm exp}}{N_{\rm obs}},
\end{equation}
(where \(N_{\rm obs}\) and \(N_{\rm exp}\) denote the observed and expected number of UHECR events within the angular window) versus the fractional contribution of  CenA  to that window. Results are shown for both the blind search (red dots) and the targeted search toward CenA (blue squares), using the 300 datasets generated for the selected realization of Model~II.

For the blind search, a distinct subset of the distribution exhibits a very small CenA contribution. These points correspond to the $\sim$1/3 of datasets whose BS maximum lies near the dipole direction rather than near CenA. In the remaining cases, CenA typically accounts for between one half and roughly 80\% of the flux excess. For the targeted search, the CenA contribution is, unsurprisingly, somewhat larger, generally lying between one half and the entirety of the excess.

These source-fraction patterns differ markedly from those found in Paper~II when using the JF12+Planck and Sun+Planck GMF models. In the former case, the flux excess was usually dominated by contributions from the Virgo cluster (see also \citealt{Ding2021}), whereas in the latter CenA contributed significantly but at a lower level than in Model~II considered here. From the point of view of identifying individual UHECR sources, Model~II is arguably more favourable; however, it is important to note that the models discussed in Paper~II—despite their different dominant contributors—also provided acceptable agreement with Auger data in both blind and targeted searches.

Finally, we tested the dependence of the BS predictions on the assumed GMF by applying the same procedures to the selected realization of Model~II under different GMF models. The BS maximum locations for each case are shown in Fig.~\ref{BS_loc_selected2} (top panel), with the corresponding cumulative distributions in the bottom panel. Relative to UF23 Spur, the predictions worsen substantially for each of the other UF23 models (base, expX, nebCor), with compatibility at best marginal. This does not necessarily imply that these GMF models cannot match Auger data, but rather that the fine tuning of the source distribution must be performed separately for each GMF model; without such tuning the agreement deteriorates. The KST24 model yields BS maxima very far from the Auger position, confirming that under this GMF model UHECRs from Cen~A experience large deflections in the relevant rigidity range \citep{KTS2025}. As already discussed by the authors of KST24, the Cen~A region lies near the transition between high- and low-deflection regimes and remains particularly uncertain. Future refinements of this GMF model may therefore alter these conclusions.

 \begin{figure*}
    \centering
   \includegraphics[width=8.5cm]{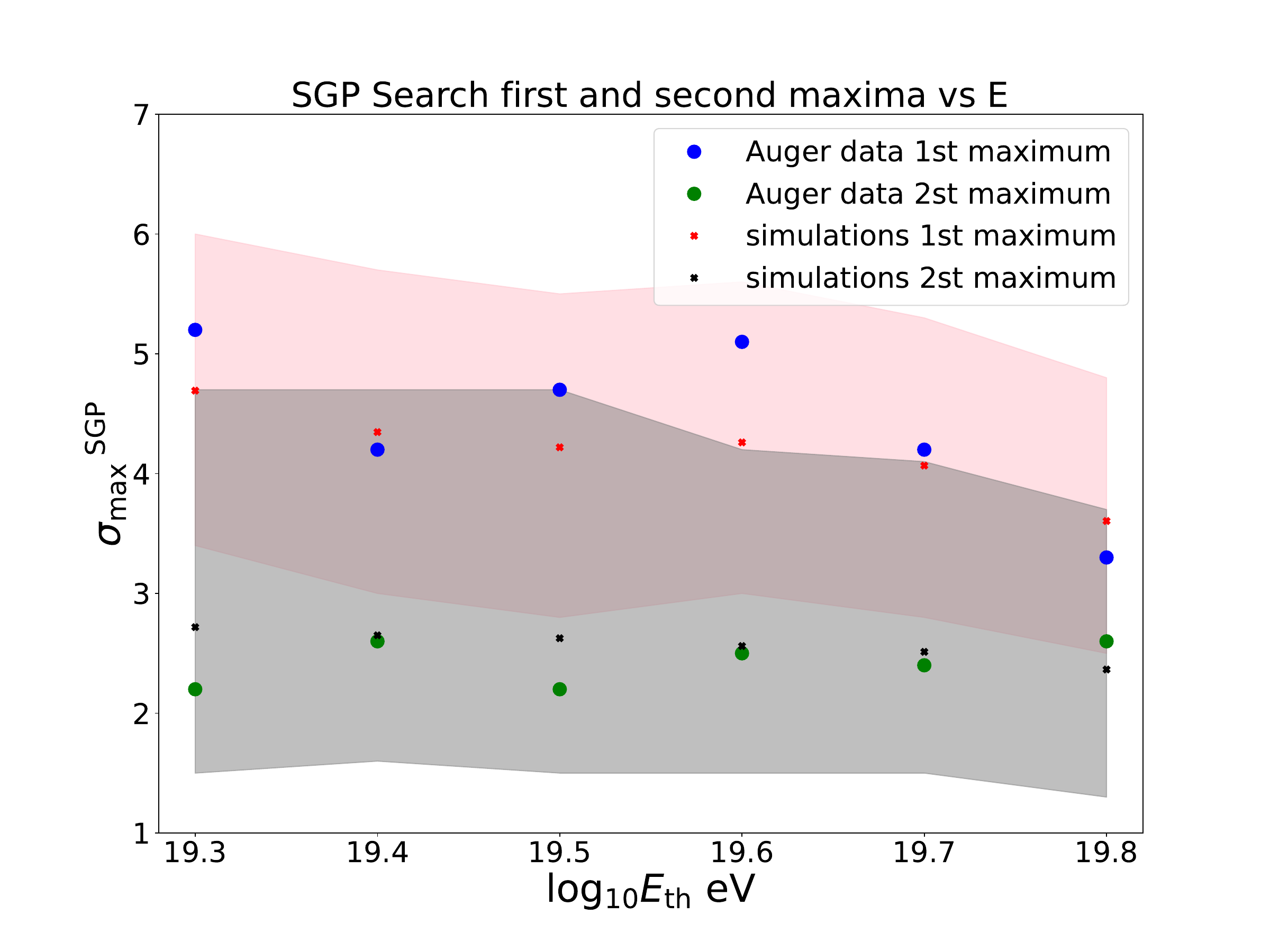}
   \includegraphics[width=8.5cm]{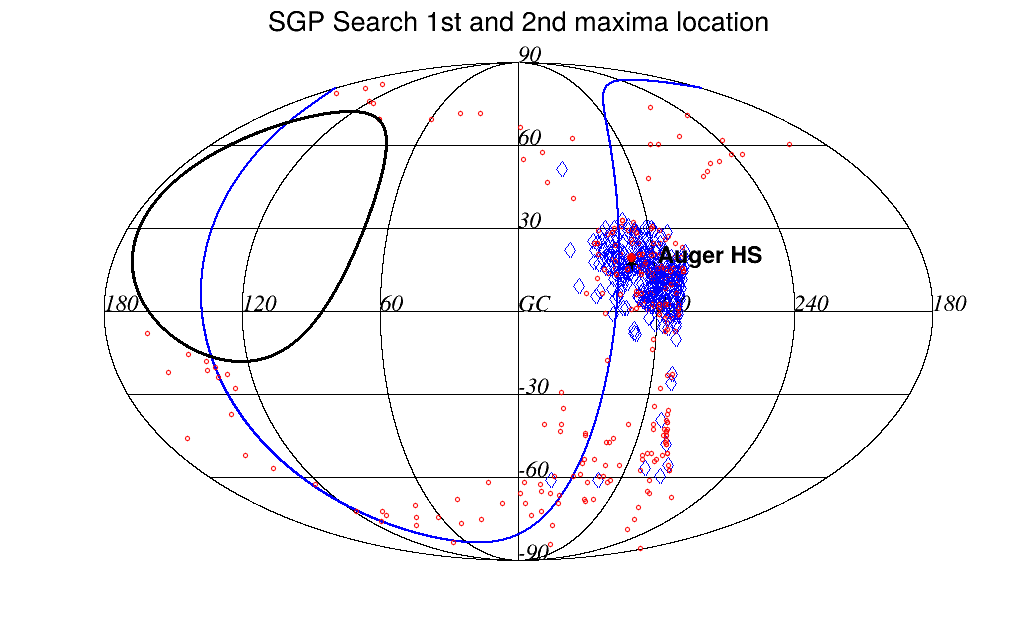}
  
      \caption{Top : Evolution of the maximum significance and of the second maximum as a function of the logarithm of the dataset energy threshold $E_{\rm th}$ for the simulated datasets computed for the selected realization of model II. The mean value obtained over 300 datasets is shown with red and black crosses for the maximum and second maximum respectively. The shaded areas show the range in which $90\%$ of the simulated datasets. The significances obtained with Auger data are show with large blue and green dots for the maxima and second maximum respectively. Bottom : Skymap in galactic coordinates of the maxima (blue open diamonds) and second maxima (red open circles) obtained for the 300 datasets generated for the selected reealization of model II.
              }
         \label{SGP_plots}
   \end{figure*}

\subsection{Search for a flux excess in the vicinity of the supergalactic plane}

We now turn to a more recent high-energy analysis released by Auger, namely the search for a flux excess in the vicinity of the supergalactic plane \citep{AugerSGP2025}. The method is very similar in spirit to the blind search: the angular window is fixed to $27^\circ$ (corresponding to the angular scale of the Auger BS maximum), and the scan over the dataset energy threshold is restricted to six values, taken as the closest integers in EeV to $\log_{10}(E/\mathrm{eV}) = 19.3$, 19.4, 19.5, 19.6, 19.7, and 19.8. In addition, the skymap scan is limited to supergalactic latitudes $-27^\circ \leq b_{\rm sgal} \leq 27^\circ$ and to declinations $\delta < 44.8^\circ$.

As in the Auger analysis, we searched our simulated datasets for the most significant excess using the above scanning procedure. After identifying the most significant window, a second maximum was determined by excluding a circular region of $27^\circ$ around the first maximum and repeating the scan. This yields, for each value of $E_{\rm th}$, both a primary and a secondary maximum within the restricted portion of the sky.

For clarity, we focus here on the predictions obtained for the selected realization of Model~II. The evolution of the significance of the first and second maxima as a function of $E_{\rm th}$, averaged over 300 datasets, is shown in the left panel of Fig.~\ref{SGP_plots}, together with the dispersion of the simulations. The mean significance decreases mildly with increasing energy threshold, although individual datasets display more irregular behaviour—much like the Auger observations (see Sects.~4 and~8 of Paper~II). Given the broad spread of the predictions, the agreement with the Auger data appears satisfactory for both the primary and secondary maxima.

The considerably lower mean significance of the second maximum in our simulations indicates that, for this particular model, only a single region in the vicinity of the supergalactic plane exhibits a strong flux excess—consistent with the Auger findings. This conclusion is reinforced by the spatial distribution of the maxima, shown in the right panel of Fig.~\ref{SGP_plots}: the first maxima cluster tightly around the region where Auger observes its excess (unsurprisingly, very close to the BS maximum), whereas the second maxima are dispersed throughout the scanned sky.

   \begin{figure*}
    \centering
   \includegraphics[width=7.5cm]{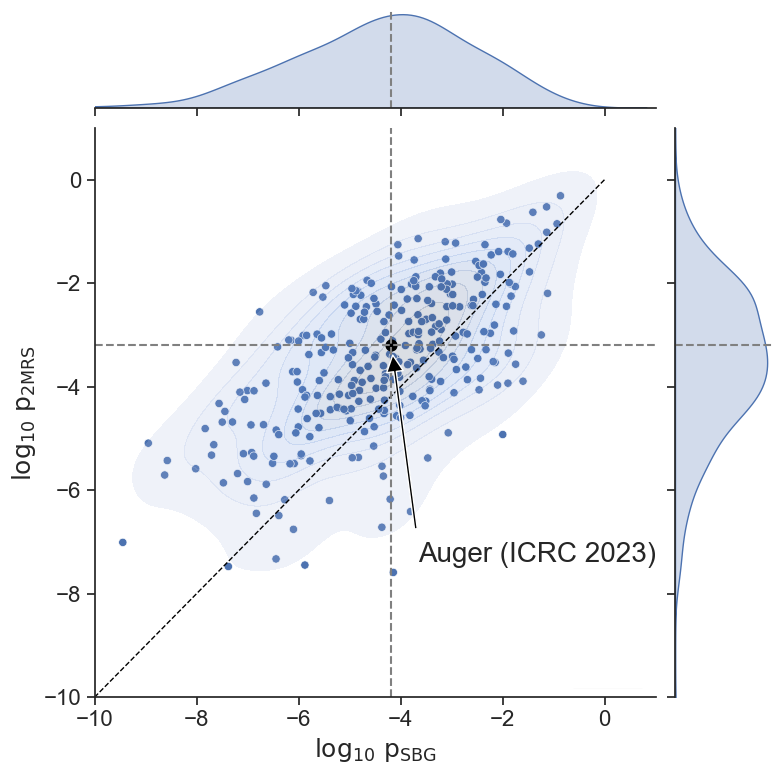}
   \includegraphics[width=7.5cm]{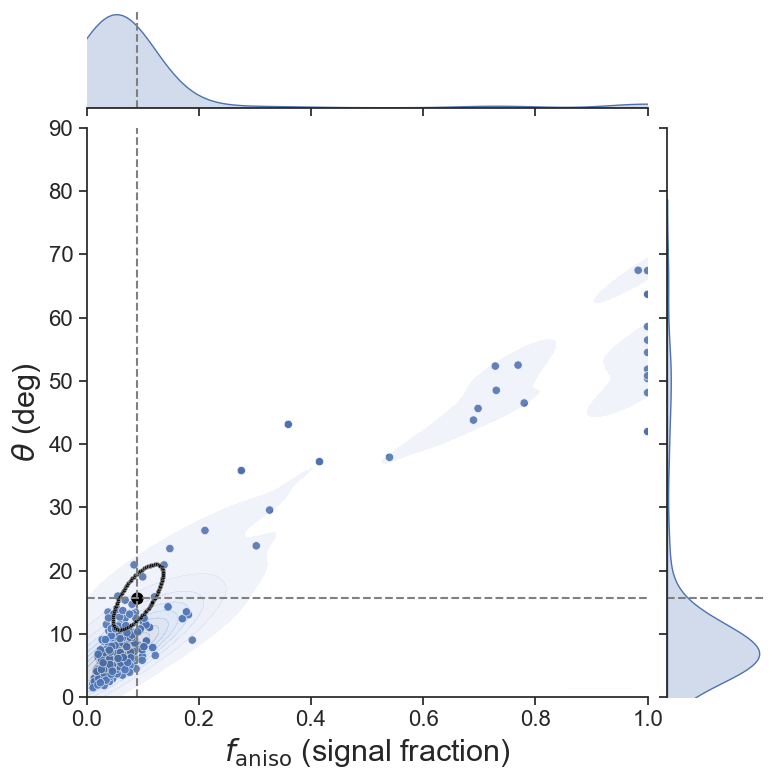}
  
      \caption{Left : p-values obtained after performing the likelihood analysis on our datasets. The logarithm of the p-value corresponding to the maximum likelihood obtained for the SBG catalog ($P_{\rm SBG}$) catalog is plotted against that obtained for 2MRS ($P_{\rm 2MRS}$) for each dataset. The astrophysical model considered to build our datasets corresponds to the selected realization of model II. The values reported by Auger in \citet{Golup2023} are shown with large black full circle. The individual distributions of the different quantities plotted are shown on top of the coordinate axis. Right : best-fit parameters obtained after performing the likelihood analysis for the SBG catalog on our datasets. The value of $f_{\rm aniso}$ and $\theta$ allowing to maximize the likelihood for each dataset are plotted against each other for the astrophysical model considered on the left panel. The 1$\sigma$ ellipse reported in \citet{Golup2023} from Auger data is shown. The individual distributions of the different quantities plotted are shown on top of the coordinate axis.
              }
         \label{likelihood_plots}
   \end{figure*}

\subsection{Likelihood analysis}

The Auger likelihood analysis \citep{AugerSFG2018,AugerAniso2022} was extensively discussed in Paper~II (see Sects.~5 and~6). In particular, we emphasized that the outcome of this analysis must be interpreted with great care when assessing which astrophysical catalog yields the lowest p-values—that is, the most significant apparent anisotropy. As shown in Paper~II, the preference for one catalog over another (the starforming/starburst galaxy catalog in the case of Auger data) can be driven not by the true nature of the UHECR sources, but rather by properties of the GMF, especially its magnification and demagnification effects across the sky.

In the present study we use exactly the same quantities and formalism as in Paper~II. We denote by $P_{\rm SBG}$ and $P_{\rm 2MRS}$ the p-values corresponding to the maximum-likelihood fits obtained using the starforming/starburst galaxy catalog and the ordinary galaxy catalog, respectively. The likelihood fit parameters, namely the anisotropic signal fraction $f_{\rm aniso}$ and the angular smearing scale $\theta$, are also extracted for each dataset (see Paper~II for details).

Here we restrict ourselves to the predictions obtained for the selected realization of Model~II, in order to assess whether this model also reproduces the Auger results for this analysis, and to examine whether the conclusions of Paper~II are modified by the use of the most recent GMF models.

The left panel of Fig.~\ref{likelihood_plots} shows the p-values associated with the maximum of the likelihood function for the SBG catalog (abscissa) and for the 2MRS catalog (ordinate). The p-values reported by Auger lie near the center of the corresponding simulated distributions for both $P_{\rm SBG}$ and $P_{\rm 2MRS}$. A large majority of the datasets ($\sim 80\%$) yield a lower value of $P_{\rm SBG}$ than of $P_{\rm 2MRS}$, in agreement with  Auger observation, even though the source distribution used in Model~II is drawn from a sampling of the 2MRS catalog (see Sect.~2.2). This behaviour is consistent with the findings of Paper~II for GMF models that predict a strong demagnification of the Virgo cluster region. This was the case for the Sun+Planck model in Paper~II, and it is also true for the UF23 spur model and for most members of the UF23 suite considered in this work.

The corresponding fit-parameter distributions are shown in the right panel of Fig.~\ref{likelihood_plots}. While many simulated datasets favor slightly smaller values of $f_{\rm aniso}$ than the Auger best fit, the predicted distributions for both $f_{\rm aniso}$ and $\theta$ remain compatible with the data. Similar agreement is obtained when using other GMF models from the UF23 suite (with the exception of the twistX model). This contrasts with Paper~II, where datasets generated with the JF12+Planck or Sun+Planck models tended to produce substantially larger best-fit values of both parameters. The overall consistency between the simulations and the Auger likelihood results is therefore improved by adopting the recent UF23 GMF models.

Finally, we note that imposing a source distribution explicitly biased toward starforming or starburst galaxies—i.e. forcing the presence of the main SBG sources used in the Auger analysis, as done in Sect.~7 of Paper~II—yields a significantly worse agreement with the data when combined with the newer GMF models. This reinforces our earlier conclusion: the apparent preference of the Auger likelihood analysis for the SBG catalog should not be interpreted as evidence regarding the true nature of the UHECR sources.

\begin{figure}
    \centering
   \includegraphics[width=8.5cm]{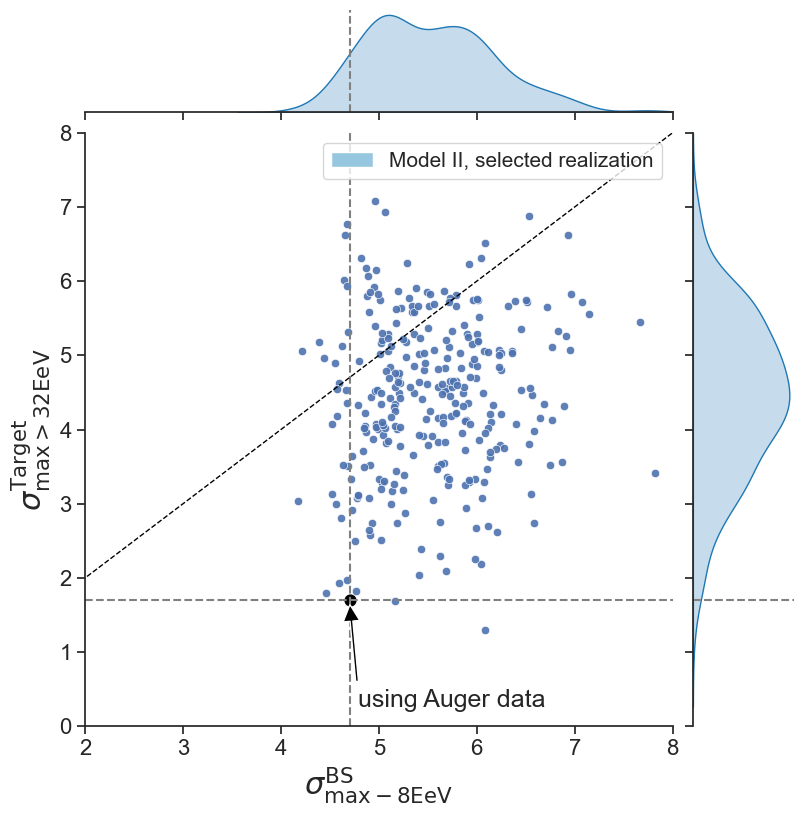}

      \caption{Scatter plot of the maximum significances for $E_{\rm th} > 32$~EeV of the flux excess in the direction of the BS maximum at 8~EeV, $\sigma^{\rm target}_{\rm max>32EeV}$ vs. the corresponding value of $\sigma^{\rm BS}_{\rm max-8EeV}$ (see text). The two values of $\sigma_{\max}$ obtained with the Auger data are indicated by the black dot. The individual distributions of the different quantities plotted are shown on top of the coordinate axis.
              }
         \label{BS_lowE}
   \end{figure}

\subsection{Energy evolution of anisotropy signals}

While the use of the most recent GMF models alleviates some of the discrepancies reported in Papers~I and~II when comparing simulations with observations, several conclusions drawn in those studies regarding the energy evolution of the anisotropy observables remain largely independent of the assumed GMF model.

In Paper~II, we emphasized the interest of extending the blind and targeted searches to energies below the conventional threshold of 32~EeV used by Auger. In models that reproduce both the UHECR spectrum and composition measured by Auger—where the rigidity evolves only mildly with energy due to the changing mass composition—and assuming standard-candle sources, we showed that more significant (and therefore better characterized) anisotropy signals are very likely to appear at lower energies (see Sect.~8 of Paper~II).

Because the Auger dataset below 32~EeV is not publicly available beyond the one used for the dipole study of \citet{AugerDip2017}, and because that dataset does not include individual event energies (only that $E_{\rm th}=8$~EeV for the full set), we could not perform a full two-parameter scan on real data. Instead, we scanned only over angular scales, $\psi \in [1^\circ, 30^\circ]$ in steps of $1^\circ$, keeping $E_{\rm th}=8$~EeV fixed. The most significant excess was found at the largest angular scale, $\psi = 30^\circ$, with a significance $\sigma^{\rm BS}_{\rm max-8EeV} \simeq 4.7$, located at $(l,b) = (267^\circ, -44^\circ)$ in Galactic coordinates. As noted in Paper~II (Sect.~9), a targeted search at this exact position using the dataset above 32~EeV yielded a strikingly small maximum significance, $\sigma^{\rm target}_{\rm max>32EeV} \sim 1.7$.

To quantify how often such a drop in significance occurs in our simulations, we repeated the BS analysis at $E_{\rm th}=8$~EeV for each dataset generated for the selected realization of Model~II, recording both the significance and location of the BS maximum. A targeted search was then performed, scanning over $E_{\rm th}$ and $\psi$ but keeping the sky position fixed to that found at 8~EeV, and the corresponding maximum significance $\sigma^{\rm target}_{\rm max>32EeV}$ was recorded. The resulting scatter plot of $\sigma^{\rm target}_{\rm max>32EeV}$ versus $\sigma^{\rm BS}_{\rm max-8EeV}$ is shown in Fig.~\ref{BS_lowE}. The Auger measurement is indicated by a black dot.

As already found in Paper~II, the very low value of $\sigma^{\rm target}_{\rm max>32EeV}$ observed by Auger is extremely uncommon in the simulations: only two out of 300 datasets yield a lower value. This result is remarkably similar to that obtained in Fig.~19 of Paper~II, despite the use of a different realization of the source distribution and different GMF models. The potential implications of this discrepancy—which include, for instance, a possible signature of source-to-source variability in UHECR output, or the disappearance of a low-energy anisotropy contribution above 32~EeV—are discussed more extensively in Paper~II and are not altered by adopting the newer GMF models.

We stress that performing this analysis with the full Auger dataset (including energies below 32~EeV) would be extremely valuable. More generally, as noted in Paper~I, the energy evolution of other anisotropy observables—such as the dipole direction—would provide powerful constraints on UHECR source models once larger statistics become available, again with only weak dependence on the assumed GMF.

\section{Discussion and conclusion}

In the present study, we have revisited the Auger anisotropy analyses in the spirit of Papers~I and~II, now incorporating the most recent GMF models available. The main new result is the significantly improved agreement between simulations and observations regarding the dipole direction when assuming that UHECR sources follow the distribution of normal galaxies and adopting the newer GMF models – most notably KST24 and several members of the UF23 family. We also find an improved description of the high-energy anisotropy signals, in particular through the likelihood-fit parameters. This overall improvement is encouraging, especially since the GMF models used here are based on more recent data and modelling techniques and are therefore expected to provide a more realistic description of the Galactic field than the previous generation of models.

Despite this progress, the main conclusions of Papers~I and~II remain essentially unchanged. With current data and current theoretical knowledge of candidate UHECR sources, it remains extremely difficult to draw robust conclusions about the origin of UHECRs or the nature of their accelerators. Although a source distribution following that of normal galaxies appears more compatible with the data than before, the existence of possible biases – potentially large ones – between the distribution of galaxies and that of UHECR sources remains nearly impossible to constrain without introducing additional (currently poorly motivated) assumptions.

The good agreement obtained with some selected realization in which Cen~A is forced to be present may be taken as mild evidence supporting a contribution of that nearby AGN to the UHECR flux. However, present data do not allow us to favour this scenario in any definitive manner over alternatives involving, for example, NGC~4945, M\,83, Circinus, or even the more distant Centaurus cluster. The first three objects could contribute in scenarios in which UHECR sources track starforming or starburst galaxies\footnote{While we argued that the Auger likelihood analysis does not provide strong evidence for such a scenario, it nevertheless remains a viable possibility.} \citep{Anc1999}. The Centaurus cluster, on the other hand, could be prominent if UHECRs are accelerated in cluster accretion shocks \citep{Norman1995,Inoue2007,Simeon2025}, especially if the Virgo region is indeed demagnified by the GMF. In our view, these persistent degeneracies arise from three principal limitations.

\smallskip
\noindent(i) Magnetic-field uncertainties.
The limited knowledge of cosmic magnetic fields, both Galactic and extragalactic, continues to prevent robust interpretation of UHECR anisotropy data. The dependence of the predictions on the assumed GMF model remains substantial in the rigidity range most relevant for UHECR anisotropies (likely between $\sim 5$ and $\sim 10$\,EV; see Paper~I), particularly when considering the UHECR image of a given source \citep{Unger2024,KTS2025}. Interpreting anisotropies or assessing possible biases between UHECR sources and galaxies requires reliable knowledge of both magnetic deflections and the magnification/demagnification patterns across the sky, especially in regions of high nearby galaxy density such as Virgo.

In this respect, the various state-of-the-art GMF models differ significantly. In the rigidity range of interest, the Virgo region is strongly demagnified in the UF23 models, whereas JF12 predicts nearly neutral magnification with moderate deflections, and KST24 predicts nearly neutral magnification but much larger deflections. Even within the UF23 suite, the rigidity at which neutral magnification is recovered varies across models (see Fig.~\ref{Magnif}), directly impacting anisotropy predictions for a fixed source-distribution hypothesis. Further narrowing the range of predictions among realistic GMF models remains critical for meaningful interpretation of UHECR anisotropy data.

Concerning extragalactic magnetic fields, we included an EGMF in our simulations, but the homogeneous and isotropic turbulent configurations we adopted (with strengths between $10^{-2}$ and 10\,nG, effectively reduced to 1\,nG for the examples shown) imply that the EGMF does not play a dominant role in shaping the anisotropy patterns (see Papers~I and~II). However, the presence of strong and structured fields of order a few tens of nG in the local ($\lesssim$ few Mpc) Universe could considerably modify the amplitude and morphology of the anisotropy. While current observations provide little evidence for such fields outside cluster environments, several recent studies explore scenarios with strong local EGMFs \citep[e.g.][]{Matthews2018,Mollerach2019,Marafico2024}. Cosmological MHD simulations \citep{Dolag2002,Sigl2002,Hack2018,Kim2019} also predict structured fields in groups and filaments, but the filling factors of high-field regions remain uncertain. If such fields exist and substantially affect UHECR propagation, strong observational constraints on them will be essential, particularly for transient-source scenarios, where the time spread of UHECR arrivals from a given source on Earth is a key ingredient.

\smallskip
\noindent(ii) Limited knowledge of the UHECR source physics.
While numerical simulations of the microphysics of cosmic-ray acceleration have made substantial progress in recent years, further advances in the theoretical modelling of potential UHECR sources remain necessary to narrow down the wide range of astrophysical free parameters that still characterize current scenarios \citep[for a recent review, see][]{No2025}. In parallel, improved multiwavelength constraints on the environments of candidate sources are required. These are expected from the next generation of observatories, for instance in $\gamma$-rays with LHAASO \citep{LHAASO2023} or CTA \citep{CTA2023}, in X-rays with ATHENA \citep{ATHENA2022}, or in radio with SKA \citep{SKA2022}. Together, these developments should help constrain the UHECR luminosity, spectrum, and composition that can be expected from the various source classes.

\smallskip
\noindent(iii) Insufficient UHECR statistics at the highest energies.
The current UHECR statistics above $\sim 50$--100~EeV remain insufficient to characterize anisotropy patterns with the level of precision needed to constrain source populations, especially given the intrinsically weak anisotropies expected for heavy-dominated compositions at the highest energies. This limitation is even more severe in the northern hemisphere, where TA observes the sky with considerably smaller exposure than Auger. The TA hotspot has not increased in significance in recent years, and its interpretation as a genuine astrophysical signal is now challenged by Auger data near the overlap region \citep{AugerSGP2025}. The commissioning of TA$\times$4 \citep{TA*4ICRC2019} will help clarify the issue, but neither Auger nor TA$\times$4 will be able to provide the statistics required to trace anisotropy evolution from 10 to above 100~EeV or to reach decisive significance at the highest energies (see Paper~II). Only a new generation of UHECR observatories \citep{Coleman2023}, with much larger exposures and ideally full-sky coverage, can meet these requirements. This could be achieved through next-generation ground-based arrays such as the proposed GCOS \citep{GCOS2021} or GRAND \citep{GRAND2020}, or through a space-based mission such as POEMMA \citep{POEMMA2021} or another mission emerging within the JEM-EUSO program \citep{JEMEUSO2023}.

In summary, the interpretation of UHECR anisotropies remains limited by magnetic-field uncertainties, source-physics ambiguities, and statistical constraints; encouraging progress has recently been achieved in several of these areas and should be further pursued through the combined development of theory, multiwavelength observations, and future large-exposure UHECR facilities.

\begin{acknowledgements}
      This work has been supported by the INTERCOS master project of IN2P3.
\end{acknowledgements}

%

\begin{thebibliography}{}
  
 
 \bibitem[Aab et al.(2015a)]{AugerDipQuad2015}
   Aab, A.,
   Abreu, P., Aglietta, M., et al. (Pierre Auger Collaboration), 2015,
   ApJ, 802, 111 

\bibitem[Aab et al.(2015b)]{AugerAni2015} Aab, A.,
   Abreu, P., Aglietta, M., et al. (Pierre Auger Collaboration), 2015, ApJ. 804, 15 [arXiv:1411.6111]

   \bibitem[Aab et al.(2017a)]{AugerMulti2017}
   Aab, A.,
   Abreu, P., Aglietta, M., et al. (Pierre Auger Collaboration) 2017d,JCAP 06 026, [arXiv: 1611.06812]


    \bibitem[Aab et al.(2017b)]{AugerDip2017}Aab, A.,
   Abreu, P., Aglietta, M., et al. (Pierre Auger Collaboration), 2017,
   Science 357, [arXiv: 1709.07321]
  
   \bibitem[Aab et al.(2018)]{AugerSFG2018}Aab, A.,
   Abreu, P., Aglietta, M., et al. (Pierre Auger Collaboration), 2018a,ApJL 853:L29, [arXiv: 1801.06160]

    \bibitem[Abdul Halim et al.(2025)]{AugerSGP2025} Abdul Halim, A., Abreu, P., Aglietta, M., et al. (Pierre Auger Collaboration), 2025, ApJ 984 2, 123


    \bibitem[Abdul Halim et al.(2024)]{AugerDip2024} Abdul Halim, A., Abreu, P., Aglietta, M., et al. (Pierre Auger Collaboration), 2024, ApJ 976 1, 48





     \bibitem[Abbasi et al.(2014)]{TASpot2014}Abbasi, R. U., Abe, M., Abu-Zayyad, T., et al. (Telescope Array Collaboration), 2014, ApJL, 790, L21  

    \bibitem[Abreu et al.(2022)]{AugerAniso2022}
   Abreu, P., Aglietta, M., Albury, J. M., et al. (Pierre Auger Collaboration), 2022, ApJ 935, 170, [arXiv:2206.13492] 
   
     \bibitem[Abraham et al.(2004)]{AugerObs} Abraham, J.  et al. (Pierre Auger Collaboration) 2004, Nucl. Instrum. Methods Phys. Res., Sect. A 523 50.

     \bibitem[Adam et al.(2016)]{GMFPlanck2016}Adam, R.  Ade, P. A. R.,  Alves, M. I. R., et al. (Planck Collaboration), 2016, A\&A 596, A103.

    
     \bibitem[Allard et al.(2022)]{Paper I}
   Allard, D.,
   Aublin, J., Baret, B.\& Parizot, E., 2022, A\&A 664, A120, [arXiv: 2110.10761]

    \bibitem[Allard et al.(2024)]{Paper II}
   Allard, D., Aublin, J., Baret, B.\& Parizot, E., 2024, A\&A 686, A292, [arXiv: 2305.17811]  

\bibitem[Alvarez-Muniz et al.(2020)]{GRAND2020} Alvarez-Muniz, J.,  Alves Batista, R., Balagopal A.~V., et al. (GRAND Collaboration), 2020, Sci. China-Phys. Mech. Astron. 63, 219501, arXiv:1810.09994  
   
\bibitem[Anchordoqui et al.(1999)]{Anc1999}Anchordoqui, L. A., Romero, G. E. \& Combi, J. A., 1999, Phys. Rev. D, 60, 103001
 \bibitem[Aublin \& Parizot(2005)]{Aublin2005}Aublin J. \& Parizot E. 2005, A\&A, 441, 407.

\bibitem[Bister \& Farrar(2025)]{Bister2025} Bister, T., \& Farrar, G. R., 2025, ApJ, 966, 71

\bibitem[Cao et al.(2022)]{LHAASO2023} Cao Z., della Volpe D., Liu S., et al. (LHAASO collaboration), 2022, Chinese Physics C , Vol. 46, No. 3, 035001-035007, arXiv:1905.02773 [astro-ph.HE]

\bibitem[Casolino et al.(2023)]{JEMEUSO2023}Casolino, M., Parizot E., et al. (JEM-EUSO collaboration), 2023, , proceedings of the 37th ICRC,  Nagoya (Japan), arXiv:2310.02624v1.

\bibitem[Coleman et al.(2023)]{Coleman2023}Coleman, A., Eser J., Mayotte E., Sarazin, F., et al., 2023, Astropart. Phys., V149, 102819 

\bibitem[Dewdney et al.(2022)]{SKA2022}Dewdney, P., Labate, M. G., Swart, G., et al., 2022, SKA1 Design Baseline Description (SKA-TEL-SKO-0001075, Revision 02). SKAO. https://doi.org/10.5281/zenodo.16895574

\bibitem[Ding et al.(2021)]{Ding2021}  Ding, C., Globus, N.,\& Farrar, G., 2021, ApJ 913, L13

\bibitem[Dolag et al.(2002)]{Dolag2002}Dolag, K., Bartelmann, M., \& Lesch, H. 2002, A\&A, 387, 383



    \bibitem[Globus et al.(2019)]{No2019}Globus, N., Piran, T.,  Hoffman, Y., et al. 2019, Mon. Not. R. Astron. Soc. 484, 4167. 

   \bibitem[Globus \& Blandford (2025)]{No2025}Globus, N. \& Blandford, R., 2025, Ann.Rev.Astron.Astrophys., 63 1, 339-377

    \bibitem[Golup et al.(2023)]{Golup2023}Golup, G., et al. (Pierre Auger Collaboration), 2023, proceedings of the 37th ICRC,  Nagoya (Japan), POS(ICRC2023), 252.

  \bibitem[Hackstein et al.(2018)]{Hack2018}Hackstein S., Vazza F., Br\"uggen M., et al. 2018, MNRAS, 475, 2519

    \bibitem[He et al.(2016)]{He16}He H. , Kusenko A.,  Nagataki S., et al.,  2016, Phys.Rev. D93, 043011


    \bibitem[Hoffman et al.(2018)]{LSSS2018}Hoffman, Y., Carlesi, E., Pomar\`ede, D., et al., 2018, Nature Astronomy, vol 2, p 680-687 

\bibitem[Hoffman \& Zanin (2023)]{CTA2023}Hofmann W. \& Zanin R., 2023, "Handbook of X-ray and Gamma-ray Astrophysics" by Springer (Eds. C. Bambi and A. Santangelo), arXiv:2305.12888 [astro-ph.IM]
    
\bibitem[H\"orandel et al.(2021)]{GCOS2021}H\"orandel, J.~R., et al., 2021, proceedings of the 37th ICRC,  Berlin (Germany), POS(ICRC2021), 027    
    \bibitem[Huchra et al.(2012)]{Huchra2012}Huchra, J. P., Macri, L. M., Masters, K. L., et al. 2012, ApJS, 199, 26

    \bibitem[Inoue et al.(2007)]{Inoue2007}Inoue, S., Sigl, G., Miniati,F., et al. 2007, astro-ph/0701167 
    
    \bibitem[Jansson \& Farrar (2012a)]{JF2012a}Jansson, R., \& Farrar, G. R., 2012, ApJ, 757, 14
    
\bibitem[Jansson \& Farrar(2012b)] {JF2012b}Jansson, R., \& Farrar, G. R., 2012, ApJL, 761, L11.


\bibitem[Kido et al.(2019)]{TA*4ICRC2019}Kido, E. et al. (Telescope Array Collaboration), 2019, proceedings of the 36th ICRC,  Madison (USA), POS(ICRC2019), 312. 

\bibitem[Kim et al.(2019)]{Kim2019}Kim, J., Ryu, D., Kang, H., Kim, S.,  Rey, S.-C.,  2019, Science Advances,  V5, issue 1

\bibitem[Kim et al.(2021)]{TAICRC2021}Kim, J., Ivanov, D., Kawata, K, et al. (Telescope Array Collaboration), 2021, proceedings of the 37th ICRC,  Berlin (Germany), POS(ICRC2021), 328. 

  \bibitem[Korochkin et al.(2024)]{KTS2024} Korochkin, A., Semikoz, D., \&  Tinyakov, P., 2024, A\&A 693, A284, [arXiv:2407.02148] 

  \bibitem[Korochkin et al.(2025)]{KTS2025} Korochkin, A., Semikoz, D., \&  Tinyakov, P., 2025, [arXiv:2501.16158]

\bibitem[Linsley (1975)]{Linsley1975}Linsley, J,. 1975, PhRvL, 34, 1530

\bibitem[Marafico etl al.(2024)]{Marafico2024} Marafico S., Biteau J., Condorelli A. , Deligny 0. , Bregeon J., 2024, ApJ, 972 1, 4

\bibitem[Matthews et al.(2018)]{Matthews2018}Matthews, J. H., Bell, A. R.,  Blundell, K. M.,  Araudo, A. T., 2018, MNRAS Let., V479, Issue 1, L76
\bibitem[Norman et al.(1995)]{Norman1995}Norman C. A.,  Melrose D. B. \& Achterberg, A., 1995,  ApJ. 454, 60

\bibitem[Mollerach \& Roulet(2019)]{Mollerach2019}Mollerach, S., \& Roulet, 2019, Phys.~Rev.~D, 99, 103010.

\bibitem[Nandra et al.(2022)]{ATHENA2022}Nandra K., Barcons X., Barret D., A. Fabian, et al., 2022, ATHENA white paper, https://api.cloud.ifca.es:8080/swift/v1/ACO/Publications/The\_Hot\_and\_Energetic\_Universe.pdf

 \bibitem[Olinto et al.(2021)]{POEMMA2021}Olinto, A. V., Krizmanic, J., Adams,  J. H., et al., 2021, JCAP 06, 007, [arXiv:2012.07945]

    \bibitem[Pfeffer et al.(2016)]{Pfeffer16}Pfeffer, D.N.,  Kovetz E.D.,  Kamionkowski M., 2015, [arXiv:1512.04959]

   \bibitem[Rouill\'e d'Orfeuil et al.(2014)]{BRDO2014}Rouill\'e d‚ÄôOrfeuil, B., Allard, D., Lachaud, C., et al., 2014, A\&A, 567, 81.

\bibitem[Schechter (1976)]{Schechter1976}Schechter P., 1976, ApJ 203, 297

 \bibitem[Sigl et al.(2002)]{Sigl2002}Sigl, G., Miniati, F., \& En{\ss}lin, T. A. 2002, Phys. Rev. D, 68, 043002

\bibitem[Simeon et al.(2025)]{Simeon2025}Simeon P., Globus N., Barrow K. \& Blandford R., 2025, arXiv : 2503.10795 [astro-ph.HE]

\bibitem[Sun et al.(2008)]{Sun2008}Sun, X.~H., Reich, W., Waelkens, A., \& En{\ss}lin, T. A. 2008, A\&A, 477, 573
    
\bibitem[Sun et al.(2010)]{Sun2010}Sun, X.~H., \& Reich, W. 2010, Res. Astron. Astrophys., 10, 1287

 \bibitem[Tully et al.(2014)]{CF22014}Tully R. B., Courtois H., Hoffman Y., Pomar\`de D., 2014, Nature, 513, 71

 \bibitem[Unger \& Farrar(2024)]{Unger2024}Unger, M. \& Farrar, G., 2024,  ApJ, 970, 95, [arXiv:2311.12120] 

 \bibitem[Xu \& Han (2024)]{Xu2024} Xu, J. \& Han, J.~L., 2024, ApJ, 966, 240
  
\end{thebibliography}
%

\end{document}